\documentclass[twocolumn,american,prx,superscriptaddress]{revtex4-2}
\usepackage[T1]{fontenc}
\usepackage[utf8]{inputenc}
\usepackage{array}
\usepackage{bm}
\usepackage{multirow}
\usepackage{amsmath}
\usepackage{amssymb}
\usepackage{graphicx}

\makeatletter

\providecommand{\tabularnewline}{\\}

\usepackage{babel}

\makeatother

\usepackage{babel}
\begin{document}
\title{Nonreciprocal constitutive laws for oriented active solids }
\author{Balázs Németh}
\email{bn273@cam.ac.uk}

\affiliation{Department of Applied Mathematics and Theoretical Physics, Centre
for Mathematical Sciences, University of Cambridge, Wilberforce Road,
Cambridge CB3 0WA, United Kingdom}
\author{Takuya Kobayashi}
\email{kobayashi@cheme.kyoto-u.ac.jp}

\affiliation{Department of Chemical Engineering, Kyoto University, Kyoto 615-8510,
Japan}
\author{Ronojoy Adhikari}
\email{ra413@cam.ac.uk}

\affiliation{Department of Applied Mathematics and Theoretical Physics, Centre
for Mathematical Sciences, University of Cambridge, Wilberforce Road,
Cambridge CB3 0WA, United Kingdom}
\begin{abstract}
We present an overdamped continuum description of oriented active
solids in which interactions respect the symmetries of space but do
not obey the principle of action and reaction. Taking position and
orientation as kinematic variables, we examine the conservation of
the linear and angular momentum variables in an elementary volume.
We find that nonreciprocal interactions yield, in addition to the
areal stresses and moment stresses of classical elasticity, volumetric
forces and torques that act as local sources of momentum and angular
momentum. Since, by symmetry, these can only depend on the strains,
nonreciprocity requires the extension of constitutive modeling to
strain-dependent volumetric forces and torques. Using Cartan's method
of moving frames and Curie's principle, we derive the materially linear
constitutive law that underpins the nonreciprocal, geometrically nonlinear
elasticity of the continuum. We study this constitutive law exhaustively
for a one-dimensional active solid and identify striking nonreciprocal
effects -- traveling waves, linear instabilities, spontaneous motion
of and about the center of mass -- that are absent in a passive,
reciprocally interacting solid. Numerical simulations of a particulate
active solid model, consisting of a linear assembly of hydrodynamically
interacting active particles, yields long-wavelength behavior that
is in excellent agreement with theory. Our study provides the foundation
for a principled macroscopic mechanics of oriented active solids with
symmetry-invariant, nonreciprocal microscopic interactions.
\end{abstract}
\maketitle

\section{Introduction}

Newton's third law of action and reaction, together with its continuum
analogue---Cauchy's fundamental lemma---forms a cornerstone of the
mechanics of particles and continua. Cauchy's lemma establishes the
existence of a stress field in a mechanical continuum and enables
the closure of the linear momentum balance equation through constitutive
laws that relate the stress to measures of deformation, such as strains
and strain rates. 

It has long been recognized, however, that particle interactions mediated
by a field need not satisfy the action--reaction principle. For example,
in electromagnetism, the force between two charges is equal and opposite
but does not act along the line joining their centers \citep{griffiths_introduction_2023}.
In low-Reynolds-number hydrodynamics, the interaction between two
spheres settling at different velocities in a viscous fluid is neither
equal nor opposite \citep{smoluchowski_mutual_1911}. Recent research
has confirmed that such violations of the third law are far from rare:
many field-mediated interactions, such as diffusiophoretic, electrophoretic,
optical or plasmic, exhibit this property \citep{buenzli_violation_2008,dholakia_colloquium_2010,ivlev_complex_2012,soto_self-assembly_2014,ivlev_statistical_2015,saha_pairing_2019,zhang_active_2021}.
For instance, the hydrodynamic forces and torques between two squirmers
generally fail to satisfy the action--reaction principle \citep{singh_universal_2016,singh_generalized_2018}.

In all these cases, however, the interactions remain invariant under
isometries of space. For example, if a pair of squirmers is translated
and rotated together---while preserving their relative separation
and orientation---their hydrodynamic interaction remains unchanged,
provided they are far from boundaries \citep{oseen_neuere_1927,happel_low_1983,pozrikidis_boundary_1992}.
Motivated by this, we focus on particulate systems whose interactions
respect spatial symmetries but violate Newton’s third law, and examine
the consequences for their coarse-grained continuum description.

Following Cauchy, we consider an elementary volume and sum the forces
arising from both inside and outside the volume. For reciprocal interactions,
internal forces cancel in pairs, leaving only the forces transmitted
across the boundary. This\emph{ areal }contribution, as Cauchy shows,
can be expressed in terms of a symmetric stress tensor \citep{love_treatise_2013}.
For nonreciprocal interactions, however, the internal forces do not
cancel, leaving an additional \emph{volumetric} contribution that
must be included in the momentum balance. We show that, when the nonreciprocal
interactions respect spatial symmetries, and a solid-like response
is assumed, this volumetric term can depend only on the strain. The
Curie principle \citep{curie_sur_1894} can then be invoked to classify
the material parameters that appear in the constitutive laws according
to their symmetry. This framework yields a coarse-grained continuum
description for elastic systems of point-like particles with nonreciprocal
interactions.

In many situations, especially when particulate degrees of freedom
encompass not only position and momentum but also orientation and
angular momentum, it is necessary to extend beyond Cauchy elasticity.
For example, in a suspension of squirmers, the orientation of each
squirmer shapes the surrounding flow and defines its hydrodynamic
interactions \citep{lighthill_squirming_1952,blake_spherical_1971,pak_generalized_2014,pedley_spherical_2016,singh_generalized_2018}.
In such settings, orientation must be treated as a kinematic variable,
and both linear and angular momentum balances must be included among
the dynamical variables. The systematic framework for such oriented
continua, pioneered by the Cosserat brothers more than a century ago
\citep{cosserat_theorie_1909}, is known today as Cosserat elasticity
\citep{ericksen_exact_1957,schaefer_cosserat_1967,truesdell_non-linear_2004,altenbach_generalized_2013}.
Recently, Cosserat theory has received renewed attention in the study
of oriented active matter \citep{chen_realization_2021,bolitho_geometric_2021,surowka_odd_2023,lee_odd_2025},
and has been realized experimentally in metamaterial structures \citep{frenzel_three-dimensional_2017,rueger_strong_2018,reasa_nonclassical_2020}.
In addition, its mathematical structure has been recognized as a Cartan
geometry \citep{nemeth_geometric_2024}.

The nonreciprocity arguments previously made for Cauchy elasticity
extend naturally to Cosserat elasticity. We consider, again, an elementary
volume and sum the torques and the moments of forces arising from
both inside and outside the volume. For reciprocal interactions, the
internal torques cancel in pairs, leaving only the torques and the
moment of the forces transmitted across the boundary. This \emph{areal
}contribution, as the classical literature shows \citep{ericksen_exact_1957,schaefer_cosserat_1967,truesdell_non-linear_2004,altenbach_generalized_2013},
can be expressed as the sum of a moment stress and an antisymmetric
stress. For nonreciprocal interactions, however, the internal torques
do not cancel, leaving a\emph{ volumetric} contribution that must
be included in angular momentum balance. We show below that, when
the nonreciprocal interactions respect spatial symmetries, and again
assuming a solid-like response in both position and orientation, this
volumetric term can depend only on the invariant measures of deformation
of the Cosserat continuum. Based on Cartan's method of moving frames
\citep{cartan_geometry_1983,flanders_differential_1989}, we formulate
constitutive laws relating the reciprocal areal contribution (stress
and moment stress) and the nonreciprocal volumetric contribution (force
and torque densities) to strains, thereby closing the balance equations
for linear and angular momentum. We exhaustively classify materially
linear constitutive laws according to the symmetry of the interactions
via Curie's principle. The use of Cartan's method of moving frames
ensures that only invariant quantities are brought into constitutive
relationship, thereby automatically respecting the spatial symmetries.
Together, these elements yield a coarse-grained continuum mechanical
description of oriented \emph{nonreciprocal} solids, which we present
as our key result.

We test our nonreciprocal Cosserat continuum approach by explicitly
coarse-graining a particulate system with nonreciprocal interactions.
Specifically, we study a model of active polymers \citep{chelakkot_flagellar_2014,winkler_physics_2020,kumar_emergent_2024,biswas_emergent_2025},
namely spherical active colloidal particles that experience both reciprocal
chaining forces and nonreciprocal, hydrodynamically mediated interactions
from friction with the fluid and slip on their surfaces \citep{jayaraman_autonomous_2012,laskar_brownian_2015,krishnamurthy_emergent_2023}.
These slip velocities induce fluid flows that, upon eliminating the
solvent degrees of freedom, yield isometry-invariant, nonreciprocal
forces and torques.\emph{ }The symmetry of these slip velocities ---
classified as apolar-achiral, polar-achiral, apolar-chiral, or polar-chiral
--- directly translates to corresponding symmetries of the induced
nonreciprocal hydrodynamic interactions. By incorporating these explicit
forces and torques into Newton’s equations, and coarse-graining the
resulting dynamics, we recover a continuum description consistent
with one-dimensional nonreciprocal Cosserat elasticity and obtain,
thereby, microscopic expressions for the phenomenological coefficients
in the constitutive laws.

Since the interactions are not derived from a potential, the symmetry
of the phenomenological coefficients is not ensured \citep{scheibner_odd_2020}.
In particular, we identify coefficients characteristic of odd Cosserat
elasticity \citep{chen_realization_2021,fruchart_odd_2023,lee_odd_2025}.
More unexpectedly, the nonreciprocal contributions generate terms
in the balance equations that are of lower order in gradients of the
deformation than those arising from reciprocal interactions \citep{beatus_phonons_2006}.
As a result, the nonreciprocal effects dominate at long wavelengths,
giving rise to striking phenomena --- such as instabilities, traveling
waves, and spontaneous center-of-mass motion --- that are absent
in reciprocal continua. Numerical solutions of Newton’s equations
in the overdamped regime confirm these coarse-grained predictions
and show quantitative agreement with the framework of nonreciprocal
Cosserat elasticity.

Our work helps to rationalize several important results in the literature
\citep{lahiri_are_1997,bolitho_geometric_2021,chajwa_waves_2020}
on driven and active colloidal suspensions which, in the context of
our study, can be understood as arising from nonreciprocity---although
they have not previously been recognized as such. Phenomenological
coarse-grained descriptions of sedimenting colloidal crystals and
disks have revealed traveling waves and instabilities.\textbf{ }Notably,
none of these studies frames the long-wavelength description within
a systematic balance-law and constitutive-law framework. A significant
step in this direction was taken in \citep{poncet_when_2022}, where
a macroscopic elastic theory for particles with nonreciprocal interactions,
that need not be invariant under isometries of space, was derived.
Our approach advances this by incorporating both linear and angular
momentum balances and by explicitly identifying the volumetric force
and torque densities arising from nonreciprocal, isometry-invariant
interactions, thus contributing to the growing fields of active solids
\citep{hawkins_stress_2014,maitra_oriented_2019,kole_layered_2021,xu_autonomous_2023,baconnier_self-aligning_2025,veenstra_adaptive_2025,sarkar_mechanochemical_2025,cocconi_mechanical_2025,welker_lattice-dependent_2025}
and nonreciprocal active matter \citep{saha_scalar_2020,you_nonreciprocity_2020,fruchart_non-reciprocal_2021,loos_long-range_2023,veenstra_nonreciprocal_2025}.

The remainder of the paper is organized as follows. In Sec. \ref{sec:illustr},
we study general properties of particle systems with nonreciprocal
and rigid-body invariant interactions. In Sec. \ref{sec:cosserat_elasticity},
we summarize the continuum theory of Cosserat rods. In Sec. \ref{sec:const_laws},
we formulate the constitutive laws and perform a symmetry classification
of constitutive moduli for a one-dimensional nonreciprocal Cosserat
continuum. In Sec. \ref{sec:squirmer_chain}, we give an explicit
microscopic realization in terms of interacting active swimmers in
Stokes flow. In Sec. \ref{sec:comparison}, we study the dynamics
of a discrete chain for various choices of swimming modes, coarse-grain
to obtain effective continuum descriptions, identify the constitutive
moduli and calculate the mode structure for linear excitations. In
Sec. \ref{sec:discussion}, we draw our conclusions and suggest further
directions of research.

\section{Isometry-invariant nonreciprocal interactions\label{sec:illustr}}

\begin{figure}
\centering
\includegraphics[width=1\columnwidth]{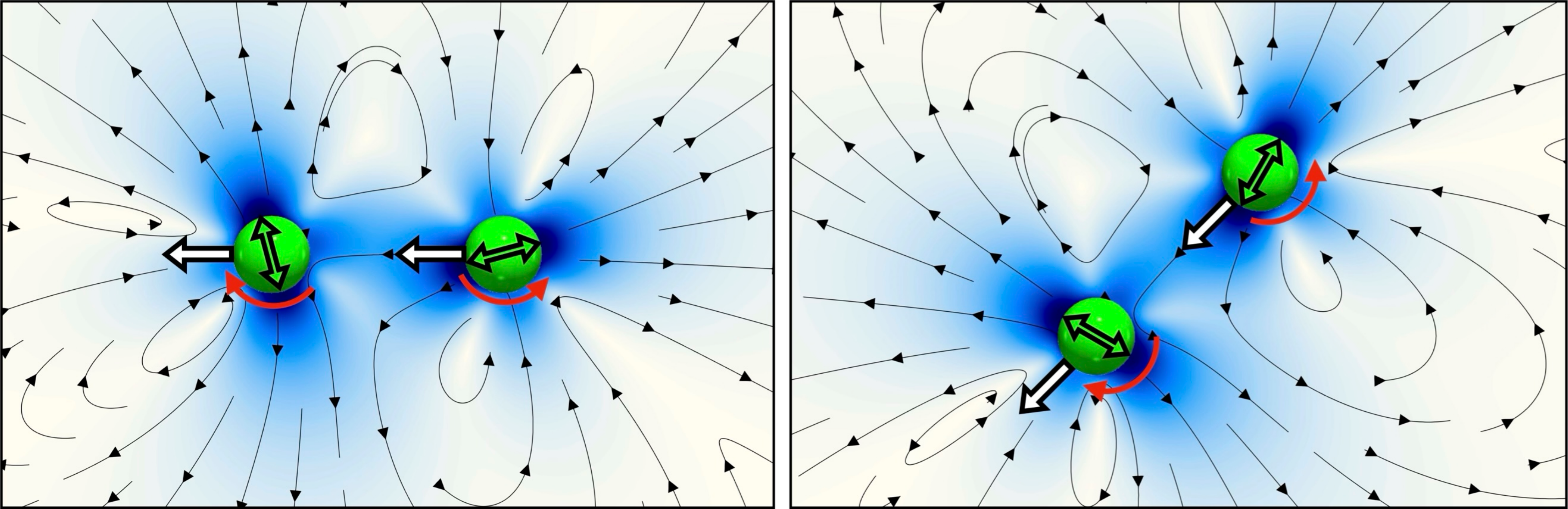}\caption{Illustration of nonreciprocity and rigid body invariance of interaction
forces and torques between active swimmers. The panels show two squirmers
(green spheres) equipped with an apolar, achiral $(2s)$ swimming
mode. The green double arrows represent their squirming axes of the
particles, while the white and red arrows show the interaction forces
and torques between the particles. The black arrows illustrate the
flow fields around the particles, where the shading indicates the
strength of the flows. Between the two panels, the positions and orientations
of the particles differ by a global isometry, under which the forces
and torques transform according to Eqs. \eqref{eq:force_inv}-\eqref{eq:torque_inv}.
This is due to the fact that the flows generated by the particles
transform in the same way under the action of the isometry.\label{fig:illustr_nonrep}}
\end{figure}
In this section we provide a precise definition of interactions that
are invariant under translations and rotations of Euclidean space
and are nonreciprocal. We consider a pair of particles in three-dimensional
Euclidean space whose centers of mass are located at $\boldsymbol{r}^{i}$
and whose orientations are determined by the orthonormal frame vectors
$\boldsymbol{e}_{a}^{i}$ rigidly attached to their centers of mass.
Here $i=1,2$ and $a=1,2,3$ are particle and frame indices respectively.
Positions are given relative to a fixed frame of reference with origin
$O$. We assume that the particles interact through pair forces $\boldsymbol{F}^{ij}$
and pair torques $\boldsymbol{T}^{ij}$ that are functions of their
positions and orientations,
\begin{equation}
\boldsymbol{F}^{ij}=\boldsymbol{F}^{ij}\left(\boldsymbol{r}^{k},\boldsymbol{e}_{a}^{k}\right),\quad\boldsymbol{T}^{ij}=\boldsymbol{T}^{ij}\left(\boldsymbol{r}^{k},\boldsymbol{e}_{a}^{k}\right),\label{eq:force_torque_def}
\end{equation}
with the shorthand notation $\boldsymbol{r}^{k}=\left\{ \boldsymbol{r}^{1},\boldsymbol{r}^{2}\right\} ,\boldsymbol{e}_{a}^{k}=\left\{ \boldsymbol{e}_{1}^{1},\dots,\boldsymbol{e}_{3}^{2}\right\} $.
Provided that the interactions between the particles are invariant
under a global rigid transformation -- an isometry of Euclidean space
-- of the form
\begin{equation}
\boldsymbol{r}^{i}\to\boldsymbol{R}\cdot\boldsymbol{r}^{i}+\boldsymbol{b},\quad\boldsymbol{e}_{a}^{i}\to\boldsymbol{R}\cdot\boldsymbol{e}_{a}^{i},\label{eq:rigid_tf}
\end{equation}
where $\boldsymbol{R}$ is an arbitrary proper orthogonal tensor and
$\boldsymbol{b}$ is an arbitrary vector, the forces and torques obey
the following transformation rule (see Fig. \ref{fig:illustr_nonrep}):
\begin{align}
\boldsymbol{F}^{ij}\left(\boldsymbol{R}\cdot\boldsymbol{r}^{k}+\boldsymbol{b},\boldsymbol{R}\cdot\boldsymbol{e}_{a}^{k}\right) & =\boldsymbol{R}\cdot\boldsymbol{F}^{ij}\left(\boldsymbol{r}^{k},\boldsymbol{e}_{a}^{k}\right),\label{eq:force_inv}\\
\boldsymbol{T}^{ij}\left(\boldsymbol{R}\cdot\boldsymbol{r}^{k}+\boldsymbol{b},\boldsymbol{R}\cdot\boldsymbol{e}_{a}^{k}\right) & =\boldsymbol{R}\cdot\boldsymbol{T}^{ij}\left(\boldsymbol{r}^{k},\boldsymbol{e}_{a}^{k}\right).\label{eq:torque_inv}
\end{align}
Eqs. \eqref{eq:force_inv}-\eqref{eq:torque_inv} imply that rigid-body
invariant interaction forces and torques have fewer degrees of freedom
than generic ones. An efficient way to deal with this degeneracy is
to consider the components of the force and torque vectors resolved
in the frame of one of the particles,
\begin{equation}
F_{a}^{ij}=\boldsymbol{e}_{a}^{i}\cdot\boldsymbol{F}^{ij},\quad T_{a}^{ij}=\boldsymbol{e}_{a}^{i}\cdot\boldsymbol{T}^{ij}.
\end{equation}
Since dot products are preserved by isometries \eqref{eq:rigid_tf},
the components $F_{a}^{ij},T_{a}^{ij}$ are invariant under such transformations.
It is convenient to introduce the notation
\begin{equation}
\boldsymbol{A}=A_{a}\boldsymbol{e}_{a},\quad A_{a}=\boldsymbol{e}_{a}\cdot\boldsymbol{A},\quad\underline{A}=(A_{1},A_{2},A_{3})
\end{equation}
for an arbitrary vector $\boldsymbol{A}$, its components $A_{a}=\boldsymbol{e}_{a}^{i}\cdot\boldsymbol{A}$
in the moving frame $\boldsymbol{e}_{a}^{i}$, and the triple $(A_{1},A_{2},A_{3})=\underline{A}$
of moving frame components. In this notation, interactions are invariant
if the components of the forces and torques, resolved in the frame
of one of the particles, are invariant under isometries, 
\begin{align}
\underline{F}^{ij}\left(\boldsymbol{R}\cdot\boldsymbol{r}^{k}+\boldsymbol{b},\boldsymbol{R}\cdot\boldsymbol{e}_{a}^{k}\right) & =\underline{F}^{ij}\left(\boldsymbol{r}^{k},\boldsymbol{e}_{a}^{k}\right),\label{eq:force_moving_frame_inv}\\
\underline{T}^{ij}\left(\boldsymbol{R}\cdot\boldsymbol{r}^{k}+\boldsymbol{b},\boldsymbol{R}\cdot\boldsymbol{e}_{a}^{k}\right) & =\underline{T}^{ij}\left(\boldsymbol{r}^{k},\boldsymbol{e}_{a}^{k}\right).\label{eq:torque_moving_frame_inv}
\end{align}

For such isometry-invariant interactions, the components of the forces
and torques can only depend on the \emph{relative} separation and
the \emph{relative }orientation of the particles, as these are invariant
under isometries. Their invariance is made explicit by resolving them
in the frame of one of the particles, 
\begin{align}
\Delta_{a}^{i} & =\boldsymbol{e}_{a}^{i}\cdot\left(\boldsymbol{r}^{i}-\boldsymbol{r}^{j}\right),\quad\Delta_{ab}^{ij}=\boldsymbol{e}_{a}^{i}\cdot\boldsymbol{e}_{b}^{j}.\label{eq:inv_scalars}
\end{align}
The $\Delta_{a}^{i}$ are the components of the relative separation
vector resolved in the frame of the $i$-th particle and the $\Delta_{ab}^{ij}$
are the direction cosines between the frames of the $i$-th and $j$-th
particles. Due to the orthogonality of the frames, the numbers $\Delta_{ab}^{ij}$
form the elements of a $3\times3$ special orthogonal matrix and hence
possess three degrees of freedom. The three components $\Delta_{a}^{i}$
and the three independent components of $\Delta_{ab}^{ij}$ are a
complete set of invariants for the relative positions and relative
orientations of the pair of particles. Given these, we can reconstruct
the position and orientation of the pair up to an isometry. The invariance
of the interactions can then be stated as 
\begin{equation}
\underline{F}^{ij}=\underline{F}^{ij}\left(\Delta_{a}^{i},\Delta_{ab}^{ij}\right),\quad\underline{T}^{ij}=\underline{T}^{ij}\left(\Delta_{a}^{i},\Delta_{ab}^{ij}\right).
\end{equation}

We now assume nonreciprocity of these isometry-invariant interactions,
in the sense that neither the forces nor the torques obey the law
of action and reaction:
\begin{align}
\boldsymbol{F}^{ij} & +\boldsymbol{F}^{ji}\neq0,\\
\left(\boldsymbol{T}^{ij}+\boldsymbol{r}^{i}\times\boldsymbol{F}^{ij}\right) & +\left(\boldsymbol{T}^{ji}+\boldsymbol{r}^{j}\times\boldsymbol{F}^{ji}\right)\neq0.
\end{align}
The total linear and angular momenta of the particles, then, are not
conserved. This can be reconciled with the assumed homogeneity and
isotropy of the interactions by recognizing that the mechanical system
must be \emph{open, }i.e., interacting with its environment. Though
the momentum and angular momentum of the system of particles are not
conserved, the \textit{sum} of the momenta and angular momenta of
the particles and the environment is conserved, as they constitute
a closed system.

Another important property of our interaction forces and torques is
that they are nonconservative, in the sense that they cannot be derived
from a potential. This is a consequence of Noether's theorem: conservative
and rigid-body invariant pair interaction forces and torques have
to reciprocal, as in a rigid-body invariant physical system they could
only follow from a rigid-body invariant pair potential, the symmetry
of which under translations and rotations results in conservation
of linear and angular momentum, respectively. Our forces and torques
are, therefore, able to inject energy into the system at the microscopic
level, which is hallmark feature of active matter \citep{ramaswamy_mechanics_2010}.

Below we shall consider an open system, consisting of of active colloids,
whose interactions are mediated through the fluid \citep{joanny_acoustic_1979,hurd_lattice_1982,dhont_introduction_1996}.
The total momentum and angular momentum of the particles and the fluid
are conserved but those of each component are not due to the exchanges
that take place during the evolution of the system. As we shall show
in the following section, the invariant scalars $\Delta_{a}^{i}$
and $\Delta_{ab}^{ij}$ become the appropriate invariant measures
of deformation in the continuum limit. The forces $\underline{F}^{ij}$
and the torques $\underline{T}^{ij}$ sum to stresses and moment stresses
and, due to their nonreciprocity, additionally to volumetric force
densities and torque densities. The isometry-invariance of the interactions
require these to depend only on the invariant measures of deformation.
The constitutive laws that emerge between stresses and strains will
generally be nonconservative, resulting from the non-potential character
of the microscopic forces and torques.

\section{Cosserat elasticity of nonreciprocal slender continua\label{sec:cosserat_elasticity}}

In this section, we consider the elasticity of a condensed phase in
which the constituents interact through nonreciprocal interactions
that respect the symmetries of space. We assume a ground state in
which all translational and rotational symmetries are broken \citep{landau_statistical_1980,chaikin_principles_1995},
therefore the long-wavelength elastic response of such a medium can
be described by Cosserat theory. We first present a brief summary
of the classical theory in this section, and then introduce the new
content in the next section, associated with the constitutive modeling
of volumetric forces and torques. We note that liquid crystal theory,
a popular framework for describing an active continuum \citep{marchetti_hydrodynamics_2013},
is not applicable here, as there are no surviving symmetries that
would give a liquid-like response to our medium. For simplicity, we
study a one-dimensional continuum, but the theory and constitutive
modeling can be extended with ease to an arbitrary number of dimensions.

\subsection{Kinematics and strain measures}

\begin{figure}
\centering
\includegraphics[width=1\columnwidth]{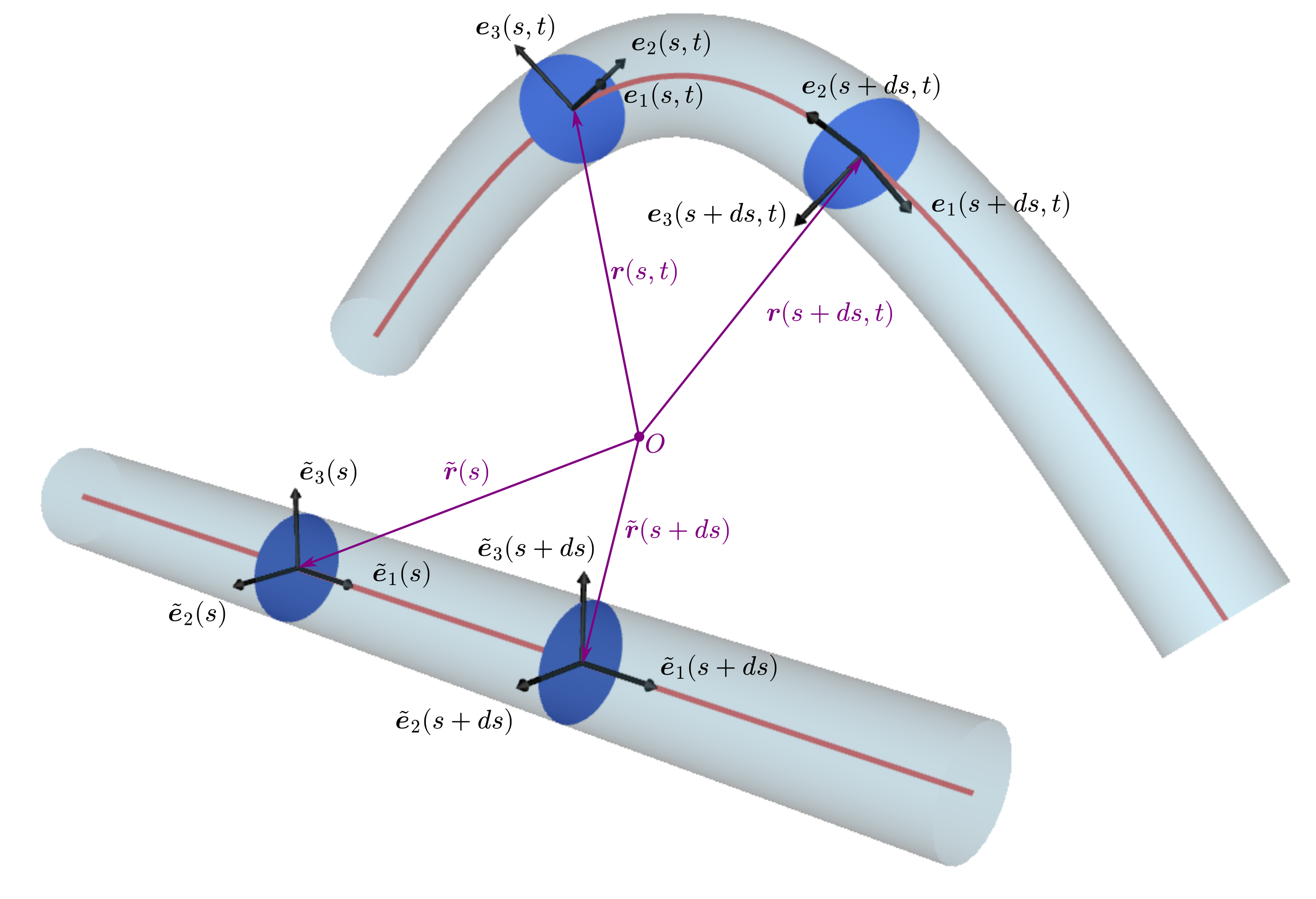}\caption{Kinematics of a Cosserat rod.\label{fig:rod_kin}}
\end{figure}
We consider the elasticity of a slender oriented continuum, described
by the position $\boldsymbol{r}(s,t)$ of its centerline and an orthonormal
moving frame $\boldsymbol{e}_{a}(s,t)$ attached to the cross-section
at material parameter $s$ and at time $t$, see Fig. \ref{fig:rod_kin}.
The deformation and motion of the rod are described by spatial and
temporal derivatives
\begin{align}
\boldsymbol{r}^{\prime} & =\boldsymbol{h},\quad\boldsymbol{e}_{a}^{\prime}=\boldsymbol{\Pi}\times\boldsymbol{e}_{a},\label{eq:spatial_def}\\
\dot{\boldsymbol{r}} & =\boldsymbol{V},\quad\dot{\boldsymbol{e}}_{a}=\boldsymbol{\Omega}\times\boldsymbol{e}_{a},\label{eq:temporal_def}
\end{align}
where differentiation with respect to $s$ and $t$ are denoted by
primes and dots respectively \citep{antman_nonlinear_2004}. The deformations
of the centerline and frame are $\boldsymbol{h}$ and $\boldsymbol{\Pi}$;
the velocity of the centerline and the angular velocity of the frame
are $\boldsymbol{V}$ and $\boldsymbol{\Omega}$. These kinematic
equations maintain the orthonormality of the frame at all materials
points and at all times. The components of the deformations $\boldsymbol{h},$
$\boldsymbol{\Pi}$ and the velocities and angular velocities in the
moving frame are 
\begin{align}
h_{a} & =\boldsymbol{e}_{a}\cdot\boldsymbol{h},\quad\Pi_{a}=\boldsymbol{e}_{a}\cdot\boldsymbol{\Pi},\label{eq:deformation_gradients}\\
V_{a} & =\boldsymbol{e}_{a}\cdot\boldsymbol{V},\quad\Omega_{a}=\boldsymbol{e}_{a}\cdot\boldsymbol{\Omega},
\end{align}
and these are invariant under isometries. Therefore, configurations
that differ from each other by a rigid motion have identical deformation
components. Conversely, given the deformation components, the configuration
can be recovered up to an isometry. Note that $h_{a}$ and $\Pi_{a}$
in Eq. \eqref{eq:deformation_gradients} are the continuum analogues
of the invariant scalars $\Delta_{a}^{i}$ and $\Delta_{ab}^{ij}$
in Eq. \eqref{eq:inv_scalars} of the previous section. The kinematic
equations are identical to Cartan's frame equations \citep{cartan_geometry_1983}.

Invariant measures of deformation are most easily defined in the moving
frame $\boldsymbol{e}_{a}\left(s,t\right)$. Since the deformation
components are invariant under isometries, a pair of configurations
related by an isometry have equal measures of deformation. Therefore,
the difference in the deformation components of a pair of configurations
provides an invariant measure of the \emph{strain }and such strain
measures vanish whenever the configurations are related by isometries.
We choose to define the strain measures with respect to the initial
configuration $\boldsymbol{r}\left(s,0\right)\equiv\tilde{\boldsymbol{r}}\left(s\right)$
and $\boldsymbol{e}_{a}\left(s,0\right)\equiv\tilde{\boldsymbol{e}}_{a}\left(s\right)$.
The translational strain measure is 
\begin{equation}
\varepsilon_{a}=h_{a}-\tilde{h}_{a}=\boldsymbol{e}_{a}\cdot\boldsymbol{r}'-\tilde{\boldsymbol{e}}_{a}\cdot\tilde{\boldsymbol{r}}'\longleftrightarrow\underline{\varepsilon}.\label{eq:tr_rod_strain}
\end{equation}
Kinematically, $\varepsilon_{a}$ measures the change in tangent vector
$\boldsymbol{r}'$ in projected on the $a$-th local frame vector.
Physically, $\varepsilon_{1}$ is a dilation and $\varepsilon_{2}$
and $\varepsilon_{3}$ are shears. The orientational strain measure
is 
\begin{equation}
\tau_{a}=\Pi_{a}-\tilde{\Pi}_{a}=\tfrac{1}{2}\epsilon_{abc}\left(\boldsymbol{e}_{c}\cdot\boldsymbol{e}_{b}'-\tilde{\boldsymbol{e}}_{c}\cdot\tilde{\boldsymbol{e}}_{b}^{\prime}\right)\longleftrightarrow\underline{\tau}.\label{eq:rot_rod_strain}
\end{equation}
Kinematically, $\tau_{a}$ measures the change in infinitesimal rotation
of the frame projected on the $a$-th frame vector. Physically, $\tau_{1}$
is a twist and $\tau_{2}$ and $\tau_{3}$ are flexures. Taken together,
$\underline{\varepsilon}$ and $\underline{\tau}$ are isometry-invariant
and geometrically nonlinear measures of strain with no restriction
to small deformations. 

The current and reference configurations are related by a translation
$\boldsymbol{u}(s,t)$ and a proper rotation \textbf{$\boldsymbol{Q}\left(s,t\right)$
}of each cross section at $s$ and $t$ and are so related as 
\begin{equation}
\boldsymbol{r}\left(s,t\right)=\tilde{\boldsymbol{r}}\left(s\right)+\boldsymbol{u}\left(s,t\right),\quad\boldsymbol{e}\left(s,t\right)=\boldsymbol{Q}\left(s,t\right)\cdot\tilde{\boldsymbol{e}}_{a}\left(s\right).
\end{equation}
From the above, we see that the strain measures vanish when configurations
are related by a rigid motion $\boldsymbol{r}(s,t)\rightarrow\boldsymbol{R}\cdot\boldsymbol{r}\left(s,t\right)+\boldsymbol{a}$
and $\boldsymbol{e}_{a}\left(s,t\right)\rightarrow\boldsymbol{R}\cdot\boldsymbol{e}_{a}\left(s,t\right)$
for constant vector $\boldsymbol{a}$ and constant proper rotation
matrix $\boldsymbol{R}$. As expected, nonzero strains indicate a
departure from isometry.

\subsection{Dynamics}

We use d'Alembert's principle of virtual work (in the inertialess
limit) to obtain the dynamics. We consider virtual displacements $\delta\boldsymbol{r}$
and $\delta\boldsymbol{\varphi}$ such that
\begin{equation}
\boldsymbol{r}\to\boldsymbol{r}+\delta\boldsymbol{r},\quad\boldsymbol{e}_{a}\to\boldsymbol{e}_{a}+\delta\boldsymbol{\varphi}\times\boldsymbol{e}_{a},\label{eq:var_rod_conf}
\end{equation}
with corresponding virtual strains 
\begin{equation}
\delta\varepsilon_{a}=\boldsymbol{e}_{a}\cdot\left(\delta\boldsymbol{r}'-\delta\boldsymbol{\varphi}\times\boldsymbol{r}'\right),\quad\delta\tau_{a}=\boldsymbol{e}_{a}\cdot\delta\boldsymbol{\varphi}'.\label{eq:var_strains}
\end{equation}
We define cross-sectional forces and moments $\boldsymbol{F}(s,t)$
and $\boldsymbol{M}(s,t)$ and force and moment densities $\boldsymbol{f}(s,t)$
and $\boldsymbol{m}(s,t)$ that are dual, respectively, to the virtual
strains and the virtual displacements. Henceforth, we refer to the
pair $\boldsymbol{F},\boldsymbol{M}$ as \textit{stresses} and to
the pair $\boldsymbol{f},\boldsymbol{m}$ as \textit{sources}. Resolving
all quantities in the moving frame, the virtual work is
\begin{equation}
\delta W=\int ds\left(\underline{f}\cdot\delta\underline{r}+\underline{m}\cdot\delta\underline{\varphi}-\underline{F}\cdot\delta\underline{\varepsilon}-\underline{M}\cdot\delta\underline{\tau}\right).\label{eq:virt_work_rod}
\end{equation}
The vanishing of the virtual work for all virtual displacements then
yields the balance laws of linear and angular momentum,\textbf{
\begin{align}
\underline{F}^{\prime}+\underline{\Pi}\times\underline{F}+\underline{f} & =0,\\
\underline{M}^{\prime}+\underline{\Pi}\times\underline{M}+\underline{h}\times\underline{F}+\underline{m} & =0,
\end{align}
}together with the boundary conditions
\begin{align}
\underline{F}\cdot\delta\underline{r}\Big|_{s=0,L}=0,\quad & \underline{M}\cdot\delta\underline{\varphi}\Big|_{s=0,L}=0.\label{eq:bc}
\end{align}
Returning to the fixed frame, recalling that $\boldsymbol{A}^{\prime}=(A_{a}\boldsymbol{e}_{a})^{\prime}=A_{a}^{\prime}\boldsymbol{e}_{a}+A_{a}\boldsymbol{e}_{a}^{\prime}$
and using the kinematic equation Eq. \eqref{eq:deformation_gradients}
to express $\boldsymbol{r}^{\prime}$ and $\boldsymbol{e}_{a}^{\prime}$
in terms of the moving frame vectors, we recover the familiar form
of the balance laws
\begin{equation}
\boldsymbol{F}'+\boldsymbol{f}=0,\quad\boldsymbol{M}'+\boldsymbol{r}'\times\boldsymbol{F}+\boldsymbol{m}=0\label{eq:balance_fixed_frame}
\end{equation}
for a slender oriented continuum, with boundary conditions $\boldsymbol{F}\cdot\delta\boldsymbol{r}=\boldsymbol{M}\cdot\delta\boldsymbol{\varphi}=0$
at $s=0,L$.

\section{Constitutive laws\label{sec:const_laws}}

The system of equations \eqref{eq:spatial_def}-\eqref{eq:temporal_def}
and \eqref{eq:balance_fixed_frame} govern the evolution of the rod,
but they cannot yet be solved. To close this system of equations,
we need to specify the relationship between stresses, sources and
strains. We shall assume that the continuum is internally elastic,
therefore stresses depend only on strains:
\[
\underline{F}=\underline{F}\left(\underline{\varepsilon},\underline{\tau}\right),\quad\underline{M}=\underline{M}\left(\underline{\varepsilon},\underline{\tau}\right).
\]
Writing the constitutive laws in the moving frame \citep{antman_nonlinear_2004}
ensures that isometric deformations induce no stresses, which is a
continuum manifestation of the isometry-invariance of the microscopic
interactions. In addition, we shall assume that sources contain dissipative
contributions $(D)$ due to the Stokes drag with the fluid, and, crucially,
that there are further contributions $\left(\star\right)$ that arise
from the nonreciprocity of the interactions (which can depend only
on strains owing to invariance of interactions under isometries):
\begin{align}
\underline{f} & =\underline{f^{D}}\left(\underline{V},\underline{\Omega}\right)+\underline{f}^{\star}\left(\underline{\varepsilon},\underline{\tau}\right),\nonumber \\
\underline{m} & =\underline{m}^{D}\left(\underline{V},\underline{\Omega}\right)+\underline{m}^{\star}\left(\underline{\varepsilon},\underline{\tau}\right).
\end{align}
This is a key contribution of our work, and we expand on this further
for the case of a materially linear continuum, where all constitutive
laws are linear or affine.

\subsection{Nonreciprocal elasticity}

A materially linear constitutive model for the volumetric forces and
torques must take the form
\begin{align}
\underline{f}^{\star} & =\underline{\tilde{f}}+\underline{\underline{H}}^{f\epsilon}\cdot\underline{\varepsilon}+\underline{\underline{H}}^{f\tau}\cdot\underline{\tau},\label{eq:rod_force_strain_const}\\
\underline{m}^{\star} & =\underline{\tilde{m}}+\underline{\underline{H}}^{m\epsilon}\cdot\underline{\varepsilon}+\underline{\underline{H}}^{m\tau}\cdot\underline{\tau},\label{eq:rod_torque_strain_const}
\end{align}
where $\tilde{\underline{f}}$ and $\tilde{\underline{m}}$ are net
force and torque densities in the reference configuration, and the
coupling matrices $\underline{\underline{H}}^{f\epsilon},\dots,\underline{\underline{H}}^{m\tau}$
relate strains to sources in a linear fashion. We shall provide a
classification of the material parameters using the Curie principle
in Sec. \ref{subsec:mat_sym}.

Apart from breaking the law of conservation of linear and angular
momentum in the medium, the constitutive sources \eqref{eq:rod_force_strain_const}-\eqref{eq:rod_torque_strain_const}
are always nonconservative. Just as in the discrete case in Sec \eqref{sec:illustr},
this is also a consequence of Noether's theorem: in a conservative
isometry-invariant system, we can only have an elastic energy that
depends solely on strains, from which we can only derive stresses
but not sources.

\subsection{Reciprocal elasticity and major symmetry}

\begin{figure*}
\centering
\includegraphics[width=1\textwidth]{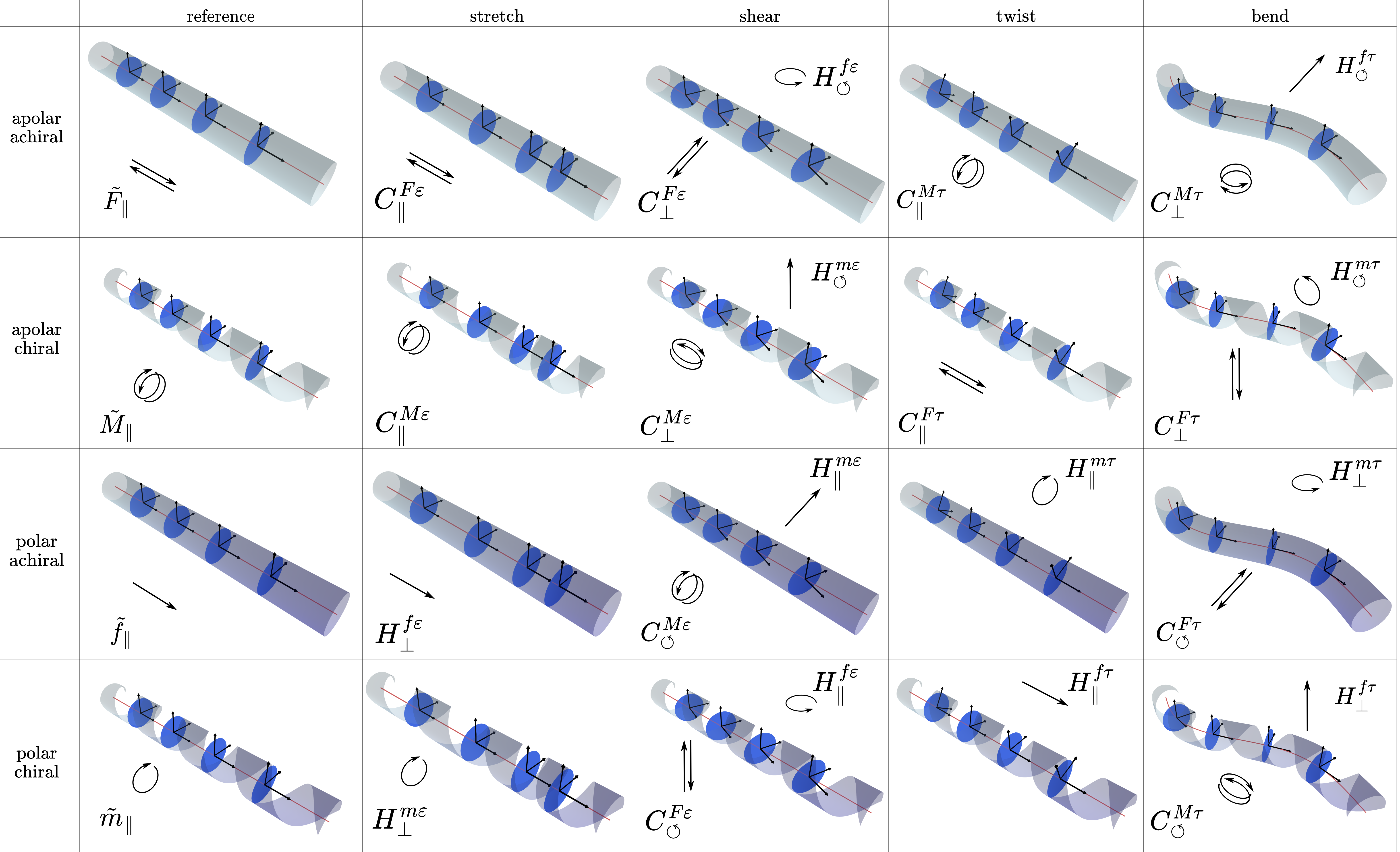}

\caption{Illustration of constitutive relations for a Cosserat rod. Stretch
deformations correspond to a nonzero $\varepsilon_{1}$, twist deformations
to a nonzero $\tau_{1}$, shear deformations to nonzero $\varepsilon_{2}$,
while bend deformations to nonzero $\tau_{3}$. Polar straight and
circular arrows represent sources of linear and angular momentum,
respectively, while dipolar straight and circular arrows represent
force and moment stresses.\label{fig:const_laws}}
\end{figure*}
Similarly as above, a materially linear constitutive model for the
stresses must take the form
\begin{align}
\underline{F}{}^{E} & =\underline{\tilde{F}}+\underline{\underline{C}}^{F\varepsilon}\cdot\underline{\varepsilon}+\underline{\underline{C}}^{F\tau}\cdot\underline{\tau},\label{eq:rod_lin_stress_const}\\
\underline{M}^{E} & =\underline{\tilde{M}}+\underline{\underline{C}}^{M\varepsilon}\cdot\underline{\varepsilon}+\underline{\underline{C}}^{M\tau}\cdot\underline{\tau},\label{eq:rod_ang_stress_const}
\end{align}

where $\tilde{\underline{F}}$ and $\tilde{\underline{M}}$ are the
prestress and pre-moment stress in the reference configuration, and
the matrices $\underline{\underline{C}}^{F\varepsilon},\dots,\underline{\underline{C}}^{M\tau}$
relate stresses to strains linearly.

Most elastic materials are hyperelastic, in which the constitutive
laws between stresses and strains can be derived from a stored elastic
energy density $\mathcal{E}\left(\underline{\varepsilon},\underline{\tau}\right)$
per unit length in the reference configuration. The energy density
can only be a function of the strains owing to rigid-body invariance.
For a hyperelastic Cosserat rod we have the following expression for
the virtual work \eqref{eq:virt_work_rod}:
\begin{equation}
\delta W=\int\delta\mathcal{E}ds=\int ds\left\{ \frac{\partial\mathcal{E}}{\partial\underline{\varepsilon}}\delta\underline{\varepsilon}+\frac{\partial\mathcal{E}}{\partial\underline{\tau}}\delta\underline{\tau}\right\} .\label{eq:hyperel_work}
\end{equation}
Combining \eqref{eq:virt_work_rod} and \eqref{eq:hyperel_work},
we can identify the constitutive stresses as
\begin{equation}
\underline{F}=\frac{\partial\mathcal{E}}{\partial\underline{\varepsilon}},\quad\underline{M}=\frac{\partial\mathcal{E}}{\partial\underline{\tau}}.\label{eq:hyperel_rod}
\end{equation}
Substituting the linear constitutive relations \eqref{eq:rod_lin_stress_const}-\eqref{eq:rod_ang_stress_const}
into \eqref{eq:hyperel_rod}, from equality of mixed partial derivatives
we obtain the following integrability conditions for the stress-strain
relations:
\begin{equation}
C_{ab}^{F\varepsilon}=C_{ba}^{F\varepsilon},\quad C_{ab}^{F\tau}=C_{ba}^{M\varepsilon},\quad C_{ab}^{M\tau}=C_{ba}^{M\tau}.\label{eq:major_sym_rod}
\end{equation}
If the conditions \eqref{eq:major_sym_rod} are satisfied, then the
constitutive relations \eqref{eq:rod_lin_stress_const}-\eqref{eq:rod_ang_stress_const}
can be derived from a quadratic stored energy density 
\begin{align}
\mathcal{E} & =\underline{\tilde{F}}\cdot\underline{\varepsilon}+\underline{\tilde{M}}\cdot\underline{\tau}+\frac{1}{2}\begin{bmatrix}\underline{\varepsilon} & \underline{\tau}\end{bmatrix}\left[\begin{array}{cc}
\underline{\underline{C}}^{F\varepsilon} & \underline{\underline{C}}^{F\tau}\\
\underline{\underline{C}}^{M\varepsilon} & \underline{\underline{C}}^{M\tau}
\end{array}\right]\left[\begin{array}{c}
\underline{\varepsilon}\\
\underline{\tau}
\end{array}\right].\label{eq:quadr_elastic_en}
\end{align}
Materials where the major symmetries \eqref{eq:major_sym_rod} are
violated have been termed odd elastic and received much attention
recently in the active matter community \citep{scheibner_odd_2020,tan_odd_2022,fruchart_odd_2023}.
As interaction forces and torques between squirmers do not derive
from a potential, the relations \eqref{eq:major_sym_rod} typically
do not hold in colloidal crystals and they are ``odd'' in this sense.

\subsection{Material symmetries\label{subsec:mat_sym}}

Using Curie's principle, further restrictions on the coupling constants
can be obtained by classifying the symmetries of the material of the
rod under the action of spatial symmetry transformations \citep{curie_sur_1894,ericksen_exact_1957,healey_material_2002}.
For simplicity, we assume that the material is isotropic in the transverse
directions $\left\{ \boldsymbol{e}_{2},\boldsymbol{e}_{3}\right\} $,
which implies two properties. First, the affine parts $\tilde{\underline{F}},\dots,\tilde{\underline{m}}$
can have nonzero components only in the $a=1$ direction along the
rod. Second, the coupling matrices $\underline{\underline{C}}^{F\varepsilon},\dots,\underline{\underline{H}}^{m\tau}$
have to be block-diagonal and isotropic in the lower-right block as
follows:

\begin{equation}
\underline{\underline{K}}=\begin{bmatrix}K_{\parallel} & 0 & 0\\
0 & K_{\perp} & K_{\circlearrowleft}\\
0 & -K_{\circlearrowleft} & K_{\perp}
\end{bmatrix},\label{eq:transverse_isotropy_moving_frame}
\end{equation}
where $\underline{\underline{K}}$ is one of the coupling matrices
$\underline{\underline{C}}^{F\varepsilon},\dots,\underline{\underline{H}}^{m\tau}$.
In tensorial form, this is equivalent to: 
\begin{equation}
\boldsymbol{K}=K_{\parallel}\boldsymbol{e}_{1}\boldsymbol{e}_{1}+K_{\perp}\left(\mathbb{I}-\boldsymbol{e}_{1}\boldsymbol{e}_{1}\right)+K_{\circlearrowleft}\boldsymbol{\epsilon}:\boldsymbol{e}_{1},\label{eq:transverse_isotropy}
\end{equation}
where $\mathbb{I}$ denotes the identity tensor, and $\boldsymbol{\epsilon}$
is the Levi-Civita tensor, so that the action of $\boldsymbol{K}$
on any vector $\boldsymbol{A}$ is given by $\boldsymbol{K}\cdot\boldsymbol{A}=K_{\parallel}A_{1}\boldsymbol{e}_{1}+K_{\perp}\left(A_{2}\boldsymbol{e}_{2}+A_{3}\boldsymbol{e}_{3}\right)+K_{\circlearrowleft}\boldsymbol{e}_{1}\times\boldsymbol{A}$.
Combining Eq. \eqref{eq:transverse_isotropy_moving_frame} with Eqs.
\eqref{eq:rod_force_strain_const}-\eqref{eq:rod_ang_stress_const},
we can write the constitutive laws in terms of the moduli $C_{\parallel}^{F\varepsilon},\dots,H_{\perp}^{m\tau}$.

By Curie's principle, the constitutive moduli have to respect the
symmetries of the constituent material of the rod. We shall consider
two types of symmetries for the constitutive laws \eqref{eq:rod_force_strain_const}-\eqref{eq:rod_ang_stress_const}:
we call a rod \textit{apolar} if it is invariant under rotations by
an angle $\pi$ about axes normal to $\boldsymbol{e}_{1}$ and \textit{achiral}
if it is invariant under mirror reflections about a plane containing
$\boldsymbol{e}_{1}$. We call rods which break these rotational or
mirror symmetries \textit{polar} and \textit{chiral}, respectively.

At a given material point $s$ on the chain, let us denote by $\mathrm{Rot}$
a rotation by an angle $\pi$ about an axis $\boldsymbol{n}=\cos\theta\boldsymbol{e}_{2}+\sin\theta\boldsymbol{e}_{3}$
perpendicular to $\boldsymbol{e}_{1}$ and by $\mathrm{Ref}$ a mirror
reflection about a plane perpendicular to $\boldsymbol{n}$. In the
moving frame, the transformations $\mathrm{Rot}$ and $\mathrm{Ref}$
are represented by the following matrices:
\[
\mathrm{Rot}=\begin{bmatrix}-1 & 0 & 0\\
0 & \cos2\theta & \sin2\theta\\
0 & \sin2\theta & -\cos2\theta
\end{bmatrix},\quad\mathrm{Ref}=-\mathrm{Rot}.
\]

Suppose we perform a symmetry transformation $\underline{\underline{S}}\in\left\{ \mathrm{Rot},\mathrm{Ref}\right\} $
on the reference configuration of the rod. Under the action of $\underline{\underline{S}}$,
each vectorial quantity $\underline{A}$ from the set of strains,
sources and stresses transforms as a pseudovector $\underline{A}\to\pm\underline{\underline{S}}\cdot\underline{A}$,
where on top of the usual tensor transformation law there might be
an extra sign associated with the transformation. This can happen
in two cases: either if $\underline{\underline{S}}=\mathrm{Ref}$
is a mirror reflection and $\underline{A}$ is an axial vector, or
if $\underline{\underline{S}}=\mathrm{Rot}$ is a rotation and $\underline{A}$
is a quantity that reverses sign under reversal of the orientation
of the chain. In the current context, we call the latter \textit{polarity-dependent}
vectors. The signs of all relevant quantities are summarized in Table
\ref{tab:signs}.\begingroup\tabcolsep=4pt\def\arraystretch{2}
\begin{table}
\centering
\begin{tabular}{|c||c|c|}
\hline 
 & Polarity-dependent & Polarity-independent\tabularnewline
\hline 
\hline 
True vector & $\underline{\varepsilon},\underline{F}$ & $\underline{f}$\tabularnewline
\hline 
Axial vector & $\underline{\tau},\underline{M}$ & $\underline{m}$\tabularnewline
\hline 
\end{tabular}

\caption{Behavior of strains, stresses and sources under mirror reflection
and reversal of chain orientation. The rotational quantities $\underline{\tau},\underline{M},\underline{m}$
all pick up a sign under mirror reflections and hence are axial vectors,
while the translational quantities $\underline{\varepsilon},\underline{F},\underline{f}$
do not. As the strains $\underline{\varepsilon},\underline{\tau}$
involve derivatives with respect to arclength, they pick up a sign
under reversal of the orientation of the rod and are polarity-dependent
vectors. The stresses $\underline{F},\underline{M}$ also depend on
the orientation of the rod as describe contact forces and torques
between different parts of the rod. The sources $\underline{f},\underline{m}$
are densities, and as such, independent of rod orientation.\label{tab:signs}}
\end{table}
\endgroup

Therefore, under the action of $\underline{\underline{S}}$, the coupling
matrix $\underline{\underline{K}}$ transforms as
\[
\underline{\underline{K}}\to\pm\underline{\underline{S}}^{T}\underline{\underline{K}}\underline{\underline{S}}=\pm\begin{pmatrix}K_{\parallel} & 0 & 0\\
0 & K_{\perp} & -K_{\circlearrowleft}\\
0 & K_{\circlearrowleft} & K_{\perp}
\end{pmatrix},
\]
where the overall sign in front is determined by the signs picked
up by the vectors coupled by $\underline{\underline{K}}$. Putting
all these considerations together, we find that each coupling constant
$C_{\perp}^{F\varepsilon},\dots,H_{\circlearrowleft}^{m\tau}$ either
stays the same or picks up a sign under $\mathrm{Rot}$ or $\mathrm{Ref}$.
We call coupling constants that pick up a sign under $\mathrm{Rot}$
or $\mathrm{Ref}$ \textit{polar} or \textit{chiral}, while those
that remain unchanged \textit{apolar} or \textit{achiral}, respectively.
The results are summarized in Table \ref{tab:syms} and illustrated
in Fig. \ref{fig:const_laws}. By Curie's principle, then, a rod made
of apolar or achiral material cannot have any nonzero polar or chiral
constitutive moduli, respectively.\begingroup\tabcolsep=3.5pt\def\arraystretch{2}
\begin{table}
\centering
\begin{tabular}{|c||c|c|c|c||c|}
\hline 
 & \multicolumn{2}{c|}{Achiral} & \multicolumn{2}{c||}{Chiral} & Interpretation\tabularnewline
\hline 
\hline 
\multirow{4}{*}{Apolar} & \multicolumn{2}{c|}{$\tilde{F}_{\parallel}$} & \multicolumn{2}{c||}{$\tilde{M}_{\parallel}$} & longitudinal prestress\tabularnewline
\cline{2-6}
 & $C_{\parallel}^{F\varepsilon}$ & $C_{\parallel}^{M\tau}$ & $C_{\parallel}^{F\tau}$ & $C_{\parallel}^{M\varepsilon}$ & longitudinal stress\tabularnewline
\cline{2-6}
 & $C_{\perp}^{F\varepsilon}$ & $C_{\perp}^{M\tau}$ & $C_{\perp}^{F\tau}$ & $C_{\perp}^{M\varepsilon}$ & transverse stress\tabularnewline
\cline{2-6}
 & $H_{\circlearrowleft}^{f\tau}$ & $H_{\circlearrowleft}^{m\varepsilon}$ & $H_{\circlearrowleft}^{f\varepsilon}$ & $H_{\circlearrowleft}^{m\tau}$ & transverse source\tabularnewline
\hline 
\multirow{4}{*}{Polar} & \multicolumn{2}{c|}{$\tilde{f}_{\parallel}$} & \multicolumn{2}{c||}{$\tilde{m}_{\parallel}$} & longitudinal pre-source\tabularnewline
\cline{2-6}
 & $H_{\parallel}^{f\varepsilon}$ & $H_{\parallel}^{m\tau}$ & $H_{\parallel}^{f\tau}$ & $H_{\parallel}^{m\varepsilon}$ & longitudinal force\tabularnewline
\cline{2-6}
 & $C_{\circlearrowleft}^{F\tau}$ & $C_{\circlearrowleft}^{M\varepsilon}$ & $C_{\circlearrowleft}^{F\varepsilon}$ & $C_{\circlearrowleft}^{M\tau}$ & transverse stress\tabularnewline
\cline{2-6}
 & $H_{\perp}^{f\varepsilon}$ & $H_{\perp}^{m\tau}$ & $H_{\perp}^{f\tau}$ & $H_{\perp}^{m\varepsilon}$ & transverse source\tabularnewline
\hline 
\end{tabular}\caption{Classification of coupling constants according to their polarity and
chirality, and their interpretation.\label{tab:syms}}
\end{table}
\endgroup

For a conservative, apolar, achiral and isotropic chain, the simplest
hyperelastic constitutive relation has 
\begin{align}
C_{\parallel}^{F\varepsilon} & =\kappa_{T}^{\parallel}, & C_{\parallel}^{M\tau} & =\kappa_{R}^{\parallel},\label{eq:hyperelastic1}\\
C_{\perp}^{F\varepsilon} & =\kappa_{T}^{\perp}, & C_{\perp}^{M\tau} & =\kappa_{R}^{\perp},\label{eq:hyperelastic2}
\end{align}
with all other moduli zero. $\kappa_{T}^{\parallel}$ is a stretching,
$\kappa_{T}^{\perp}$ is a shearing, $\kappa_{R}^{\parallel}$ is
a twisting and $\kappa_{R}^{\perp}$ is a bending modulus of the chain.
This follows from a transversely isotropic quadratic elastic energy
\eqref{eq:quadr_elastic_en}. A further simplification can be obtained
by setting $\kappa_{T}^{\parallel}=\kappa_{T}^{\perp}\equiv\kappa_{T}$
and $\kappa_{R}^{\parallel}=\kappa_{R}^{\perp}\equiv\kappa_{R}$,
which we assume in the remainder of the paper.

\subsection{Dissipation}

We model dissipative effects on the rod by introducing additional
constitutive sources that depend on the velocities of the constituents
of the solid. We assume that dissipative sources, coming from Stokes
drag with the fluid, are linear and local in the velocities
\begin{align}
\underline{f}^{D} & =-\underline{\underline{\Gamma}}^{TT}\cdot\underline{V}-\underline{\underline{\Gamma}}^{TR}\cdot\underline{\Omega},\nonumber \\
\underline{m}^{D} & =-\underline{\underline{\Gamma}}^{RT}\cdot\underline{V}-\underline{\underline{\Gamma}}^{RR}\cdot\underline{\Omega},\label{eq:substr_friction}
\end{align}
For simplicity, we will assume that there is no translation-rotation
coupling in the friction tensors and they are fully diagonal: $\underline{\underline{\Gamma}}^{TT}=\Gamma^{T}\mathbb{I},\quad\underline{\underline{\Gamma}}^{TT}=\Gamma^{R}\mathbb{I},$
with all other components zero.

\section{Active colloidal chains\label{sec:squirmer_chain}}

We now demonstrate explicit microscopic realizations of the above
continuum theory by means of a one-dimensional chain of $N$ squirmers
equipped with various swimming modes in an unbounded Stokesian fluid.
We are going to look at four different swimming modes that exhibit
all four possible apolar versus polar and achiral versus chiral symmetry
combinations at lowest order. We will show that the elastic response
of the chain to small deformations about its steady state is governed
by the effective theory described in the previous section, and compute
the elastic moduli in terms of the microscopic parameters. We perform
our calculations in the presence a stable background elastic potential,
thus providing further intuition to Cosserat rod theory via a discrete
mechanical approach.

We model squirmer $i=1,\dots,N$ as a rigid sphere of radius $a$
with center located at $\boldsymbol{r}^{i}$ and an orthonormal body
frame $\boldsymbol{e}_{a}^{i}$ rotating with it. We denote the translational
and angular velocity of swimmer $i$ by $\boldsymbol{v}^{i},\boldsymbol{\omega}^{i}$,
respectively. The equations of motion of squirmer $i$ read:
\begin{align}
\dot{\boldsymbol{r}}^{i} & =\boldsymbol{v}^{i}, & m\dot{\boldsymbol{v}}^{i} & =\boldsymbol{F}_{C}^{i}+\boldsymbol{F}_{D}^{i}+\boldsymbol{F}_{A}^{i},\label{eq:pos_eom}\\
\dot{\boldsymbol{e}}_{a}^{i} & =\boldsymbol{\omega}^{i}\times\boldsymbol{e}_{a}^{i}, & I\dot{\boldsymbol{\omega}}^{i} & =\boldsymbol{T}_{C}^{i}+\boldsymbol{T}_{D}^{i}+\boldsymbol{T}_{A}^{i},\label{eq:rot_eom}
\end{align}
where $m$ is the mass of the squirmers, $I=2ma^{2}/5$ is their moment
of inertia, and we have split the total force $\boldsymbol{F}^{i}$
and torque $\boldsymbol{T}^{i}$ acting on squirmer $i$ as a sum
of three contributions: a conservative force and torque $\boldsymbol{F}_{C}^{i},\boldsymbol{T}_{C}^{i}$
coming from springlike interactions between neighboring squirmers,
a dissipative drag force and torque $\boldsymbol{F}_{D}^{i},\boldsymbol{T}_{D}^{i}$
representing friction between the squirmers and the surrounding flow
and the active forces $\boldsymbol{F}_{A}^{i},\boldsymbol{T}_{A}^{i}$
that result from the slip velocities on the surfaces of the squirmers.

The conservative forces and torques follow by differentiating the
potential $V\left(\boldsymbol{r}^{i},\dots,\boldsymbol{r}^{N},\boldsymbol{e}_{1}^{a},\dots,\boldsymbol{e}_{b}^{N}\right)$,
representing elastic interactions along the chain (its detailed expression
given in Appendix B) with respect to $\boldsymbol{r}_{i}$ and $\boldsymbol{e}_{i}^{a}$:
\begin{equation}
\boldsymbol{F}_{C}^{i}=-\frac{\partial V}{\partial\boldsymbol{r}^{i}},\quad\boldsymbol{T}_{C}^{i}=-\sum_{a=1}^{3}\boldsymbol{e}_{a}^{i}\times\frac{\partial V}{\partial\boldsymbol{e}_{a}^{i}}.\label{eq:cons_deriv}
\end{equation}

By linearity of Stokes flow, the dissipative forces and torques are
linear functions of the velocities (summation over $j=1,\dots,N$
implicit): 
\begin{align}
\boldsymbol{F}_{D}^{i} & =-\boldsymbol{\gamma}_{ij}^{TT}\cdot\boldsymbol{v}^{j}-\boldsymbol{\gamma}_{ij}^{TR}\cdot\boldsymbol{\omega}^{j},\label{eq:tr_drag}\\
\boldsymbol{T}_{D}^{i} & =-\boldsymbol{\gamma}_{ij}^{RT}\cdot\boldsymbol{v}^{j}-\boldsymbol{\gamma}_{ij}^{RR}\cdot\boldsymbol{\omega}^{j},\label{eq:rot_drag}
\end{align}
where $\boldsymbol{\gamma}_{ij}\left(\boldsymbol{r}^{1},\dots,\boldsymbol{r}^{N},\boldsymbol{e}_{a}^{1},\dots,\boldsymbol{e}_{b}^{N}\right)$
is the friction tensor, with the superscript $T,R$ denoting translational
and rotational components, respectively.

The active forces and torques are obtained by solving for the Stokes
flow surrounding the particles with the slip boundary conditions on
the surfaces of the squirmers and integrating the tractions (see Appendix
A for details). We restrict our attention to a scenario when the active
forces and torques can be well approximated as a sum of pairwise interactions
between particles (using notation of Sec. \ref{sec:illustr}):
\begin{equation}
\boldsymbol{F}_{A}^{i}=\sum_{j=1}^{N}\boldsymbol{F}_{A}^{ij}\left(\boldsymbol{r}^{k},\boldsymbol{e}_{a}^{k}\right),\quad\boldsymbol{T}_{A}^{i}=\sum_{j=1}^{N}\boldsymbol{T}_{A}^{ij}\left(\boldsymbol{r}^{k},\boldsymbol{e}_{a}^{k}\right).\label{eq:act_forces_torques}
\end{equation}
As argued in Sec. \ref{sec:illustr}, the pair interaction forces
and torques $\boldsymbol{F}_{ij}^{A}\left(\boldsymbol{r}^{k},\boldsymbol{e}_{a}^{k}\right),\boldsymbol{T}_{ij}^{A}\left(\boldsymbol{r}^{k},\boldsymbol{e}_{a}^{k}\right)$
are taken to be rigid-body invariant but are nonreciprocal in general.

We focus on small displacements $\boldsymbol{u}^{i}$ and rotations
$\boldsymbol{\varphi}^{i}$ about a (possibly time-dependent) steady
state $\tilde{\boldsymbol{r}}^{i}\left(t\right),\tilde{\boldsymbol{e}}_{a}^{i}\left(t\right)$
and linearize: 
\begin{align}
\boldsymbol{r}^{i}\left(t\right) & \approx\tilde{\boldsymbol{r}}^{i}\left(t\right)+\boldsymbol{u}^{i}\left(t\right),\label{eq:tr_perturb}\\
\boldsymbol{e}_{a}^{i}\left(t\right) & \approx\tilde{\boldsymbol{e}}_{a}^{i}\left(t\right)+\boldsymbol{\varphi}^{i}\left(t\right)\times\tilde{\boldsymbol{e}}_{a}^{i}\left(t\right).\label{eq:rot_perturbation}
\end{align}
Substituting into the equations of motion \eqref{eq:pos_eom}-\eqref{eq:rot_eom}
and taking the overdamped limit, we obtain a $6N$-dimensional linear
system \citep{bolitho_geometric_2021} 
\begin{equation}
\left[\begin{array}{cc}
\boldsymbol{\gamma}_{ij}^{TT} & \boldsymbol{\gamma}_{ij}^{TR}\\
\boldsymbol{\gamma}_{ij}^{RT} & \boldsymbol{\gamma}_{ij}^{RR}
\end{array}\right]\left[\begin{array}{c}
\dot{\boldsymbol{u}}^{j}\\
\dot{\boldsymbol{\varphi}}^{j}
\end{array}\right]=\boldsymbol{J}\left[\begin{array}{c}
\boldsymbol{u}^{i}\\
\boldsymbol{\varphi}^{i}
\end{array}\right],\label{eq:lin_dyn}
\end{equation}
with a $6N\times6N$ dimensional Jacobian matrix $\boldsymbol{J}$
that receives three contributions: one from the linearization of the
elastic potential, one from the linearization of the active forces
and finally one from the position- and orientation-dependence of the
friction tensors if we linearize about a non-stationary steady state.

Eq. \eqref{eq:lin_dyn} is the discretization of linearized Cosserat
equations of motion under a short-range hydrodynamics approximation.
Indeed, all friction terms and elastic terms in the continuum description
are local in the sense that they only depend on the configuration
variables and their derivatives at a point. Therefore, in order to
make contact with the continuum theory, we will consider hydrodynamic
interactions only at the nearest-neighbor level $j=i\pm1$. In this
limit, fluid friction will follow from the diagonal terms in the friction
tensors of the left hand side of Eq. \eqref{eq:lin_dyn}. The position-
and orientation-dependence of the friction tensors will lead to additional
elastic terms in $\boldsymbol{J}$ if we linearize about a non-stationary
steady state \citep{lahiri_are_1997,bolitho_geometric_2021}. In the
sequel, we will neglect these terms as we would like to focus on the
contributions of active forces and torques. While long-range hydrodynamics
\citep{felderhof_mesoscopic_2003,singh_universal_2016,bolitho_geometric_2021}
becomes important for larger deformations of the chain, we expect
that the qualitative features of the dynamics will be captured in
the short-ranged limit as well. In the next section, we populate the
entries of $\boldsymbol{J}$ and show how they correspond to the nonreciprocal
elastic moduli predicted by the continuum theory.

\section{Coarse-graining and comparison to nonreciprocal Cosserat elasticity\label{sec:comparison}}

\subsection{Linearized elastohydrodynamics}

\begingroup\tabcolsep=4pt\def\arraystretch{2}
\begin{table}
\centering
\begin{tabular}{|c||c|c|c|c|}
\hline 
 & \multicolumn{2}{c|}{Achiral} & \multicolumn{2}{c|}{Chiral}\tabularnewline
\hline 
\hline 
\multirow{3}{*}{Apolar} & $\propto\boldsymbol{u}_{\perp}''$ & $\sim k^{2}$ & $\propto\hat{\boldsymbol{x}}\times\boldsymbol{u}_{\perp}'''$ & $\sim k^{3}$\tabularnewline
\cline{2-5}
 & \multicolumn{2}{c|}{$\propto\int ds\left|\boldsymbol{u}_{\perp}'\right|^{2}$} & \multicolumn{2}{c|}{$\propto\int ds\left|\boldsymbol{u}'_{\perp}\times\boldsymbol{u}_{\perp}''\right|$}\tabularnewline
\cline{2-5}
 & \multicolumn{2}{c|}{bending instability} & \multicolumn{2}{c|}{helical instability}\tabularnewline
\hline 
\multirow{3}{*}{Polar} & $\propto\boldsymbol{u}_{\perp}',\boldsymbol{u}_{\perp}'''$ & $\sim ik,ik^{3}$ & $\propto\hat{\boldsymbol{x}}\times\boldsymbol{u}_{\perp}''$ & $\sim ik^{2}$\tabularnewline
\cline{2-5}
 & \multicolumn{2}{c|}{none} & \multicolumn{2}{c|}{none}\tabularnewline
\cline{2-5}
 & \multicolumn{2}{c|}{traveling waves} & \multicolumn{2}{c|}{rotating waves}\tabularnewline
\hline 
\end{tabular}\caption{Leading order force terms, corresponding elastic energy terms and
transverse modes from activity in the beam limit. For the polar and
achiral chain, one can transform to a comoving frame to eliminate
the $\boldsymbol{u}_{\perp}'$ term, in which case the leading force
term from activity becomes $\propto\boldsymbol{u}_{\perp}'''$ and
leads to a dispersion relation of the form $\sim ik^{3}$. \label{tab:beam_lim}}
\end{table}
\endgroup In this section, we introduce some notation for and highlight
general features of the linearized continuum dynamics of Cosserat
rods, which shall be compared to the discrete particulate model outlined
in the previous section.

In the continuum limit, we look at the linearized evolution of small
displacements about a steady state solution $\tilde{\boldsymbol{r}},\tilde{\boldsymbol{e}}_{a}$
of the equations of motion \eqref{eq:balance_fixed_frame}. We will
focus on small deformations of the rod that satisfy $\left|\boldsymbol{u}'\right|\ll1$
and $\left|\boldsymbol{Q}-\boldsymbol{I}\right|\ll1$. In this case,
we can represent $\boldsymbol{Q}$ by an infinitesimal rotation field
$\boldsymbol{\varphi}\left(s,t\right)$ defined via $\boldsymbol{e}_{a}\approx\tilde{\boldsymbol{e}}_{a}+\boldsymbol{\varphi}\times\tilde{\boldsymbol{e}}_{a}$.
To first order in $\boldsymbol{u}'$ and $\boldsymbol{\varphi}$,
the strain measures become
\begin{equation}
\varepsilon_{a}\approx\tilde{\boldsymbol{e}}_{a}\cdot\left(\boldsymbol{u}'+\tilde{\boldsymbol{r}}'\times\boldsymbol{\varphi}\right),\quad\tau_{a}\approx\tilde{\boldsymbol{e}}_{a}\cdot\boldsymbol{\varphi}'.\label{eq:rod_strain_lin}
\end{equation}
We will linearize about steady states with zero strain, but they might
be translating or rotating with constant velocity $\tilde{\boldsymbol{V}}$
or angular velocity $\tilde{\boldsymbol{\Omega}}$. The linear and
angular velocities of the chain to leading order in displacements
are given by:
\begin{equation}
\boldsymbol{V}\approx\tilde{\boldsymbol{V}}+\dot{\boldsymbol{u}},\quad\boldsymbol{\Omega}\approx\tilde{\boldsymbol{\Omega}}+\dot{\boldsymbol{\varphi}}-\tilde{\boldsymbol{\Omega}}\times\boldsymbol{\varphi}.\label{eq:vel_pert}
\end{equation}
To obtain the linearized equations of motion, the linearized quantities
\eqref{eq:rod_strain_lin}-\eqref{eq:vel_pert} have to be substituted
into the balance laws \eqref{eq:balance_fixed_frame}, supplemented
with the appropriate constitutive relations. Neglecting viscosity
terms, the general form of the linearized equations of motion reads
\begin{equation}
\left[\begin{array}{cc}
\boldsymbol{\Gamma}^{TT} & \boldsymbol{\Gamma}^{TR}\\
\boldsymbol{\Gamma}^{RT} & \boldsymbol{\Gamma}^{RR}
\end{array}\right]\left[\begin{array}{c}
\dot{\boldsymbol{u}}\\
\dot{\boldsymbol{\varphi}}
\end{array}\right]=\mathcal{L}\left[\begin{array}{c}
\boldsymbol{u}\\
\boldsymbol{\varphi}
\end{array}\right]\label{eq:rod_gen_lin}
\end{equation}
for a linear differential operator $\mathcal{L}$ that we specify
in the sequel for each symmetry combination.

Equation \eqref{eq:rod_gen_lin} is of the form of a linear advection-diffusion
equation for the displacements $\boldsymbol{u},\boldsymbol{\varphi}$.
The exact form of the operator $\mathcal{L}$ is constrained by the
structure of the equations of motion \eqref{eq:balance_fixed_frame}.
We compute the spectrum of the $\mathcal{L}$ by a Fourier transform.
While $\mathcal{L}$ in general does not depend on $\boldsymbol{u}$
(only its derivatives), it typically depends on $\boldsymbol{\varphi}$,
which implies that spectrum of the operator will be gapped: not all
modes will relax slowly in the limit of a perturbation of long wavelength.
This is a general feature of Cosserat theory, and implies that one
can often eliminate the angle $\boldsymbol{\varphi}$ as it is a fast
variable compared to the displacement $\boldsymbol{u}$. We focus
on the ungapped acoustic modes, and derive their dispersion relations
in Sec. \ref{sec:comparison}. We expect that for apolar systems,
the spectrum remains real, while for polar systems it acquires an
imaginary part, leading to traveling waves. Table \ref{tab:beam_lim}
summarizes our results, which we elaborate in the following sections.

\subsection{Passive chain}

The continuum theory in the linearized regime \eqref{eq:rod_gen_lin}
about a straight stationary reference configuration with parallel
frames, hyperelastic constitutive relation \eqref{eq:hyperelastic1}-\eqref{eq:hyperelastic2}
and fluid friction is governed by the equations of motion
\begin{align}
\Gamma^{T}\dot{\bm{u}} & =\kappa_{T}\left(\bm{u}''+\hat{\bm{x}}\times\bm{\varphi}'\right),\label{eq:passive_cont_tr}\\
\Gamma^{R}\dot{\bm{\varphi}} & =\kappa_{T}\hat{\bm{x}}\times\left(\bm{u}'+\hat{\bm{x}}\times\bm{\varphi}_{\perp}\right)+\kappa_{R}\bm{\varphi}''.\label{eq:passive_cont_rot}
\end{align}
Longitudinal (along $\boldsymbol{\hat{x}}$) and transverse (perpendicular
to $\hat{\boldsymbol{x}}$) parts of the dynamics decouple, and we
will mainly be concerned with the evolution of transverse perturbations.
Introducing the transverse displacements
\begin{equation}
\boldsymbol{u}_{\perp}:=\boldsymbol{u}-\left(\boldsymbol{u}\cdot\hat{\boldsymbol{x}}\right)\hat{\boldsymbol{x}},\quad\boldsymbol{\varphi}_{\perp}:=\boldsymbol{\varphi}-\left(\boldsymbol{\varphi}\cdot\hat{\boldsymbol{x}}\right)\hat{\boldsymbol{x}},\label{eq:compl_tr_def}
\end{equation}
the transverse part of the dynamics \eqref{eq:passive_cont_tr}-\eqref{eq:passive_cont_rot}
reads: 
\begin{align}
\Gamma^{T}\dot{\boldsymbol{u}}_{\perp} & =\kappa_{T}\left(\boldsymbol{u}_{\perp}''+\hat{\boldsymbol{x}}\times\boldsymbol{\varphi}_{\perp}'\right),\label{eq:tr_part}\\
\Gamma^{R}\dot{\boldsymbol{\varphi}}_{\perp} & =-\kappa_{T}\boldsymbol{\varphi}_{\perp}'+\kappa_{T}\hat{\boldsymbol{x}}\times\boldsymbol{u}_{\perp}'+\kappa_{R}\boldsymbol{\varphi}_{\perp}''.\label{eq:rot_part}
\end{align}

To obtain the mode structure for the conservative chain, we perform
a discrete Fourier transform by substituting plane-wave solutions
of the form $\sim e^{iqx}$ into Eqs. \eqref{eq:tr_part}-\eqref{eq:rot_part}.
For transverse perturbations, we get two modes, both with a twofold
degeneracy, whose dispersion relations in the long-wavelength limit
read
\begin{align}
\Lambda_{ac}\left(q\right) & \approx-\frac{\kappa_{R}}{\Gamma^{T}}q^{4}+\mathcal{O}\left(q^{6}\right),\label{eq:pass_ac}\\
\Lambda_{op}\left(q\right) & \approx-\frac{\kappa_{T}}{\Gamma^{R}}+\mathcal{O}\left(q^{2}\right),\label{eq:pass_op}
\end{align}
consistently with \eqref{eq:tr_part}-\eqref{eq:rot_part}. The growth
rate of the acoustic bending mode $\Lambda_{ac}\left(q\right)$ goes
to zero as $q\to0$ and its scaling is consistent with standard Euler-Bernoulli
beam theory. On the other hand, the growth rate of the optical mode
$\Lambda_{op}\left(q\right)$ has a finite gap at zero wavenumber.
Therefore, in what follows, we will focus on how activity changes
the exponent and also the stability of the acoustic mode since the
optical mode can only receive higher-order corrections from active
effects.

To gain further insight into the scaling of the acoustic mode, it
is instructive to derive the beam limit $\kappa_{T}\to\infty$ of
the transverse parts of Eqs. \eqref{eq:tr_part}-\eqref{eq:rot_part}.
Physically, this limit corresponds to an inextensible and unshearable
rod. As the equation for the angle $\boldsymbol{\varphi}_{\perp}$
contains a decay term $-\kappa_{T}\boldsymbol{\varphi}_{\perp}'$,
in this limit it will quickly relax to a value dictated by the perpendicular
displacement $\boldsymbol{u}_{\perp}$ and can be adiabatically eliminated
from the equations of motion (see Appendix C for details). We get:
\begin{equation}
\Gamma^{T}\dot{\boldsymbol{u}}_{\perp}=-\kappa_{R}\boldsymbol{u}_{\perp}'''',\label{eq:eb}
\end{equation}
which are the overdamped equations of Euler-Bernoulli beam theory
\citep{landau_theory_1984,wiggins_flexive_1998}, consistently with
the transverse acoustic mode \eqref{eq:pass_ac}.

We now provide a discrete realization of the passive continuum dynamics
\eqref{eq:passive_cont_tr}-\eqref{eq:passive_cont_rot}, obtained
from the overdamped limit of the discrete dynamical equations of motion
\eqref{eq:pos_eom}-\eqref{eq:rot_eom} in the absence of activity
\begin{equation}
\gamma_{ij}^{TT}\bm{v}^{j}+\gamma_{ij}^{TR}\boldsymbol{\omega}^{j}=\bm{F}_{C}^{i},\quad\gamma_{ij}^{RT}\bm{v}^{j}+\gamma_{ij}^{RR}\boldsymbol{\omega}^{j}=\bm{T}_{C}^{i},\label{eq:passive_eoms}
\end{equation}
and a suitable potential $V$ in \eqref{eq:cons_deriv}. We choose
$V$ to correspond to a discrete finite difference approximation of
the hyperelastic constitutive law \eqref{eq:hyperelastic1}-\eqref{eq:hyperelastic2},
and give its full expression in Appendix B. With this choice, the
straight configuration $\tilde{\boldsymbol{r}}^{i}=\begin{pmatrix}id & 0 & 0\end{pmatrix}^{T}$
with parallel frames $\tilde{\boldsymbol{e}}_{1}^{i}=\hat{\boldsymbol{x}},\tilde{\boldsymbol{e}}_{2}^{i}=\hat{\boldsymbol{y}},\tilde{\boldsymbol{e}}_{3}^{i}=\hat{\boldsymbol{z}}$
is a stationary solution of \eqref{eq:passive_eoms}. Linearizing
about this configuration, keeping only the leading one-body friction
terms $\gamma_{ii}^{TT}=\gamma^{T}=6\pi\eta a,\gamma_{ii}^{RR}=\gamma^{R}=8\pi\eta a^{3},\gamma_{ii}^{TR}=\gamma_{ii}^{RT}=0$
(no summation on $i$), we get the following equations of motion
\begin{align}
\gamma^{T}\dot{\bm{u}}^{i} & =\lambda\mathcal{D}_{u}^{2}\bm{u}^{i}+\lambda d\hat{\bm{x}}\times\mathcal{D}_{u\varphi}^{1}\bm{\varphi}^{i},\label{eq:passive_discr_tr}\\
\gamma^{R}\dot{\bm{\varphi}}^{i} & =\lambda d\hat{\bm{x}}\times\left(\mathcal{D}_{\varphi u}^{1}\boldsymbol{u}^{i}+d\hat{\bm{x}}\times\mathcal{D}^{0}\bm{\varphi}^{i}\right)+\mu\mathcal{D}_{\varphi}^{2}\boldsymbol{\varphi}^{i},\label{eq:passive_discr_rot}
\end{align}
where $\lambda$ and $\mu$ are discrete elastic moduli, and $\mathcal{D}_{u}^{2},\mathcal{D}_{\varphi}^{2},\mathcal{D}_{u\varphi}^{1},\mathcal{D}_{\varphi u}^{1},\mathcal{D}^{0}$
are finite difference operators that follow from the linearization
of conservative forces and torques (for their precise definition,
see Appendix B). In the continuum or long-wavelength limit, $\mathcal{D}_{u}^{2},\mathcal{D}_{\varphi}^{2}$
become second derivatives, while $\mathcal{D}_{u\varphi}^{1},\mathcal{D}_{\varphi u}^{1}$
first derivatives with respect to $s$, consistently with the continuum
equations \eqref{eq:tr_part}-\eqref{eq:rot_part}. We can then identify
the material parameters as
\begin{equation}
\Gamma^{T}=\frac{\gamma^{T}}{d},\quad\Gamma^{R}=\frac{\gamma^{R}}{d},\quad\kappa_{T}=\lambda d,\quad\kappa_{R}=\mu d.\label{eq:cons_id}
\end{equation}

\subsection{Apolar, achiral chain}

\begin{figure*}
\centering
\includegraphics[width=1\textwidth]{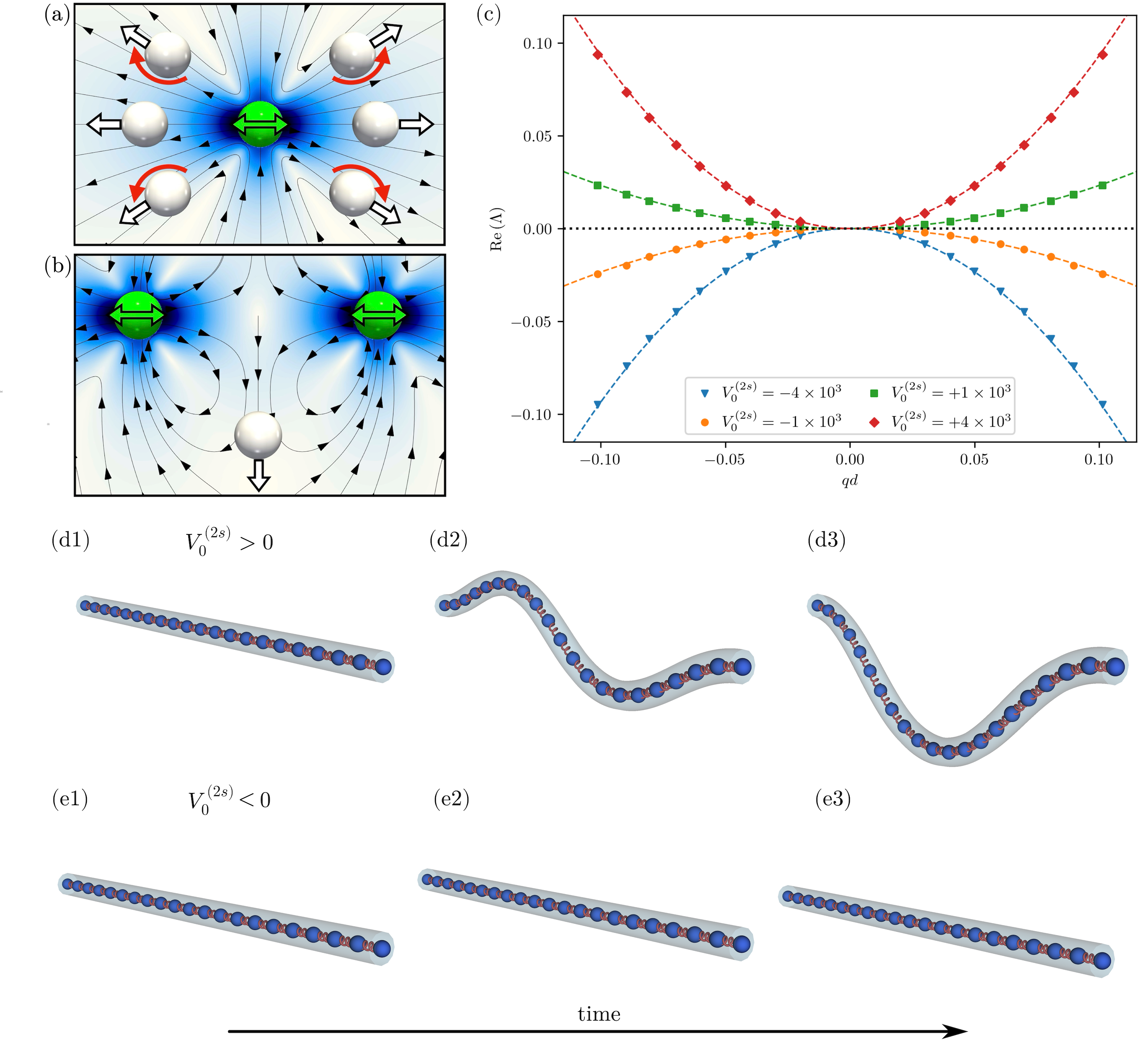}

\caption{(a)-(e) Phenomenology of an apolar, achiral squirmer chain and comparison
to continuum theory. (a) Flow field around a $(2s)$ squirmer (green
sphere) and forces and torques on tracer particles (white spheres)
around it. The green double arrow represents the squirming axis of
the active particle, white arrows the forces and red circular arrows
the torques on tracers. The flow field is shaded according to flow
velocity. (b) Response of apolar, achiral chain to a transverse perturbation:
pusher-type $V_{0}^{(2s)}>0$ squirmers generate a net transverse
force in the plane of deformation, leading to a buckling instability.
(c) Growth rates of transverse perturbations for apolar, achiral chains
for small wavenumbers. Dashed lines are theoretical predictions from
Eq. \eqref{eq:2s_tr_displ}, markers are results of numerical simulations.
(d) Time evolution of a chain of pusher-type $V_{0}^{(2s)}>0$ squirmers
under clamped boundary conditions, starting from random initial conditions.
The chain buckles due to the presence of activity. (e) Time evolution
of a chain of puller-type $V_{0}^{(2s)}<0$ squirmers under clamped
boundary conditions, starting from random initial conditions. The
chain is stable against transverse perturbations.\label{fig:2s}}
\end{figure*}
We now turn to investigate the effects of activity on the discrete
chain and compare it with an effective continuum theory. In this section,
we look at an apolar, achiral chain, where each squirmer is endowed
with a $(2s)$ swimming mode (a force dipole) of strength $V_{0}^{(2s)}$
along its $\boldsymbol{e}_{1}$ body frame vector, see Fig. \ref{fig:2s}.
The active force and torque arising from this apolar, achiral squirming
mode, exerted by squirmer $j$ on squirmer $i$ are given by (to leading
order in the ratio $a/d$) \citep{singh_generalized_2018}:
\begin{align}
\bm{F}_{A}^{ij} & =\frac{7a^{2}\gamma^{T}V_{0}^{(2s)}}{6r^{2}}\left(3\left(\boldsymbol{e}_{1}^{j}\cdot\hat{\boldsymbol{r}}^{ij}\right)^{2}-1\right)\hat{\boldsymbol{r}}^{ij},\label{eq:2s_force}\\
\boldsymbol{T}_{A}^{ij} & =\frac{7a^{2}\gamma^{R}V_{0}^{(2s)}}{2r^{3}}\left(\boldsymbol{e}_{1}^{j}\cdot\hat{\boldsymbol{r}}^{ij}\right)\left(\boldsymbol{e}_{1}^{j}\times\hat{\boldsymbol{r}}^{ij}\right),\label{eq:2s_torque}
\end{align}
where $r=\left|\boldsymbol{r}^{i}-\boldsymbol{r}^{j}\right|$ is the
distance between squirmers $i$ and $j$ and $\hat{\boldsymbol{r}}^{ij}=\left|\boldsymbol{r}^{i}-\boldsymbol{r}^{j}\right|/r$
is the normalized relative separation between the centers of the swimmers. 

Observe that the active forces and torques in Eqs. \eqref{eq:2s_force}-\eqref{eq:2s_torque}
are invariant under isometries. This can be seen for example by looking
at the body frame components $\boldsymbol{e}_{a}^{j}\cdot\bm{F}_{A}^{ij}$
of the active force, which depend only on the components of the relative
displacement $\Delta_{a}^{j}=\boldsymbol{e}_{1}^{j}\cdot\left(\boldsymbol{r}^{i}-\boldsymbol{r}^{j}\right)$
and distance $r=\sqrt{\sum_{a=1}^{3}\left(\Delta_{a}^{j}\right)^{2}}$
between the particles. The active forces and torques in Eqs. \eqref{eq:2s_force}-\eqref{eq:2s_torque}
are also nonreciprocal: for instance, $\bm{F}_{A}^{ij}+\bm{F}_{A}^{ji}\neq\boldsymbol{0}$
generally, as $\bm{F}_{A}^{ij}$ depends on the orientation of particle
$j$ but not of particle $i$, and vice versa. Therefore, we expect
that the long-wavelength dynamics of the chain is governed by an effective
continuum elastic theory with nonreciprocal constitutive laws.

The straight configuration $\tilde{\boldsymbol{r}}^{i}=\begin{pmatrix}id & 0 & 0\end{pmatrix}^{T}$
with parallel frames $\tilde{\boldsymbol{e}}_{1}^{i}=\hat{\boldsymbol{x}},\tilde{\boldsymbol{e}}_{2}^{i}=\hat{\boldsymbol{y}},\tilde{\boldsymbol{e}}_{3}^{i}=\hat{\boldsymbol{z}}$
is a stationary solution of the equations of motion even in the presence
of the $(2s)$ swimming mode, as the active torques \eqref{eq:2s_torque}
vanish in this state, while the active forces \eqref{eq:2s_force}
cancel each other. Nevertheless, the particles are not force-free
in this configuration, only the net force on them is zero. At the
continuum level, this manifests in a nonzero prestress $\tilde{F}_{\parallel}$
in the constitutive relations, whose magnitude can be computed by
looking at the force one side of the chain exerts on the other in
the reference state. In the nearest-neighbor approximation, we get:
\begin{equation}
\tilde{F}_{\parallel}=-\frac{7a^{2}\gamma^{T}V_{0}^{(2s)}}{3d^{2}}=-\frac{7a^{2}}{3d}\Gamma^{T}V_{0}^{(2s)}\label{eq:2s_prestress}
\end{equation}

We linearize the discrete equations of motion around the reference
configuration and compare it with the linearized continuum equations
of motion for a general apolar, achiral rod. By matching the coefficients
in the microscopic and continuum models, we identify the elastic moduli,
listed in Table \ref{tab:moduli} in Appendix C. Focussing on transverse
perturbations, we look at the beam limit $\kappa_{T}\to\infty$ of
the continuum equations of motion by adiabatically eliminating $\boldsymbol{\varphi}_{\perp}$.
We find 
\begin{equation}
\Gamma^{T}\dot{\boldsymbol{u}}_{\perp}=\left(\tilde{F}_{\parallel}-H_{\circlearrowleft}^{f\tau}\right)\boldsymbol{u}_{\perp}''-\left(\kappa_{R}+C_{\perp}^{M\tau}\right)\boldsymbol{u}_{\perp}''''.\label{eq:2s_beam_lim}
\end{equation}
In this limit, activity changes the bending modulus $\kappa_{R}$
by $C_{\perp}^{M\tau}$ and introduces a term proportional to the
(linearized) curvature of the rod $\boldsymbol{u}_{\perp}''$ that
originates from prestress. From Eq. \eqref{eq:2s_beam_lim}, the acoustic
branch of the transverse mode in the long-wavelength limit $qd\ll1$
follows (to leading order in activity): 
\begin{equation}
\Lambda^{(2s)}\left(q\right)=\frac{7a^{2}}{3d^{2}}\frac{V_{0}^{(2s)}}{d}\left(qd\right)^{2}-\frac{\kappa_{R}}{\Gamma^{T}d^{4}}\left(qd\right)^{4}.\label{eq:2s_tr_displ}
\end{equation}

For chains of puller-type $V_{0}^{(2s)}<0$ squirmers, $\Lambda_{}^{(2s)}<0$
and the chain is transversely stable. In contrast, for chains of pusher-type
$V_{0}^{(2s)}>0$ squirmers, for an infinite chain the transverse
dispersion relation $\Lambda^{(2s)}$ is positive for arbitrary small
wavenumbers, leading a transverse instability \citep{jayaraman_autonomous_2012,lauga_zigzag_2021},
although for finite chains its exact nature depends on the boundary
conditions. The instability can be interpreted as an analogue of Euler
buckling: pusher-type swimmers exert extensile stresses along the
chain, which will buckle for sufficiently high activity. The analogy
is explicit in \eqref{eq:2s_beam_lim} since that equation governs
the time evolution of the transverse displacement for a prestressed
overdamped Euler-Bernoulli beam. As a curiosity, we remark that while
the full continuum equations of motion cannot be derived from an elastic
energy (as they do not conserve linear and angular momentum), the
right hand side of \eqref{eq:2s_beam_lim} follows from the negative
gradient of a standard elastic energy of the form
\[
E=\frac{1}{2}\int ds\left\{ \left(\tilde{F}_{\parallel}-H_{\circlearrowleft}^{f\tau}\right)\left|\boldsymbol{u}_{\perp}'\right|^{2}+\left(\kappa_{R}+C_{\perp}^{M\tau}\right)\left|\boldsymbol{u}_{\perp}''\right|^{2}\right\} .
\]
The long-wavelength buckling instability can be interpreted as the
energy functional losing positive definiteness for negative $\tilde{F}_{\parallel}-H_{\circlearrowleft}^{f\tau}$.

\subsection{Apolar, chiral chain}

\begin{figure*}
\centering
\includegraphics[width=1\textwidth]{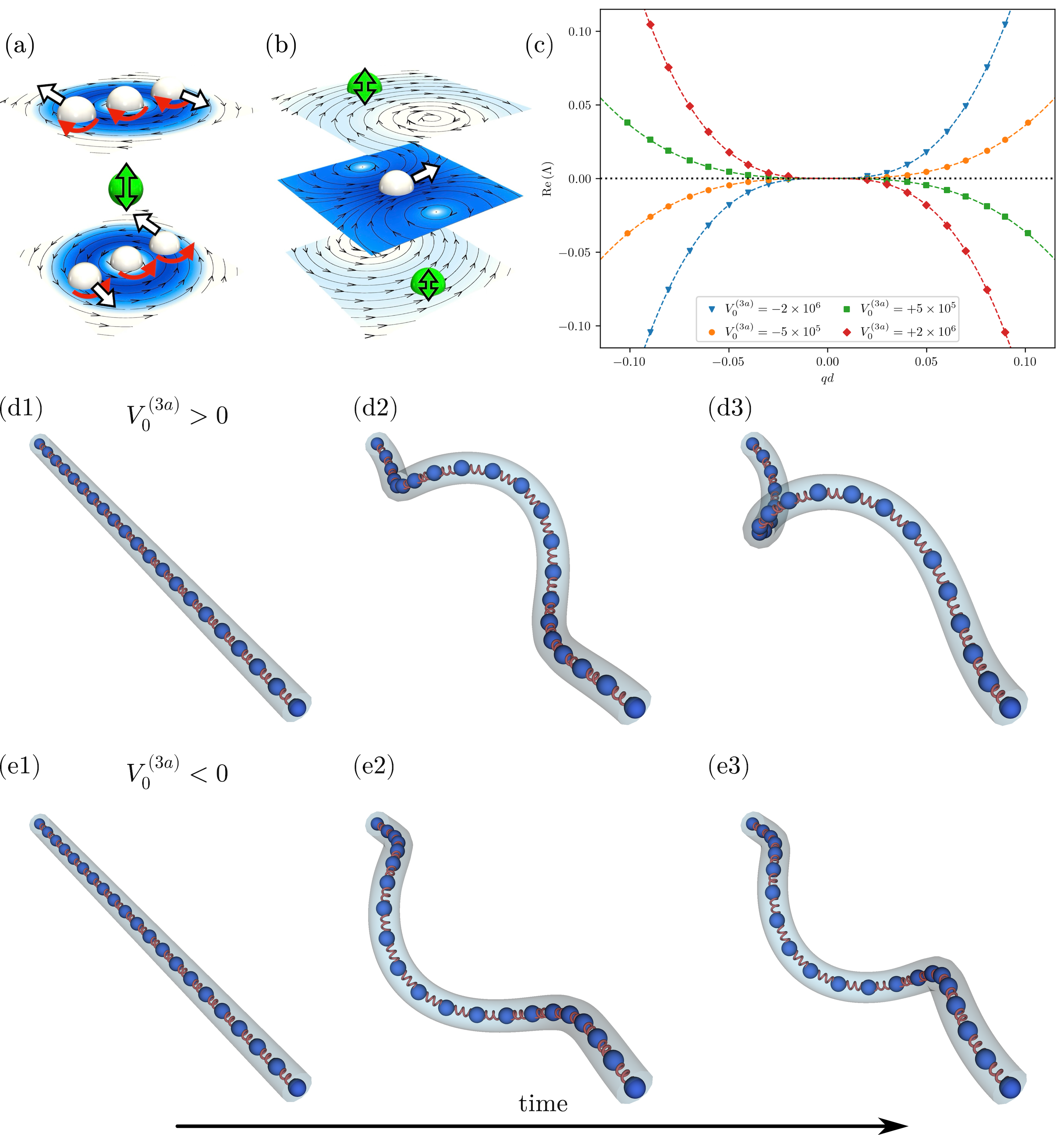}

\caption{(a)-(e) Phenomenology of an apolar, chiral squirmer chain and comparison
to continuum theory. (a) Flow field around a $(3a)$ squirmer (green
sphere) and forces and torques on tracer particles (white spheres)
around it. The green double arrow represents the squirming axis of
the active particle, white arrows the forces and red circular arrows
the torques on tracers. The flow field is shaded according to flow
velocity. (b) Response of apolar, chiral chain to a transverse perturbation:
an out-of-plane transverse force is generated, leading to a helical
buckling instability. (c) Growth rates of transverse perturbations
for apolar, chiral chains for small wavenumbers. Dashed lines are
theoretical predictions from Eq. \eqref{eq:3a_trans_ac_disp}, markers
are results of numerical simulations. (d) Time evolution of a chain
of $(3a)$ squirmers with $V_{0}^{(3a)}$ positive under clamped boundary
conditions, starting from random initial conditions. The chain buckles
into a chiral helical state due to the presence of activity. (e) Time
evolution of a chain of $(3a)$ squirmers with $V_{0}^{(3a)}$ negative
under clamped boundary conditions, starting from random initial conditions.
The chain buckles into a chiral helical state of opposite helicity
due to the presence of activity.\label{fig:3a}}
\end{figure*}
In this section, we look at an apolar, chiral chain, where each squirmer
is endowed with a $(3a)$ swimming mode (a torque dipole) along its
$\boldsymbol{e}_{1}$ body frame vector, see Fig. \ref{fig:3a}. The
active force and torque arising from this apolar, chiral mode exerted
by squirmer $j$ on squirmer $i$ are given by \citep{singh_generalized_2018}:
\begin{widetext}
\begin{align}
\bm{F}_{A}^{ij} & =\frac{13a^{3}}{12r^{3}}\gamma_{T}V_{0}^{(3a)}\left(\boldsymbol{e}_{1}^{j}\cdot\hat{\boldsymbol{r}}^{ij}\right)\left(\boldsymbol{e}_{1}^{j}\times\hat{\boldsymbol{r}}^{ij}\right),\label{eq:3a_force}\\
\bm{T}_{A}^{ij} & =\frac{13a^{3}}{24r^{4}}\gamma_{R}V_{0}^{(3a)}\left(\left[5\left(\boldsymbol{e}_{1}^{j}\cdot\hat{\boldsymbol{r}}^{ij}\right)^{2}-1\right]\hat{\boldsymbol{r}}^{ij}-2\left(\boldsymbol{e}_{1}^{j}\cdot\hat{\boldsymbol{r}}^{ij}\right)\boldsymbol{e}_{1}^{j}\right).\label{eq:3a_torque}
\end{align}
\end{widetext}

The straight configuration $\tilde{\boldsymbol{r}}^{i}=\begin{pmatrix}id & 0 & 0\end{pmatrix}^{T}$
with parallel frames $\tilde{\boldsymbol{e}}_{1}^{i}=\hat{\boldsymbol{x}},\tilde{\boldsymbol{e}}_{2}^{i}=\hat{\boldsymbol{y}},\tilde{\boldsymbol{e}}_{3}^{i}=\hat{\boldsymbol{z}}$
is again a stationary solution of the equations of motion even in
the presence of the $(3a)$ swimming mode, as the active forces vanish
in this state, while the active torques cancel each other. Nevertheless,
the particles are not torque-free in this configuration, only the
net torque upon them is zero. At the continuum level, this manifests
in a nonzero pre-moment stress $\tilde{M}_{\parallel}$ in the constitutive
relations, the magnitude of which can be computed by looking at the
torque one side of the chain exerts on the other in the reference
state. Restricting our attention to nearest-neighbor interactions,
we get:
\begin{equation}
\tilde{M}_{\parallel}=\frac{13a^{3}}{12d^{4}}\gamma^{R}V_{0}^{(3a)}=\frac{13a^{3}}{12d^{3}}\Gamma^{R}V_{0}^{(3a)}\label{eq:3a_pretorque}
\end{equation}

Chirality also allows for twist-stretch coupling in the longitudinal
direction, thereby providing a hydrodynamic analogue to chiral metamaterial
structures \citep{frenzel_three-dimensional_2017,montazeri_non-reciprocity_2025}.

We linearize the discrete equations of motion and and the continuum
ones for a chiral, apolar rod. By comparing the two, we identify the
constitutive moduli to leading order in the ratio $a/d$, listed in
Table \ref{tab:moduli} in Appendix C. Focussing on transverse perturbations,
we again look at the beam limit $\kappa_{T}\to\infty$ of the continuum
equations of motion by adiabatically eliminating $\boldsymbol{\varphi}_{\perp}$.
We find the following equation of motion:
\begin{equation}
\Gamma^{T}\dot{\boldsymbol{u}}_{\perp}=\left(\tilde{M}_{\parallel}-H_{\circlearrowleft}^{m\tau}\right)\hat{\boldsymbol{x}}\times\boldsymbol{u}_{\perp}'''-\kappa_{R}\boldsymbol{u}_{\perp}''''.\label{eq:3a_beam_lim}
\end{equation}
In this limit, activity introduces a term proportional to $\hat{\boldsymbol{x}}\times\boldsymbol{u}_{\perp}'''$
that originates from the preexisting moment stress $\tilde{M}_{\parallel}$.
From Eq. \eqref{eq:3a_beam_lim}, the acoustic branch of the transverse
mode in the long-wavelength limit $qd\ll1$ follows (to leading order
in activity, wavenumber and the ratio $a/d$):
\begin{equation}
\Lambda^{(3a)}\left(q\right)=-\frac{13a^{3}}{18d^{3}}\frac{V_{0}^{(3a)}}{d}\left(qd\right)^{3}-\frac{\kappa_{R}}{\Gamma^{T}d^{4}}\left(qd\right)^{4}.\label{eq:3a_trans_ac_disp}
\end{equation}
This implies that an infinite chain of torque dipoles always develops
a helical instability with the sign of helicity selected by the activity
such that $V_{0}^{(3a)}k<0$. This can be interpreted as an active
helical buckling due to a preexisting twist in the material because
of the active torque dipoles.

Note that there is a slight difference between the rate of the instability
as predicted by the continuum theory and the actual rate from the
discrete equations. This is due to the fact that the instability is
of third order in gradients, while the effective continuum theory
is only accurate up to second order in gradients. This subtle effect
only modifies the rate of the instability but does not alter it qualitatively.

Just as in the case of an active apolar, achiral beam, the full continuum
equations of motion cannot be derived from an elastic energy, but
in the beam limit the right hand side of \eqref{eq:3a_beam_lim} does
follow from the negative gradient of an elastic energy of the form
\[
E=\frac{1}{2}\int ds\left\{ \left(H_{\circlearrowleft}^{m\tau}-\tilde{M}_{\parallel}\right)\hat{\boldsymbol{x}}\cdot\left(\boldsymbol{u}_{\perp}'\times\boldsymbol{u}_{\perp}''\right)+\kappa_{R}\left|\boldsymbol{u}_{\perp}''\right|^{2}\right\} .
\]
The helical buckling instability can then be interpreted as the energy
functional never being positive definite owing to the chiral term
$\hat{\boldsymbol{x}}\cdot\left(\boldsymbol{u}_{\perp}'\times\boldsymbol{u}_{\perp}''\right)$.

\subsection{Polar, achiral chain}

\begin{figure*}
\centering
\includegraphics[width=1\textwidth]{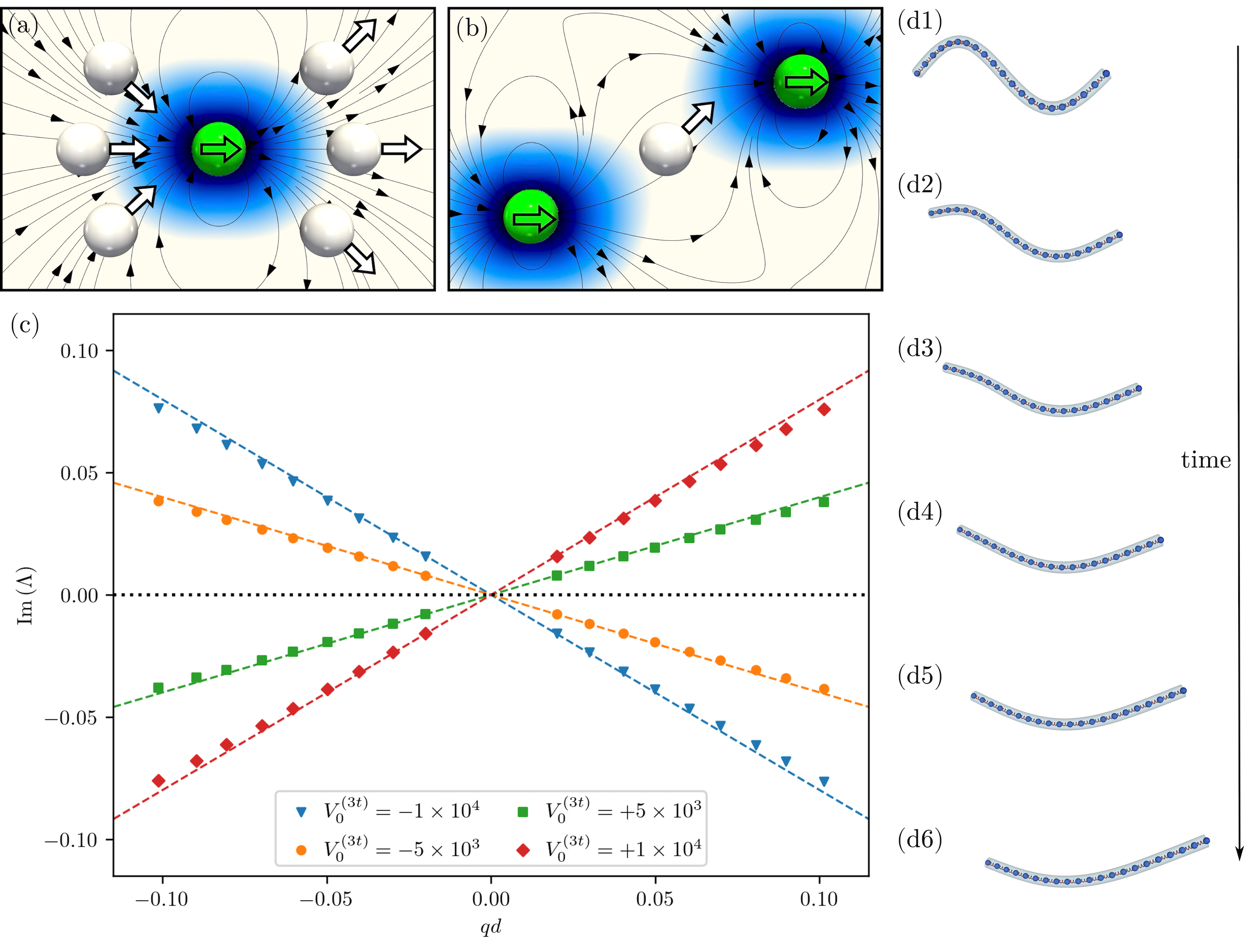}

\caption{(a)-(d) Phenomenology of a polar, achiral squirmer chain and comparison
to continuum theory. (a) Flow field around a $(3t)$ squirmer (green
sphere) and forces and torques on tracer particles (white spheres)
around it. The green arrow represents the squirming axis of the active
particle, white arrows the forces and red circular arrows the torques
on tracers. The flow field is shaded according to flow velocity. (b)
Response of polar, achiral chain to a transverse perturbation: a transverse
displacement generates forces along the chain, leading to advection.
(c) Phase velocity of transverse waves for polar, achiral chains for
small wavenumbers. Dashed lines are theoretical predictions from Eq.
\eqref{eq:3t_trans_ac_disp}, markers are results of numerical simulations.
(d) Time evolution of a chain of $(3t)$ squirmers under free boundary
conditions, starting from an initial sinusoidal chain. The chain propels
itself and the waveform travels together with it. The chain is eventually
straightened due to the bending rigidity.\label{fig:3t}}
\end{figure*}
In this section, we consider a polar, achiral chain, where each squirmer
is endowed with a $(3t)$ squirming mode (a source dipole) along its
$\boldsymbol{e}_{1}$ body frame vector, see Fig. \ref{fig:3t}. The
active force and torque exerted by squirmer $j$ on squirmer $i$
due to this mode are given by \citep{singh_generalized_2018}:
\begin{align}
\bm{F}_{A}^{ij} & =-\frac{a^{3}}{5r^{3}}\gamma^{T}V_{0}^{(3t)}\left(\bm{e}_{1}^{j}-3\left(\bm{e}_{1}^{j}\cdot\hat{\bm{r}}^{ij}\right)\hat{\bm{r}}^{ij}\right),\label{eq:3t_force}\\
\bm{T}_{A}^{ij} & =\bm{0}.\label{eq:3t_torque}
\end{align}
The torque is zero as the (3t) squirming mode produces a potential
flow with zero vorticity \citep{pozrikidis_boundary_1992}.

The polar, achiral chain generates a net force even in a straight
configuration $\left(\tilde{\bm{r}}_{i},\tilde{\bm{e}}_{a}^{i}\right)$,
leading to a uniformly translating steady state. The pre-force $\tilde{f}_{\parallel}$
can be computed by looking at the net force exerted on particle $i$
by its neighbors and is given by
\begin{equation}
\tilde{f}_{\parallel}=\frac{4a^{3}}{5d^{4}}\gamma^{T}V_{0}^{(3t)}=\frac{4a^{3}}{5d^{3}}\Gamma^{T}V_{0}^{(3t)}.\label{eq:3t_pre_force}
\end{equation}
This means that the steady state translates with a constant velocity
\begin{align}
\bm{V}^{(0)} & =\frac{4a^{3}}{5d^{3}}V_{0}^{(3t)}\hat{\bm{x}}\equiv V^{(0)}\hat{\boldsymbol{x}}.\label{eq:3t_vel}
\end{align}
We linearize both the discrete equations of motion and the continuum
equations for a polar, achiral chain. By matching the coefficients
from the microscopic and continuum descriptions, we identify the elastic
moduli, listed in Table \ref{tab:moduli} in Appendix C.

The dispersion relations for longitudinal displacement waves in the
long-wavelength limit $qd\ll1$, derived from Eqs. \eqref{eq:3t_discr_tr}
and \eqref{eq:3t_discr_rot}, are: 
\begin{align}
\Lambda_{\parallel}^{(3t)}\left(q\right) & =-\frac{12ia^{3}}{5d^{3}}\frac{V_{0}^{(3t)}}{d}\left(qd\right)-\frac{\kappa_{T}}{\Gamma^{T}d^{2}}\left(qd\right)^{2},\label{eq:3t_long_tr_disp}
\end{align}
The chain remains stable with respect to longitudinal translational
perturbations, as the active contribution is purely imaginary and
thus modifies only the frequency of the modes without affecting their
stability. The appearance of a nonzero imaginary part in Eq. \eqref{eq:3t_long_tr_disp}
implies that there are longitudinal traveling waves on a polar chain.
These are analogous to waves previously seen in nonreciprocal robotic
metamaterials \citep{brandenbourger_non-reciprocal_2019}, where linear
momentum conservation was explicitly broken by springs that exerted
different forces when displaced to the left or right.

For transverse perturbations, we look at the beam limit $\kappa_{T}\to\infty$
of the continuum equations of motion and eliminate $\boldsymbol{\varphi}_{\perp}$
adiabatically. We find the following equation of motion:
\begin{equation}
\Gamma^{T}\dot{\boldsymbol{u}}_{\perp}=\Gamma^{T}V^{(0)}\boldsymbol{u}_{\perp}'-H_{\perp}^{m\tau}\boldsymbol{u}_{\perp}'''-\kappa_{R}\boldsymbol{u}_{\perp}''''.\label{eq:3t_beam_lim}
\end{equation}
In this limit, activity introduces advective terms proportional to
$\boldsymbol{u}_{\perp}'$ and $\boldsymbol{u}_{\perp}'''$ into the
beam equation. From Eq. \eqref{eq:3t_beam_lim}, the acoustic branch
of the transverse mode in the long-wavelength limit $qd\ll1$ follows
(to leading order in activity and wavenumber):
\begin{equation}
\Lambda^{(3t)}\left(q\right)=\frac{iV^{(0)}}{d}\left(qd\right)-\frac{\kappa_{R}}{\Gamma^{T}d^{4}}\left(qd\right)^{4}.\label{eq:3t_trans_ac_disp}
\end{equation}
Therefore a polar achiral chain hosts stable transverse traveling
waves, even in the overdamped limit. To leading order in wavenumber,
the waves travel with the speed of chain given in Eq. \eqref{eq:3t_vel}.
This term can be removed by going into a reference frame co-moving
with the chain, in which case the leading order contribution from
activity will be proportional to $\left(qd\right)^{3}$.

\subsection{Polar, chiral chain}

\begin{figure*}
\centering
\includegraphics[width=1\textwidth]{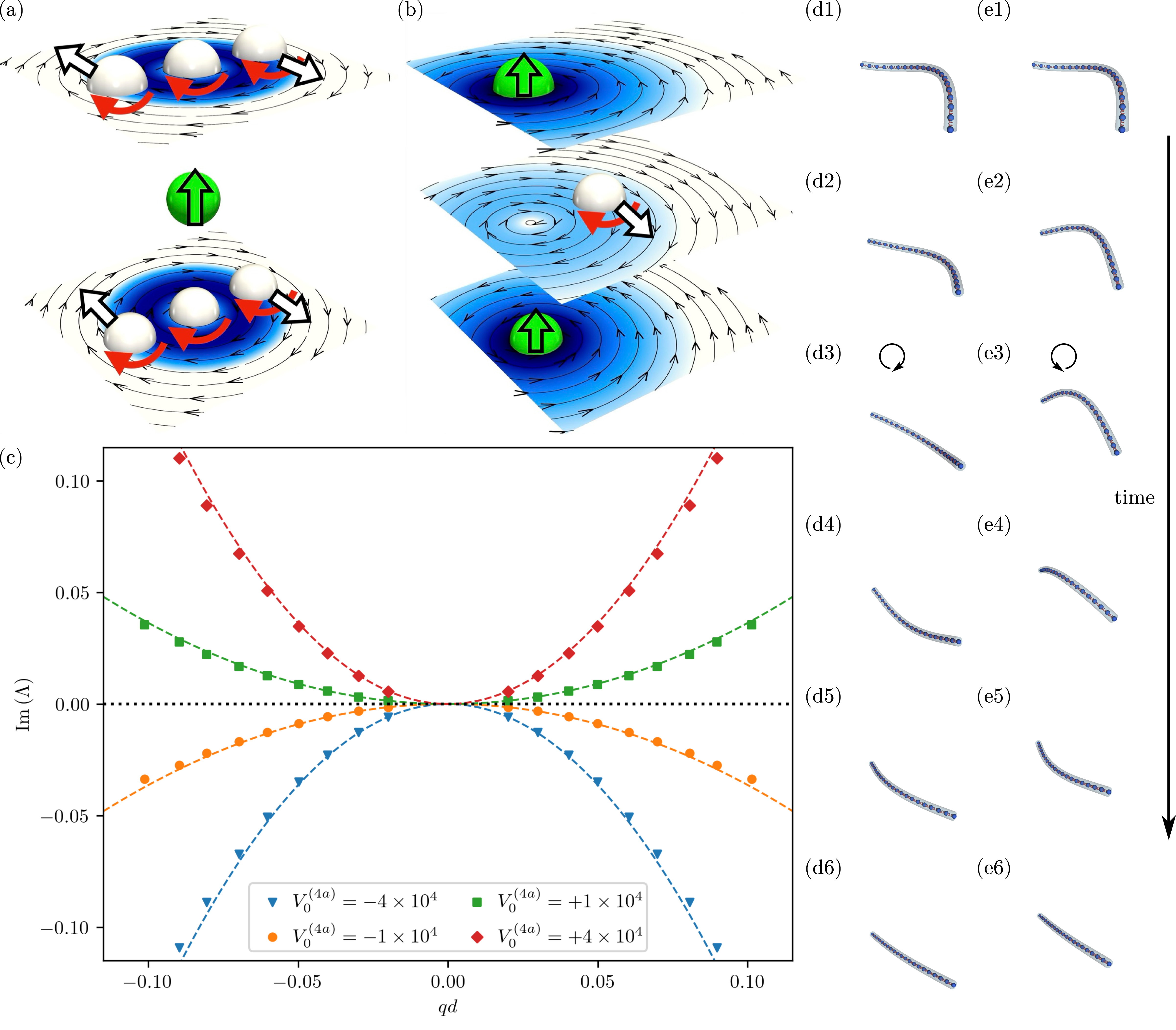}

\caption{(a)-(e) Phenomenology of a polar, chiral squirmer chain and comparison
to continuum theory. (a) Flow field around a $(4a)$ squirmer (green
sphere) and forces and torques on tracer particles (white spheres)
around it. The green arrow represents the squirming axis of the active
particle, white arrows the forces and red circular arrows the torques
on tracers. The flow field is shaded according to flow velocity. (b)
Response of polar, chiral chain to a transverse perturbation: a transverse
displacement generates forces out of the plane of deformation, leading
to a transverse force perpendicular to the displacement. (c) Phase
velocity of transverse waves for polar, chiral chains for small wavenumbers.
Dashed lines are theoretical predictions from Eq. \eqref{eq:4a_trans_ac_disp},
markers are results of numerical simulations. (d) Time evolution of
a chain of $(4a)$ squirmers under free boundary conditions, starting
from an initial helical chain of positive helicity. The chain rotates
in a spiral fashion and straightens out due to the bending rigidity.
(e) Time evolution of a chain of $(4a)$ squirmers under free boundary
conditions, starting from an initial helical chain of negative helicity.
The chain rotates in the opposite direction. Circular arrows near
(d3) and (e3) illustrate the direction of rotation of the chains.\label{fig:4a}}
\end{figure*}
We now examine a polar, chiral chain, where each squirmer is endowed
with a $\left(4a\right)$ squirming mode (a chiral octupole) along
its $\boldsymbol{e}_{1}$ body frame vector, see Fig. \ref{fig:4a}.
The active force and torque arising from this mode exerted by squirmer
$j$ on squirmer $i$ are given by \citep{singh_generalized_2018}:
\begin{widetext}
\begin{align}
\bm{F}_{A}^{ij} & =\frac{363a^{4}}{40r^{4}}\gamma^{T}V_{0}^{(4a)}\left(1-5\left(\bm{e}_{1}^{j}\cdot\hat{\bm{r}}^{ij}\right)^{2}\right)\left(\bm{e}_{j}^{1}\times\hat{\bm{r}}^{ij}\right),\label{eq:4a_force}\\
\bm{T}_{A}^{ij} & =\frac{363a^{4}}{80r^{5}}\gamma^{R}V_{0}^{(4a)}\left(-3\bm{e}_{j}^{1}+15\left[\left(\bm{e}_{1}^{j}\cdot\hat{\bm{r}}^{ij}\right)\hat{\bm{r}}^{ij}+\left(\bm{e}_{1}^{j}\cdot\hat{\bm{r}}^{ij}\right)^{2}\right]\bm{e}_{1}^{j}-35\left(\bm{e}_{1}^{j}\cdot\hat{\bm{r}}^{ij}\right)^{3}\hat{\bm{r}}^{ij}\right).\label{eq:4a_torque}
\end{align}
\end{widetext}

Even in a straight configuration $\left(\tilde{\bm{r}}_{i},\tilde{\bm{e}}_{a}^{i}\right)$,
the polar and chiral chain experiences a net torque, leading to a
uniformly rotating steady state. The pre-torque $\tilde{m}_{\parallel}$
can be computed by looking at the net torque exerted on particle $i$
by its neighbors and is given by
\begin{equation}
\tilde{m}_{\parallel}=-\frac{363a^{4}}{5d^{6}}\gamma^{R}V_{0}^{(4a)}=-\frac{363a^{4}}{5d^{5}}\Gamma^{R}V_{0}^{(4a)}.\label{eq:4a_pre_torque}
\end{equation}
This means that in the steady state, the chain rotates with a constant
angular velocity
\begin{align}
\bm{\Omega}^{(0)} & =-\frac{363a^{4}}{5d^{4}}\frac{V_{0}^{(4a)}}{d}\hat{\bm{x}}.\label{eq:4a_ang_vel}
\end{align}
We linearize both the discrete equations of motion and the continuum
equations for a polar, chiral chain. The moduli identified from matching
discrete and continuum models are, to leading order in the ratio $a/d$,
listed in Table \ref{tab:moduli} in Appendix C. Focussing on transverse
perturbations, we look at the beam limit $\kappa_{T}\to\infty$ of
the continuum equations of motion. We find the following equation
of motion:
\begin{equation}
\Gamma^{T}\dot{\boldsymbol{u}}_{\perp}=H_{\perp}^{f\tau}\hat{\boldsymbol{x}}\times\boldsymbol{u}_{\perp}''-C_{\circlearrowleft}^{M\tau}\hat{\boldsymbol{x}}\times\boldsymbol{u}_{\perp}''''-\kappa_{R}\boldsymbol{u}_{\perp}''''.\label{eq:4a_beam_lim}
\end{equation}
In a polar, chiral chain, activity introduces a term proportional
to $\hat{\boldsymbol{x}}\times\boldsymbol{u}_{\perp}''$ at lowest
order in gradients and an odd bending modulus $C_{\circlearrowleft}^{M\tau}$
\citep{chen_realization_2021}. The acoustic branch of the transverse
modes in the long-wavelength limit $qd\to0$ follows (to leading order
in activity, wavenumber and the ratio $a/d$):
\begin{equation}
\Lambda^{(4a)}\left(q\right)=\frac{363ia^{4}}{10d^{4}}\frac{V_{0}^{(4a)}}{d}\left(qd\right)^{2}-\frac{\kappa_{R}}{\Gamma^{T}d^{4}}\left(qd\right)^{4}.\label{eq:4a_trans_ac_disp}
\end{equation}
Thus, a polar and chiral chain hosts stable propagating chiral transverse
spiral waves. As the imaginary part of the growth rate in Eq. \eqref{eq:4a_trans_ac_disp}
is even in wavenumber, on a polar and chiral chain, circularly polarized
waves of opposite handedness rotate in opposite directions.

\section{Discussion\label{sec:discussion}}

In this paper we have outlined a framework to study nonreciprocal
oriented active solids by augmenting the constitutive laws of Cosserat
elasticity to include sources of linear and angular momentum. Assuming
invariance of interactions under rigid transformations, we have shown
that these constitutive sources can only depend on strains. We then
specialized to a one-dimensional Cosserat rod model, where we have
classified the coupling constants based on the symmetry of the material
under rotations and mirror reflections. By means of a chain of squirmers
in a Stokesian fluid, we have provided an explicit discrete realization
of the continuum model and calculated the effective elastic moduli
by coarse-graining. The polarity and chirality of the elastic moduli
were consistent with the continuum predictions based on Curie's principle.
We have also shown how odd elastic moduli naturally emerge from the
active interactions between squirmers. We have obtained effective
active beam equations for each symmetry combination and the linearized
mode structure of the active chains. We demonstrated that the elastic
terms coming from activity are lower order in gradients than passive
elasticity, giving rise to rich long-wavelength phenomena such as
bending and helical instabilities, traveling waves and spiral waves.
Our numerical simulations have shown excellent agreement with the
theoretical predictions.

Our work can be continued and extended in numerous directions. For
simplicity, we have only considered the small-displacement dynamics
of the active chains, however, the nonreciprocal constitutive relations
are still applicable when the displacements are large, as they only
require strain to be small. It would be interesting to consider geometrically
nonlinear systems where the constitutive relations are kept linear
in the strains but the displacements are not linearized. As we have
argued, apolar modes typically give rise to instabilities, while polar
ones to traveling waves, raising the question whether nonlinear self-sustained
oscillations are possible in the presence of both apolar and polar
modes. Incorporating thermal noise into our description could also
lead to exciting phenomena, as the eigenvalues of the linearized dynamics
can acquire imaginary parts, which can result in stochastic limit
cycles. 

Another promising continuation of our work could be a more careful
study of nonreciprocal Cosserat solids in dimensions greater than
one. The nonreciprocal terms could again give contributions lower
order in gradients than passive elasticity and potentially dramatically
alter the long-wavelength dynamics of the solid. Perhaps more interestingly,
though, in dimensions greater than one the theory allows for topological
defects, whose behavior is fundamentally altered by the presence of
active and nonreciprocal forces and moments \citep{braverman_topological_2021,bililign_motile_2022,poncet_when_2022}.
For a Cosserat solid, dislocations and disclinations are independent
and interact nontrivially with each other \citep{schaefer_cosserat_1967}.
We expect that topological defects become motile and might rotate
or translate owing to the active forces and moments around them. We
hope that our work will provide inspiration for metamaterial design
and pave the way for constitutive modeling of oriented active solids.

\section*{Acknowledgements}

We thank Rajesh Singh for stimulating discussions. This research was
supported by Engineering and Physical Sciences Research Council Grant
No. EP/W524141/1 (B.N.), by the JSPS Core-to-Core Program ``Advanced
core-to-core network for the physics of self-organizing active matter
Grant No. JPJSCCA20230002'' (T.K.), by JST SPRING, Grant No. JPMJSP2110
(T.K.), and in part by NSF Grant No. PHY-2309135 to the Kavli Institute
for Theoretical Physics.

\bibliographystyle{apsrev4-2}
\bibliography{manuscript}

\begin{thebibliography}{86}%
\makeatletter
\providecommand \@ifxundefined [1]{%
 \@ifx{#1\undefined}
}%
\providecommand \@ifnum [1]{%
 \ifnum #1\expandafter \@firstoftwo
 \else \expandafter \@secondoftwo
 \fi
}%
\providecommand \@ifx [1]{%
 \ifx #1\expandafter \@firstoftwo
 \else \expandafter \@secondoftwo
 \fi
}%
\providecommand \natexlab [1]{#1}%
\providecommand \enquote  [1]{``#1''}%
\providecommand \bibnamefont  [1]{#1}%
\providecommand \bibfnamefont [1]{#1}%
\providecommand \citenamefont [1]{#1}%
\providecommand \href@noop [0]{\@secondoftwo}%
\providecommand \href [0]{\begingroup \@sanitize@url \@href}%
\providecommand \@href[1]{\@@startlink{#1}\@@href}%
\providecommand \@@href[1]{\endgroup#1\@@endlink}%
\providecommand \@sanitize@url [0]{\catcode `\\12\catcode `\$12\catcode
  `\&12\catcode `\#12\catcode `\^12\catcode `\_12\catcode `\%12\relax}%
\providecommand \@@startlink[1]{}%
\providecommand \@@endlink[0]{}%
\providecommand \url  [0]{\begingroup\@sanitize@url \@url }%
\providecommand \@url [1]{\endgroup\@href {#1}{\urlprefix }}%
\providecommand \urlprefix  [0]{URL }%
\providecommand \Eprint [0]{\href }%
\providecommand \doibase [0]{https://doi.org/}%
\providecommand \selectlanguage [0]{\@gobble}%
\providecommand \bibinfo  [0]{\@secondoftwo}%
\providecommand \bibfield  [0]{\@secondoftwo}%
\providecommand \translation [1]{[#1]}%
\providecommand \BibitemOpen [0]{}%
\providecommand \bibitemStop [0]{}%
\providecommand \bibitemNoStop [0]{.\EOS\space}%
\providecommand \EOS [0]{\spacefactor3000\relax}%
\providecommand \BibitemShut  [1]{\csname bibitem#1\endcsname}%
\let\auto@bib@innerbib\@empty
\bibitem [{\citenamefont {Griffiths}(2023)}]{griffiths_introduction_2023}%
  \BibitemOpen
  \bibfield  {author} {\bibinfo {author} {\bibfnamefont {D.~J.}\ \bibnamefont
  {Griffiths}},\ }\href@noop {} {\emph {\bibinfo {title} {Introduction to
  {Electrodynamics}}}},\ \bibinfo {edition} {5th}\ ed.\ (\bibinfo  {publisher}
  {Cambridge University Press},\ \bibinfo {address} {Cambridge},\ \bibinfo
  {year} {2023})\BibitemShut {NoStop}%
\bibitem [{\citenamefont {Smoluchowski}(1911)}]{smoluchowski_mutual_1911}%
  \BibitemOpen
  \bibfield  {author} {\bibinfo {author} {\bibfnamefont {M.}~\bibnamefont
  {Smoluchowski}},\ }\href@noop {} {\bibfield  {journal} {\bibinfo  {journal}
  {Bull. Acad. Sci. Cracovie A}\ }\textbf {\bibinfo {volume} {1}},\ \bibinfo
  {pages} {23} (\bibinfo {year} {1911})}\BibitemShut {NoStop}%
\bibitem [{\citenamefont {Buenzli}\ and\ \citenamefont
  {Soto}(2008)}]{buenzli_violation_2008}%
  \BibitemOpen
  \bibfield  {author} {\bibinfo {author} {\bibfnamefont {P.~R.}\ \bibnamefont
  {Buenzli}}\ and\ \bibinfo {author} {\bibfnamefont {R.}~\bibnamefont {Soto}},\
  }\href {https://doi.org/10.1103/PhysRevE.78.020102} {\bibfield  {journal}
  {\bibinfo  {journal} {Physical Review E}\ }\textbf {\bibinfo {volume} {78}},\
  \bibinfo {pages} {020102} (\bibinfo {year} {2008})}\BibitemShut {NoStop}%
\bibitem [{\citenamefont {Dholakia}\ and\ \citenamefont
  {Zemánek}(2010)}]{dholakia_colloquium_2010}%
  \BibitemOpen
  \bibfield  {author} {\bibinfo {author} {\bibfnamefont {K.}~\bibnamefont
  {Dholakia}}\ and\ \bibinfo {author} {\bibfnamefont {P.}~\bibnamefont
  {Zemánek}},\ }\href {https://doi.org/10.1103/RevModPhys.82.1767} {\bibfield
  {journal} {\bibinfo  {journal} {Reviews of Modern Physics}\ }\textbf
  {\bibinfo {volume} {82}},\ \bibinfo {pages} {1767} (\bibinfo {year}
  {2010})}\BibitemShut {NoStop}%
\bibitem [{\citenamefont {Ivlev}\ \emph {et~al.}(2012)\citenamefont {Ivlev},
  \citenamefont {Löwen}, \citenamefont {Morfill},\ and\ \citenamefont
  {Royall}}]{ivlev_complex_2012}%
  \BibitemOpen
  \bibfield  {author} {\bibinfo {author} {\bibfnamefont {A.}~\bibnamefont
  {Ivlev}}, \bibinfo {author} {\bibfnamefont {H.}~\bibnamefont {Löwen}},
  \bibinfo {author} {\bibfnamefont {G.}~\bibnamefont {Morfill}},\ and\ \bibinfo
  {author} {\bibfnamefont {C.~P.}\ \bibnamefont {Royall}},\ }\href
  {https://doi.org/10.1142/8139} {\emph {\bibinfo {title} {Complex {Plasmas}
  and {Colloidal} {Dispersions}: {Particle}-{Resolved} {Studies} of {Classical}
  {Liquids} and {Solids}}}},\ \bibinfo {series} {Series in {Soft} {Condensed}
  {Matter}}, Vol.~\bibinfo {volume} {5}\ (\bibinfo  {publisher} {World
  Scientific},\ \bibinfo {year} {2012})\BibitemShut {NoStop}%
\bibitem [{\citenamefont {Soto}\ and\ \citenamefont
  {Golestanian}(2014)}]{soto_self-assembly_2014}%
  \BibitemOpen
  \bibfield  {author} {\bibinfo {author} {\bibfnamefont {R.}~\bibnamefont
  {Soto}}\ and\ \bibinfo {author} {\bibfnamefont {R.}~\bibnamefont
  {Golestanian}},\ }\href {https://doi.org/10.1103/PhysRevLett.112.068301}
  {\bibfield  {journal} {\bibinfo  {journal} {Physical Review Letters}\
  }\textbf {\bibinfo {volume} {112}},\ \bibinfo {pages} {068301} (\bibinfo
  {year} {2014})}\BibitemShut {NoStop}%
\bibitem [{\citenamefont {Ivlev}\ \emph {et~al.}(2015)\citenamefont {Ivlev},
  \citenamefont {Bartnick}, \citenamefont {Heinen}, \citenamefont {Du},
  \citenamefont {Nosenko},\ and\ \citenamefont
  {Löwen}}]{ivlev_statistical_2015}%
  \BibitemOpen
  \bibfield  {author} {\bibinfo {author} {\bibfnamefont {A.}~\bibnamefont
  {Ivlev}}, \bibinfo {author} {\bibfnamefont {J.}~\bibnamefont {Bartnick}},
  \bibinfo {author} {\bibfnamefont {M.}~\bibnamefont {Heinen}}, \bibinfo
  {author} {\bibfnamefont {C.-R.}\ \bibnamefont {Du}}, \bibinfo {author}
  {\bibfnamefont {V.}~\bibnamefont {Nosenko}},\ and\ \bibinfo {author}
  {\bibfnamefont {H.}~\bibnamefont {Löwen}},\ }\href
  {https://doi.org/10.1103/PhysRevX.5.011035} {\bibfield  {journal} {\bibinfo
  {journal} {Physical Review X}\ }\textbf {\bibinfo {volume} {5}},\ \bibinfo
  {pages} {011035} (\bibinfo {year} {2015})}\BibitemShut {NoStop}%
\bibitem [{\citenamefont {Saha}\ \emph {et~al.}(2019)\citenamefont {Saha},
  \citenamefont {Ramaswamy},\ and\ \citenamefont
  {Golestanian}}]{saha_pairing_2019}%
  \BibitemOpen
  \bibfield  {author} {\bibinfo {author} {\bibfnamefont {S.}~\bibnamefont
  {Saha}}, \bibinfo {author} {\bibfnamefont {S.}~\bibnamefont {Ramaswamy}},\
  and\ \bibinfo {author} {\bibfnamefont {R.}~\bibnamefont {Golestanian}},\
  }\href {https://doi.org/10.1088/1367-2630/ab20fd} {\bibfield  {journal}
  {\bibinfo  {journal} {New Journal of Physics}\ }\textbf {\bibinfo {volume}
  {21}},\ \bibinfo {pages} {063006} (\bibinfo {year} {2019})}\BibitemShut
  {NoStop}%
\bibitem [{\citenamefont {Zhang}\ \emph {et~al.}(2021)\citenamefont {Zhang},
  \citenamefont {Alert}, \citenamefont {Yan}, \citenamefont {Wingreen},\ and\
  \citenamefont {Granick}}]{zhang_active_2021}%
  \BibitemOpen
  \bibfield  {author} {\bibinfo {author} {\bibfnamefont {J.}~\bibnamefont
  {Zhang}}, \bibinfo {author} {\bibfnamefont {R.}~\bibnamefont {Alert}},
  \bibinfo {author} {\bibfnamefont {J.}~\bibnamefont {Yan}}, \bibinfo {author}
  {\bibfnamefont {N.~S.}\ \bibnamefont {Wingreen}},\ and\ \bibinfo {author}
  {\bibfnamefont {S.}~\bibnamefont {Granick}},\ }\href
  {https://doi.org/10.1038/s41567-021-01238-8} {\bibfield  {journal} {\bibinfo
  {journal} {Nature Physics}\ }\textbf {\bibinfo {volume} {17}},\ \bibinfo
  {pages} {961} (\bibinfo {year} {2021})}\BibitemShut {NoStop}%
\bibitem [{\citenamefont {Singh}\ and\ \citenamefont
  {Adhikari}(2016)}]{singh_universal_2016}%
  \BibitemOpen
  \bibfield  {author} {\bibinfo {author} {\bibfnamefont {R.}~\bibnamefont
  {Singh}}\ and\ \bibinfo {author} {\bibfnamefont {R.}~\bibnamefont
  {Adhikari}},\ }\href {https://doi.org/10.1103/PhysRevLett.117.228002}
  {\bibfield  {journal} {\bibinfo  {journal} {Physical Review Letters}\
  }\textbf {\bibinfo {volume} {117}},\ \bibinfo {pages} {228002} (\bibinfo
  {year} {2016})}\BibitemShut {NoStop}%
\bibitem [{\citenamefont {Singh}\ and\ \citenamefont
  {Adhikari}(2018)}]{singh_generalized_2018}%
  \BibitemOpen
  \bibfield  {author} {\bibinfo {author} {\bibfnamefont {R.}~\bibnamefont
  {Singh}}\ and\ \bibinfo {author} {\bibfnamefont {R.}~\bibnamefont
  {Adhikari}},\ }\href {https://doi.org/10.1088/2399-6528/aaab0d} {\bibfield
  {journal} {\bibinfo  {journal} {Journal of Physics Communications}\ }\textbf
  {\bibinfo {volume} {2}},\ \bibinfo {pages} {025025} (\bibinfo {year}
  {2018})}\BibitemShut {NoStop}%
\bibitem [{\citenamefont {Oseen}(1927)}]{oseen_neuere_1927}%
  \BibitemOpen
  \bibfield  {author} {\bibinfo {author} {\bibfnamefont {C.~W.}\ \bibnamefont
  {Oseen}},\ }\href@noop {} {\emph {\bibinfo {title} {Neuere {Methoden} und
  {Ergebnisse} in der {Hydrodynamik}}}}\ (\bibinfo  {publisher} {Akademische
  Verlagsgesellschaft},\ \bibinfo {address} {Leipzig},\ \bibinfo {year}
  {1927})\BibitemShut {NoStop}%
\bibitem [{\citenamefont {Happel}\ and\ \citenamefont
  {Brenner}(1983)}]{happel_low_1983}%
  \BibitemOpen
  \bibfield  {author} {\bibinfo {author} {\bibfnamefont {J.}~\bibnamefont
  {Happel}}\ and\ \bibinfo {author} {\bibfnamefont {H.}~\bibnamefont
  {Brenner}},\ }\href {https://doi.org/10.1007/978-94-009-8352-6} {\emph
  {\bibinfo {title} {Low {Reynolds} number hydrodynamics: with special
  applications to particulate media}}},\ edited by\ \bibinfo {editor}
  {\bibfnamefont {R.~J.}\ \bibnamefont {Moreau}},\ \bibinfo {series} {Mechanics
  of fluids and transport processes}, Vol.~\bibinfo {volume} {1}\ (\bibinfo
  {publisher} {Springer Netherlands},\ \bibinfo {address} {Dordrecht},\
  \bibinfo {year} {1983})\BibitemShut {NoStop}%
\bibitem [{\citenamefont {Pozrikidis}(1992)}]{pozrikidis_boundary_1992}%
  \BibitemOpen
  \bibfield  {author} {\bibinfo {author} {\bibfnamefont {C.}~\bibnamefont
  {Pozrikidis}},\ }\href {https://doi.org/10.1017/CBO9780511624124} {\emph
  {\bibinfo {title} {Boundary {Integral} and {Singularity} {Methods} for
  {Linearized} {Viscous} {Flow}}}},\ \bibinfo {edition} {1st}\ ed.\ (\bibinfo
  {publisher} {Cambridge University Press},\ \bibinfo {year}
  {1992})\BibitemShut {NoStop}%
\bibitem [{\citenamefont {Love}(2013)}]{love_treatise_2013}%
  \BibitemOpen
  \bibfield  {author} {\bibinfo {author} {\bibfnamefont {A.~E.~H.}\
  \bibnamefont {Love}},\ }\href@noop {} {\emph {\bibinfo {title} {A treatise on
  the mathematical theory of elasticity}}},\ \bibinfo {edition} {4th}\ ed.\
  (\bibinfo  {publisher} {Cambridge Univ. Press},\ \bibinfo {address}
  {Cambridge},\ \bibinfo {year} {2013})\BibitemShut {NoStop}%
\bibitem [{\citenamefont {Curie}(1894)}]{curie_sur_1894}%
  \BibitemOpen
  \bibfield  {author} {\bibinfo {author} {\bibfnamefont {P.}~\bibnamefont
  {Curie}},\ }\href {https://doi.org/10.1051/jphystap:018940030039300}
  {\bibfield  {journal} {\bibinfo  {journal} {Journal de Physique Théorique et
  Appliquée}\ }\textbf {\bibinfo {volume} {3}},\ \bibinfo {pages} {393}
  (\bibinfo {year} {1894})}\BibitemShut {NoStop}%
\bibitem [{\citenamefont {Lighthill}(1952)}]{lighthill_squirming_1952}%
  \BibitemOpen
  \bibfield  {author} {\bibinfo {author} {\bibfnamefont {M.~J.}\ \bibnamefont
  {Lighthill}},\ }\href {https://doi.org/10.1002/cpa.3160050201} {\bibfield
  {journal} {\bibinfo  {journal} {Communications on Pure and Applied
  Mathematics}\ }\textbf {\bibinfo {volume} {5}},\ \bibinfo {pages} {109}
  (\bibinfo {year} {1952})}\BibitemShut {NoStop}%
\bibitem [{\citenamefont {Blake}(1971)}]{blake_spherical_1971}%
  \BibitemOpen
  \bibfield  {author} {\bibinfo {author} {\bibfnamefont {J.~R.}\ \bibnamefont
  {Blake}},\ }\href {https://doi.org/10.1017/S002211207100048X} {\bibfield
  {journal} {\bibinfo  {journal} {Journal of Fluid Mechanics}\ }\textbf
  {\bibinfo {volume} {46}},\ \bibinfo {pages} {199} (\bibinfo {year}
  {1971})}\BibitemShut {NoStop}%
\bibitem [{\citenamefont {Pak}\ and\ \citenamefont
  {Lauga}(2014)}]{pak_generalized_2014}%
  \BibitemOpen
  \bibfield  {author} {\bibinfo {author} {\bibfnamefont {O.~S.}\ \bibnamefont
  {Pak}}\ and\ \bibinfo {author} {\bibfnamefont {E.}~\bibnamefont {Lauga}},\
  }\href {https://doi.org/10.1007/s10665-014-9690-9} {\bibfield  {journal}
  {\bibinfo  {journal} {Journal of Engineering Mathematics}\ }\textbf {\bibinfo
  {volume} {88}},\ \bibinfo {pages} {1} (\bibinfo {year} {2014})}\BibitemShut
  {NoStop}%
\bibitem [{\citenamefont {Pedley}(2016)}]{pedley_spherical_2016}%
  \BibitemOpen
  \bibfield  {author} {\bibinfo {author} {\bibfnamefont {T.~J.}\ \bibnamefont
  {Pedley}},\ }\href {https://doi.org/10.1093/imamat/hxw030} {\bibfield
  {journal} {\bibinfo  {journal} {IMA Journal of Applied Mathematics}\ }\textbf
  {\bibinfo {volume} {81}},\ \bibinfo {pages} {488} (\bibinfo {year}
  {2016})}\BibitemShut {NoStop}%
\bibitem [{\citenamefont {Cosserat}\ and\ \citenamefont
  {Cosserat}(1909)}]{cosserat_theorie_1909}%
  \BibitemOpen
  \bibfield  {author} {\bibinfo {author} {\bibfnamefont {E.}~\bibnamefont
  {Cosserat}}\ and\ \bibinfo {author} {\bibfnamefont {F.}~\bibnamefont
  {Cosserat}},\ }\href@noop {} {\emph {\bibinfo {title} {Théorie des {Corps}
  {Deformables}}}}\ (\bibinfo  {publisher} {Hermann},\ \bibinfo {year}
  {1909})\BibitemShut {NoStop}%
\bibitem [{\citenamefont {Ericksen}\ and\ \citenamefont
  {Truesdell}(1957)}]{ericksen_exact_1957}%
  \BibitemOpen
  \bibfield  {author} {\bibinfo {author} {\bibfnamefont {J.~L.}\ \bibnamefont
  {Ericksen}}\ and\ \bibinfo {author} {\bibfnamefont {C.}~\bibnamefont
  {Truesdell}},\ }\href {https://doi.org/10.1007/BF00298012} {\bibfield
  {journal} {\bibinfo  {journal} {Archive for Rational Mechanics and Analysis}\
  }\textbf {\bibinfo {volume} {1}},\ \bibinfo {pages} {295} (\bibinfo {year}
  {1957})}\BibitemShut {NoStop}%
\bibitem [{\citenamefont {Schaefer}(1967)}]{schaefer_cosserat_1967}%
  \BibitemOpen
  \bibfield  {author} {\bibinfo {author} {\bibfnamefont {H.}~\bibnamefont
  {Schaefer}},\ }\href {https://doi.org/10.1002/zamm.19670470802} {\bibfield
  {journal} {\bibinfo  {journal} {ZAMM - Journal of Applied Mathematics and
  Mechanics / Zeitschrift für Angewandte Mathematik und Mechanik}\ }\textbf
  {\bibinfo {volume} {47}},\ \bibinfo {pages} {485} (\bibinfo {year}
  {1967})}\BibitemShut {NoStop}%
\bibitem [{\citenamefont {Truesdell}\ and\ \citenamefont
  {Noll}(2004)}]{truesdell_non-linear_2004}%
  \BibitemOpen
  \bibfield  {author} {\bibinfo {author} {\bibfnamefont {C.}~\bibnamefont
  {Truesdell}}\ and\ \bibinfo {author} {\bibfnamefont {W.}~\bibnamefont
  {Noll}},\ }\href@noop {} {\emph {\bibinfo {title} {The {Non}-{Linear} {Field}
  {Theories} of {Mechanics}}}},\ \bibinfo {edition} {3rd}\ ed.,\ edited by\
  \bibinfo {editor} {\bibfnamefont {S.~S.}\ \bibnamefont {Antman}}\ (\bibinfo
  {publisher} {Springer},\ \bibinfo {address} {Berlin},\ \bibinfo {year}
  {2004})\BibitemShut {NoStop}%
\bibitem [{\citenamefont {Altenbach}\ \emph {et~al.}(2013)\citenamefont
  {Altenbach}, \citenamefont {Eremeyev}, \citenamefont {Pfeiffer},
  \citenamefont {Rammerstorfer}, \citenamefont {Salençon}, \citenamefont
  {Schrefler},\ and\ \citenamefont {Serafini}}]{altenbach_generalized_2013}%
  \BibitemOpen
  \bibinfo {editor} {\bibfnamefont {H.}~\bibnamefont {Altenbach}}, \bibinfo
  {editor} {\bibfnamefont {V.~A.}\ \bibnamefont {Eremeyev}}, \bibinfo {editor}
  {\bibfnamefont {F.}~\bibnamefont {Pfeiffer}}, \bibinfo {editor}
  {\bibfnamefont {F.~G.}\ \bibnamefont {Rammerstorfer}}, \bibinfo {editor}
  {\bibfnamefont {J.}~\bibnamefont {Salençon}}, \bibinfo {editor}
  {\bibfnamefont {B.}~\bibnamefont {Schrefler}},\ and\ \bibinfo {editor}
  {\bibfnamefont {P.}~\bibnamefont {Serafini}},\ eds.,\ \href
  {https://doi.org/10.1007/978-3-7091-1371-4} {\emph {\bibinfo {title}
  {Generalized {Continua} -- from the {Theory} to {Engineering}
  {Applications}}}},\ \bibinfo {series} {{CISM} {International} {Centre} for
  {Mechanical} {Sciences}}, Vol.\ \bibinfo {volume} {541}\ (\bibinfo
  {publisher} {Springer},\ \bibinfo {address} {Vienna},\ \bibinfo {year}
  {2013})\BibitemShut {NoStop}%
\bibitem [{\citenamefont {Chen}\ \emph {et~al.}(2021)\citenamefont {Chen},
  \citenamefont {Li}, \citenamefont {Scheibner}, \citenamefont {Vitelli},\ and\
  \citenamefont {Huang}}]{chen_realization_2021}%
  \BibitemOpen
  \bibfield  {author} {\bibinfo {author} {\bibfnamefont {Y.}~\bibnamefont
  {Chen}}, \bibinfo {author} {\bibfnamefont {X.}~\bibnamefont {Li}}, \bibinfo
  {author} {\bibfnamefont {C.}~\bibnamefont {Scheibner}}, \bibinfo {author}
  {\bibfnamefont {V.}~\bibnamefont {Vitelli}},\ and\ \bibinfo {author}
  {\bibfnamefont {G.}~\bibnamefont {Huang}},\ }\href
  {https://doi.org/10.1038/s41467-021-26034-z} {\bibfield  {journal} {\bibinfo
  {journal} {Nature Communications}\ }\textbf {\bibinfo {volume} {12}},\
  \bibinfo {pages} {5935} (\bibinfo {year} {2021})}\BibitemShut {NoStop}%
\bibitem [{\citenamefont {Bolitho}(2021)}]{bolitho_geometric_2021}%
  \BibitemOpen
  \bibfield  {author} {\bibinfo {author} {\bibfnamefont {A.}~\bibnamefont
  {Bolitho}},\ }\emph {\bibinfo {title} {Geometric {Mechanics} of {Active}
  {Particles}}},\ \href {https://doi.org/10.17863/CAM.83455} {Ph.D. thesis},\
  \bibinfo  {school} {Apollo - University of Cambridge Repository} (\bibinfo
  {year} {2021})\BibitemShut {NoStop}%
\bibitem [{\citenamefont {Surówka}\ \emph {et~al.}(2023)\citenamefont
  {Surówka}, \citenamefont {Souslov}, \citenamefont {Jülicher},\ and\
  \citenamefont {Banerjee}}]{surowka_odd_2023}%
  \BibitemOpen
  \bibfield  {author} {\bibinfo {author} {\bibfnamefont {P.}~\bibnamefont
  {Surówka}}, \bibinfo {author} {\bibfnamefont {A.}~\bibnamefont {Souslov}},
  \bibinfo {author} {\bibfnamefont {F.}~\bibnamefont {Jülicher}},\ and\
  \bibinfo {author} {\bibfnamefont {D.}~\bibnamefont {Banerjee}},\ }\href
  {https://doi.org/10.1103/PhysRevE.108.064609} {\bibfield  {journal} {\bibinfo
   {journal} {Physical Review E}\ }\textbf {\bibinfo {volume} {108}},\ \bibinfo
  {pages} {064609} (\bibinfo {year} {2023})}\BibitemShut {NoStop}%
\bibitem [{\citenamefont {Lee}\ \emph {et~al.}(2025)\citenamefont {Lee},
  \citenamefont {Lubensky},\ and\ \citenamefont {Markovich}}]{lee_odd_2025}%
  \BibitemOpen
  \bibfield  {author} {\bibinfo {author} {\bibfnamefont {C.-T.}\ \bibnamefont
  {Lee}}, \bibinfo {author} {\bibfnamefont {T.~C.}\ \bibnamefont {Lubensky}},\
  and\ \bibinfo {author} {\bibfnamefont {T.}~\bibnamefont {Markovich}},\ }\href
  {https://doi.org/10.48550/arXiv.2508.04468} {\bibinfo {title} {Odd elasticity
  in disordered chiral active materials}} (\bibinfo {year} {2025}),\ \bibinfo
  {note} {arXiv:2508.04468 [cond-mat]}\BibitemShut {NoStop}%
\bibitem [{\citenamefont {Frenzel}\ \emph {et~al.}(2017)\citenamefont
  {Frenzel}, \citenamefont {Kadic},\ and\ \citenamefont
  {Wegener}}]{frenzel_three-dimensional_2017}%
  \BibitemOpen
  \bibfield  {author} {\bibinfo {author} {\bibfnamefont {T.}~\bibnamefont
  {Frenzel}}, \bibinfo {author} {\bibfnamefont {M.}~\bibnamefont {Kadic}},\
  and\ \bibinfo {author} {\bibfnamefont {M.}~\bibnamefont {Wegener}},\ }\href
  {https://doi.org/10.1126/science.aao4640} {\bibfield  {journal} {\bibinfo
  {journal} {Science}\ }\textbf {\bibinfo {volume} {358}},\ \bibinfo {pages}
  {1072} (\bibinfo {year} {2017})}\BibitemShut {NoStop}%
\bibitem [{\citenamefont {Rueger}\ and\ \citenamefont
  {Lakes}(2018)}]{rueger_strong_2018}%
  \BibitemOpen
  \bibfield  {author} {\bibinfo {author} {\bibfnamefont {Z.}~\bibnamefont
  {Rueger}}\ and\ \bibinfo {author} {\bibfnamefont {R.}~\bibnamefont {Lakes}},\
  }\href {https://doi.org/10.1103/PhysRevLett.120.065501} {\bibfield  {journal}
  {\bibinfo  {journal} {Physical Review Letters}\ }\textbf {\bibinfo {volume}
  {120}},\ \bibinfo {pages} {065501} (\bibinfo {year} {2018})}\BibitemShut
  {NoStop}%
\bibitem [{\citenamefont {Reasa}\ and\ \citenamefont
  {Lakes}(2020)}]{reasa_nonclassical_2020}%
  \BibitemOpen
  \bibfield  {author} {\bibinfo {author} {\bibfnamefont {D.}~\bibnamefont
  {Reasa}}\ and\ \bibinfo {author} {\bibfnamefont {R.}~\bibnamefont {Lakes}},\
  }\href {https://doi.org/10.1103/PhysRevLett.125.205502} {\bibfield  {journal}
  {\bibinfo  {journal} {Physical Review Letters}\ }\textbf {\bibinfo {volume}
  {125}},\ \bibinfo {pages} {205502} (\bibinfo {year} {2020})}\BibitemShut
  {NoStop}%
\bibitem [{\citenamefont {Németh}\ and\ \citenamefont
  {Adhikari}(2024)}]{nemeth_geometric_2024}%
  \BibitemOpen
  \bibfield  {author} {\bibinfo {author} {\bibfnamefont {B.}~\bibnamefont
  {Németh}}\ and\ \bibinfo {author} {\bibfnamefont {R.}~\bibnamefont
  {Adhikari}},\ }\href {https://doi.org/10.1063/5.0182475} {\bibfield
  {journal} {\bibinfo  {journal} {Journal of Mathematical Physics}\ }\textbf
  {\bibinfo {volume} {65}},\ \bibinfo {pages} {061902} (\bibinfo {year}
  {2024})}\BibitemShut {NoStop}%
\bibitem [{\citenamefont {Cartan}(1983)}]{cartan_geometry_1983}%
  \BibitemOpen
  \bibfield  {author} {\bibinfo {author} {\bibfnamefont {E.}~\bibnamefont
  {Cartan}},\ }\href@noop {} {\emph {\bibinfo {title} {Geometry of {Riemannian}
  spaces}}},\ \bibinfo {series} {Lie groups}\ No.~\bibinfo {number} {13}\
  (\bibinfo  {publisher} {Math Sci Press},\ \bibinfo {address} {Brookline,
  Mass},\ \bibinfo {year} {1983})\ \bibinfo {note} {translated by James
  Glazebrook, notes and appendices by R. Hermann}\BibitemShut {NoStop}%
\bibitem [{\citenamefont {Flanders}(1989)}]{flanders_differential_1989}%
  \BibitemOpen
  \bibfield  {author} {\bibinfo {author} {\bibfnamefont {H.}~\bibnamefont
  {Flanders}},\ }\href@noop {} {\emph {\bibinfo {title} {Differential forms
  with applications to the physical sciences}}},\ Dover {Books} on
  {Mathematics}\ (\bibinfo  {publisher} {Dover Publications},\ \bibinfo
  {address} {Mineola, NY},\ \bibinfo {year} {1989})\BibitemShut {NoStop}%
\bibitem [{\citenamefont {Chelakkot}\ \emph {et~al.}(2014)\citenamefont
  {Chelakkot}, \citenamefont {Gopinath}, \citenamefont {Mahadevan},\ and\
  \citenamefont {Hagan}}]{chelakkot_flagellar_2014}%
  \BibitemOpen
  \bibfield  {author} {\bibinfo {author} {\bibfnamefont {R.}~\bibnamefont
  {Chelakkot}}, \bibinfo {author} {\bibfnamefont {A.}~\bibnamefont {Gopinath}},
  \bibinfo {author} {\bibfnamefont {L.}~\bibnamefont {Mahadevan}},\ and\
  \bibinfo {author} {\bibfnamefont {M.~F.}\ \bibnamefont {Hagan}},\ }\href
  {https://doi.org/10.1098/rsif.2013.0884} {\bibfield  {journal} {\bibinfo
  {journal} {Journal of The Royal Society Interface}\ }\textbf {\bibinfo
  {volume} {11}},\ \bibinfo {pages} {20130884} (\bibinfo {year}
  {2014})}\BibitemShut {NoStop}%
\bibitem [{\citenamefont {Winkler}\ and\ \citenamefont
  {Gompper}(2020)}]{winkler_physics_2020}%
  \BibitemOpen
  \bibfield  {author} {\bibinfo {author} {\bibfnamefont {R.~G.}\ \bibnamefont
  {Winkler}}\ and\ \bibinfo {author} {\bibfnamefont {G.}~\bibnamefont
  {Gompper}},\ }\href {https://doi.org/10.1063/5.0011466} {\bibfield  {journal}
  {\bibinfo  {journal} {The Journal of Chemical Physics}\ }\textbf {\bibinfo
  {volume} {153}},\ \bibinfo {pages} {040901} (\bibinfo {year}
  {2020})}\BibitemShut {NoStop}%
\bibitem [{\citenamefont {Kumar}\ \emph {et~al.}(2024)\citenamefont {Kumar},
  \citenamefont {Murali}, \citenamefont {Subramaniam}, \citenamefont {Singh},\
  and\ \citenamefont {Thutupalli}}]{kumar_emergent_2024}%
  \BibitemOpen
  \bibfield  {author} {\bibinfo {author} {\bibfnamefont {M.}~\bibnamefont
  {Kumar}}, \bibinfo {author} {\bibfnamefont {A.}~\bibnamefont {Murali}},
  \bibinfo {author} {\bibfnamefont {A.~G.}\ \bibnamefont {Subramaniam}},
  \bibinfo {author} {\bibfnamefont {R.}~\bibnamefont {Singh}},\ and\ \bibinfo
  {author} {\bibfnamefont {S.}~\bibnamefont {Thutupalli}},\ }\href
  {https://doi.org/10.1038/s41467-024-49155-7} {\bibfield  {journal} {\bibinfo
  {journal} {Nature Communications}\ }\textbf {\bibinfo {volume} {15}},\
  \bibinfo {pages} {4903} (\bibinfo {year} {2024})}\BibitemShut {NoStop}%
\bibitem [{\citenamefont {Biswas}\ \emph {et~al.}(2025)\citenamefont {Biswas},
  \citenamefont {More},\ and\ \citenamefont {Kandula}}]{biswas_emergent_2025}%
  \BibitemOpen
  \bibfield  {author} {\bibinfo {author} {\bibfnamefont {B.}~\bibnamefont
  {Biswas}}, \bibinfo {author} {\bibfnamefont {P.}~\bibnamefont {More}},\ and\
  \bibinfo {author} {\bibfnamefont {H.~N.}\ \bibnamefont {Kandula}},\ }\href
  {https://doi.org/10.1021/acsnano.5c08920} {\bibfield  {journal} {\bibinfo
  {journal} {ACS Nano}\ }\textbf {\bibinfo {volume} {19}},\ \bibinfo {pages}
  {31038} (\bibinfo {year} {2025})}\BibitemShut {NoStop}%
\bibitem [{\citenamefont {Jayaraman}\ \emph {et~al.}(2012)\citenamefont
  {Jayaraman}, \citenamefont {Ramachandran}, \citenamefont {Ghose},
  \citenamefont {Laskar}, \citenamefont {Bhamla}, \citenamefont {Kumar},\ and\
  \citenamefont {Adhikari}}]{jayaraman_autonomous_2012}%
  \BibitemOpen
  \bibfield  {author} {\bibinfo {author} {\bibfnamefont {G.}~\bibnamefont
  {Jayaraman}}, \bibinfo {author} {\bibfnamefont {S.}~\bibnamefont
  {Ramachandran}}, \bibinfo {author} {\bibfnamefont {S.}~\bibnamefont {Ghose}},
  \bibinfo {author} {\bibfnamefont {A.}~\bibnamefont {Laskar}}, \bibinfo
  {author} {\bibfnamefont {M.~S.}\ \bibnamefont {Bhamla}}, \bibinfo {author}
  {\bibfnamefont {P.~B.~S.}\ \bibnamefont {Kumar}},\ and\ \bibinfo {author}
  {\bibfnamefont {R.}~\bibnamefont {Adhikari}},\ }\href
  {https://doi.org/10.1103/PhysRevLett.109.158302} {\bibfield  {journal}
  {\bibinfo  {journal} {Physical Review Letters}\ }\textbf {\bibinfo {volume}
  {109}},\ \bibinfo {pages} {158302} (\bibinfo {year} {2012})}\BibitemShut
  {NoStop}%
\bibitem [{\citenamefont {Laskar}\ and\ \citenamefont
  {Adhikari}(2015)}]{laskar_brownian_2015}%
  \BibitemOpen
  \bibfield  {author} {\bibinfo {author} {\bibfnamefont {A.}~\bibnamefont
  {Laskar}}\ and\ \bibinfo {author} {\bibfnamefont {R.}~\bibnamefont
  {Adhikari}},\ }\href {https://doi.org/10.1039/C5SM02021B} {\bibfield
  {journal} {\bibinfo  {journal} {Soft Matter}\ }\textbf {\bibinfo {volume}
  {11}},\ \bibinfo {pages} {9073} (\bibinfo {year} {2015})}\BibitemShut
  {NoStop}%
\bibitem [{\citenamefont {Krishnamurthy}\ and\ \citenamefont
  {Prakash}(2023)}]{krishnamurthy_emergent_2023}%
  \BibitemOpen
  \bibfield  {author} {\bibinfo {author} {\bibfnamefont {D.}~\bibnamefont
  {Krishnamurthy}}\ and\ \bibinfo {author} {\bibfnamefont {M.}~\bibnamefont
  {Prakash}},\ }\href {https://doi.org/10.1073/pnas.2304981120} {\bibfield
  {journal} {\bibinfo  {journal} {Proceedings of the National Academy of
  Sciences}\ }\textbf {\bibinfo {volume} {120}},\ \bibinfo {pages}
  {e2304981120} (\bibinfo {year} {2023})}\BibitemShut {NoStop}%
\bibitem [{\citenamefont {Scheibner}\ \emph {et~al.}(2020)\citenamefont
  {Scheibner}, \citenamefont {Souslov}, \citenamefont {Banerjee}, \citenamefont
  {Surówka}, \citenamefont {Irvine},\ and\ \citenamefont
  {Vitelli}}]{scheibner_odd_2020}%
  \BibitemOpen
  \bibfield  {author} {\bibinfo {author} {\bibfnamefont {C.}~\bibnamefont
  {Scheibner}}, \bibinfo {author} {\bibfnamefont {A.}~\bibnamefont {Souslov}},
  \bibinfo {author} {\bibfnamefont {D.}~\bibnamefont {Banerjee}}, \bibinfo
  {author} {\bibfnamefont {P.}~\bibnamefont {Surówka}}, \bibinfo {author}
  {\bibfnamefont {W.~T.~M.}\ \bibnamefont {Irvine}},\ and\ \bibinfo {author}
  {\bibfnamefont {V.}~\bibnamefont {Vitelli}},\ }\href
  {https://doi.org/10.1038/s41567-020-0795-y} {\bibfield  {journal} {\bibinfo
  {journal} {Nature Physics}\ }\textbf {\bibinfo {volume} {16}},\ \bibinfo
  {pages} {475} (\bibinfo {year} {2020})}\BibitemShut {NoStop}%
\bibitem [{\citenamefont {Fruchart}\ \emph {et~al.}(2023)\citenamefont
  {Fruchart}, \citenamefont {Scheibner},\ and\ \citenamefont
  {Vitelli}}]{fruchart_odd_2023}%
  \BibitemOpen
  \bibfield  {author} {\bibinfo {author} {\bibfnamefont {M.}~\bibnamefont
  {Fruchart}}, \bibinfo {author} {\bibfnamefont {C.}~\bibnamefont
  {Scheibner}},\ and\ \bibinfo {author} {\bibfnamefont {V.}~\bibnamefont
  {Vitelli}},\ }\href
  {https://doi.org/10.1146/annurev-conmatphys-040821-125506} {\bibfield
  {journal} {\bibinfo  {journal} {Annual Review of Condensed Matter Physics}\
  }\textbf {\bibinfo {volume} {14}},\ \bibinfo {pages} {471} (\bibinfo {year}
  {2023})}\BibitemShut {NoStop}%
\bibitem [{\citenamefont {Beatus}\ \emph {et~al.}(2006)\citenamefont {Beatus},
  \citenamefont {Tlusty},\ and\ \citenamefont {Bar-Ziv}}]{beatus_phonons_2006}%
  \BibitemOpen
  \bibfield  {author} {\bibinfo {author} {\bibfnamefont {T.}~\bibnamefont
  {Beatus}}, \bibinfo {author} {\bibfnamefont {T.}~\bibnamefont {Tlusty}},\
  and\ \bibinfo {author} {\bibfnamefont {R.}~\bibnamefont {Bar-Ziv}},\ }\href
  {https://doi.org/10.1038/nphys432} {\bibfield  {journal} {\bibinfo  {journal}
  {Nature Physics}\ }\textbf {\bibinfo {volume} {2}},\ \bibinfo {pages} {743}
  (\bibinfo {year} {2006})}\BibitemShut {NoStop}%
\bibitem [{\citenamefont {Lahiri}\ and\ \citenamefont
  {Ramaswamy}(1997)}]{lahiri_are_1997}%
  \BibitemOpen
  \bibfield  {author} {\bibinfo {author} {\bibfnamefont {R.}~\bibnamefont
  {Lahiri}}\ and\ \bibinfo {author} {\bibfnamefont {S.}~\bibnamefont
  {Ramaswamy}},\ }\href {https://doi.org/10.1103/PhysRevLett.79.1150}
  {\bibfield  {journal} {\bibinfo  {journal} {Physical Review Letters}\
  }\textbf {\bibinfo {volume} {79}},\ \bibinfo {pages} {1150} (\bibinfo {year}
  {1997})}\BibitemShut {NoStop}%
\bibitem [{\citenamefont {Chajwa}\ \emph {et~al.}(2020)\citenamefont {Chajwa},
  \citenamefont {Menon}, \citenamefont {Ramaswamy},\ and\ \citenamefont
  {Govindarajan}}]{chajwa_waves_2020}%
  \BibitemOpen
  \bibfield  {author} {\bibinfo {author} {\bibfnamefont {R.}~\bibnamefont
  {Chajwa}}, \bibinfo {author} {\bibfnamefont {N.}~\bibnamefont {Menon}},
  \bibinfo {author} {\bibfnamefont {S.}~\bibnamefont {Ramaswamy}},\ and\
  \bibinfo {author} {\bibfnamefont {R.}~\bibnamefont {Govindarajan}},\ }\href
  {https://doi.org/10.1103/PhysRevX.10.041016} {\bibfield  {journal} {\bibinfo
  {journal} {Physical Review X}\ }\textbf {\bibinfo {volume} {10}},\ \bibinfo
  {pages} {041016} (\bibinfo {year} {2020})}\BibitemShut {NoStop}%
\bibitem [{\citenamefont {Poncet}\ and\ \citenamefont
  {Bartolo}(2022)}]{poncet_when_2022}%
  \BibitemOpen
  \bibfield  {author} {\bibinfo {author} {\bibfnamefont {A.}~\bibnamefont
  {Poncet}}\ and\ \bibinfo {author} {\bibfnamefont {D.}~\bibnamefont
  {Bartolo}},\ }\href {https://doi.org/10.1103/PhysRevLett.128.048002}
  {\bibfield  {journal} {\bibinfo  {journal} {Physical Review Letters}\
  }\textbf {\bibinfo {volume} {128}},\ \bibinfo {pages} {048002} (\bibinfo
  {year} {2022})}\BibitemShut {NoStop}%
\bibitem [{\citenamefont {Hawkins}\ and\ \citenamefont
  {Liverpool}(2014)}]{hawkins_stress_2014}%
  \BibitemOpen
  \bibfield  {author} {\bibinfo {author} {\bibfnamefont {R.~J.}\ \bibnamefont
  {Hawkins}}\ and\ \bibinfo {author} {\bibfnamefont {T.~B.}\ \bibnamefont
  {Liverpool}},\ }\href {https://doi.org/10.1103/PhysRevLett.113.028102}
  {\bibfield  {journal} {\bibinfo  {journal} {Physical Review Letters}\
  }\textbf {\bibinfo {volume} {113}},\ \bibinfo {pages} {028102} (\bibinfo
  {year} {2014})}\BibitemShut {NoStop}%
\bibitem [{\citenamefont {Maitra}\ and\ \citenamefont
  {Ramaswamy}(2019)}]{maitra_oriented_2019}%
  \BibitemOpen
  \bibfield  {author} {\bibinfo {author} {\bibfnamefont {A.}~\bibnamefont
  {Maitra}}\ and\ \bibinfo {author} {\bibfnamefont {S.}~\bibnamefont
  {Ramaswamy}},\ }\href {https://doi.org/10.1103/PhysRevLett.123.238001}
  {\bibfield  {journal} {\bibinfo  {journal} {Physical Review Letters}\
  }\textbf {\bibinfo {volume} {123}},\ \bibinfo {pages} {238001} (\bibinfo
  {year} {2019})}\BibitemShut {NoStop}%
\bibitem [{\citenamefont {Kole}\ \emph {et~al.}(2021)\citenamefont {Kole},
  \citenamefont {Alexander}, \citenamefont {Ramaswamy},\ and\ \citenamefont
  {Maitra}}]{kole_layered_2021}%
  \BibitemOpen
  \bibfield  {author} {\bibinfo {author} {\bibfnamefont {S.}~\bibnamefont
  {Kole}}, \bibinfo {author} {\bibfnamefont {G.~P.}\ \bibnamefont {Alexander}},
  \bibinfo {author} {\bibfnamefont {S.}~\bibnamefont {Ramaswamy}},\ and\
  \bibinfo {author} {\bibfnamefont {A.}~\bibnamefont {Maitra}},\ }\href
  {https://doi.org/10.1103/PhysRevLett.126.248001} {\bibfield  {journal}
  {\bibinfo  {journal} {Physical Review Letters}\ }\textbf {\bibinfo {volume}
  {126}},\ \bibinfo {pages} {248001} (\bibinfo {year} {2021})}\BibitemShut
  {NoStop}%
\bibitem [{\citenamefont {Xu}\ \emph {et~al.}(2023)\citenamefont {Xu},
  \citenamefont {Huang}, \citenamefont {Zhang},\ and\ \citenamefont
  {Wu}}]{xu_autonomous_2023}%
  \BibitemOpen
  \bibfield  {author} {\bibinfo {author} {\bibfnamefont {H.}~\bibnamefont
  {Xu}}, \bibinfo {author} {\bibfnamefont {Y.}~\bibnamefont {Huang}}, \bibinfo
  {author} {\bibfnamefont {R.}~\bibnamefont {Zhang}},\ and\ \bibinfo {author}
  {\bibfnamefont {Y.}~\bibnamefont {Wu}},\ }\href
  {https://doi.org/10.1038/s41567-022-01836-0} {\bibfield  {journal} {\bibinfo
  {journal} {Nature Physics}\ }\textbf {\bibinfo {volume} {19}},\ \bibinfo
  {pages} {46} (\bibinfo {year} {2023})}\BibitemShut {NoStop}%
\bibitem [{\citenamefont {Baconnier}\ \emph {et~al.}(2025)\citenamefont
  {Baconnier}, \citenamefont {Dauchot}, \citenamefont {Démery}, \citenamefont
  {Düring}, \citenamefont {Henkes}, \citenamefont {Huepe},\ and\ \citenamefont
  {Shee}}]{baconnier_self-aligning_2025}%
  \BibitemOpen
  \bibfield  {author} {\bibinfo {author} {\bibfnamefont {P.}~\bibnamefont
  {Baconnier}}, \bibinfo {author} {\bibfnamefont {O.}~\bibnamefont {Dauchot}},
  \bibinfo {author} {\bibfnamefont {V.}~\bibnamefont {Démery}}, \bibinfo
  {author} {\bibfnamefont {G.}~\bibnamefont {Düring}}, \bibinfo {author}
  {\bibfnamefont {S.}~\bibnamefont {Henkes}}, \bibinfo {author} {\bibfnamefont
  {C.}~\bibnamefont {Huepe}},\ and\ \bibinfo {author} {\bibfnamefont
  {A.}~\bibnamefont {Shee}},\ }\href
  {https://doi.org/10.1103/RevModPhys.97.015007} {\bibfield  {journal}
  {\bibinfo  {journal} {Reviews of Modern Physics}\ }\textbf {\bibinfo {volume}
  {97}},\ \bibinfo {pages} {015007} (\bibinfo {year} {2025})}\BibitemShut
  {NoStop}%
\bibitem [{\citenamefont {Veenstra}\ \emph
  {et~al.}(2025{\natexlab{a}})\citenamefont {Veenstra}, \citenamefont
  {Scheibner}, \citenamefont {Brandenbourger}, \citenamefont {Binysh},
  \citenamefont {Souslov}, \citenamefont {Vitelli},\ and\ \citenamefont
  {Coulais}}]{veenstra_adaptive_2025}%
  \BibitemOpen
  \bibfield  {author} {\bibinfo {author} {\bibfnamefont {J.}~\bibnamefont
  {Veenstra}}, \bibinfo {author} {\bibfnamefont {C.}~\bibnamefont {Scheibner}},
  \bibinfo {author} {\bibfnamefont {M.}~\bibnamefont {Brandenbourger}},
  \bibinfo {author} {\bibfnamefont {J.}~\bibnamefont {Binysh}}, \bibinfo
  {author} {\bibfnamefont {A.}~\bibnamefont {Souslov}}, \bibinfo {author}
  {\bibfnamefont {V.}~\bibnamefont {Vitelli}},\ and\ \bibinfo {author}
  {\bibfnamefont {C.}~\bibnamefont {Coulais}},\ }\href
  {https://doi.org/10.1038/s41586-025-08646-3} {\bibfield  {journal} {\bibinfo
  {journal} {Nature}\ }\textbf {\bibinfo {volume} {639}},\ \bibinfo {pages}
  {935} (\bibinfo {year} {2025}{\natexlab{a}})}\BibitemShut {NoStop}%
\bibitem [{\citenamefont {Sarkar}\ \emph {et~al.}(2025)\citenamefont {Sarkar},
  \citenamefont {Ash}, \citenamefont {Wu}, \citenamefont {Boechler},
  \citenamefont {Shankar},\ and\ \citenamefont
  {Mao}}]{sarkar_mechanochemical_2025}%
  \BibitemOpen
  \bibfield  {author} {\bibinfo {author} {\bibfnamefont {S.}~\bibnamefont
  {Sarkar}}, \bibinfo {author} {\bibfnamefont {B.}~\bibnamefont {Ash}},
  \bibinfo {author} {\bibfnamefont {Y.}~\bibnamefont {Wu}}, \bibinfo {author}
  {\bibfnamefont {N.}~\bibnamefont {Boechler}}, \bibinfo {author}
  {\bibfnamefont {S.}~\bibnamefont {Shankar}},\ and\ \bibinfo {author}
  {\bibfnamefont {X.}~\bibnamefont {Mao}},\ }\href
  {https://doi.org/10.48550/arXiv.2505.18272} {\bibinfo {title}
  {Mechanochemical feedback drives complex inertial dynamics in active solids}}
  (\bibinfo {year} {2025}),\ \bibinfo {note} {arXiv:2505.18272
  [cond-mat]}\BibitemShut {NoStop}%
\bibitem [{\citenamefont {Cocconi}\ \emph {et~al.}(2025)\citenamefont
  {Cocconi}, \citenamefont {Chatzittofi},\ and\ \citenamefont
  {Golestanian}}]{cocconi_mechanical_2025}%
  \BibitemOpen
  \bibfield  {author} {\bibinfo {author} {\bibfnamefont {L.}~\bibnamefont
  {Cocconi}}, \bibinfo {author} {\bibfnamefont {M.}~\bibnamefont
  {Chatzittofi}},\ and\ \bibinfo {author} {\bibfnamefont {R.}~\bibnamefont
  {Golestanian}},\ }\href {https://doi.org/10.48550/arXiv.2506.18000} {\bibinfo
  {title} {Mechanical inhibition of dissipation in a thermodynamically
  consistent active solid}} (\bibinfo {year} {2025}),\ \bibinfo {note}
  {arXiv:2506.18000 [cond-mat]}\BibitemShut {NoStop}%
\bibitem [{\citenamefont {Welker}\ and\ \citenamefont
  {Alert}(2025)}]{welker_lattice-dependent_2025}%
  \BibitemOpen
  \bibfield  {author} {\bibinfo {author} {\bibfnamefont {T.}~\bibnamefont
  {Welker}}\ and\ \bibinfo {author} {\bibfnamefont {R.}~\bibnamefont {Alert}},\
  }\href {https://doi.org/10.1039/D5SM00627A} {\bibfield  {journal} {\bibinfo
  {journal} {Soft Matter}\ ,\ \bibinfo {pages} {10.1039.D5SM00627A}} (\bibinfo
  {year} {2025})}\BibitemShut {NoStop}%
\bibitem [{\citenamefont {Saha}\ \emph {et~al.}(2020)\citenamefont {Saha},
  \citenamefont {Agudo-Canalejo},\ and\ \citenamefont
  {Golestanian}}]{saha_scalar_2020}%
  \BibitemOpen
  \bibfield  {author} {\bibinfo {author} {\bibfnamefont {S.}~\bibnamefont
  {Saha}}, \bibinfo {author} {\bibfnamefont {J.}~\bibnamefont
  {Agudo-Canalejo}},\ and\ \bibinfo {author} {\bibfnamefont {R.}~\bibnamefont
  {Golestanian}},\ }\href {https://doi.org/10.1103/PhysRevX.10.041009}
  {\bibfield  {journal} {\bibinfo  {journal} {Physical Review X}\ }\textbf
  {\bibinfo {volume} {10}},\ \bibinfo {pages} {041009} (\bibinfo {year}
  {2020})}\BibitemShut {NoStop}%
\bibitem [{\citenamefont {You}\ \emph {et~al.}(2020)\citenamefont {You},
  \citenamefont {Baskaran},\ and\ \citenamefont
  {Marchetti}}]{you_nonreciprocity_2020}%
  \BibitemOpen
  \bibfield  {author} {\bibinfo {author} {\bibfnamefont {Z.}~\bibnamefont
  {You}}, \bibinfo {author} {\bibfnamefont {A.}~\bibnamefont {Baskaran}},\ and\
  \bibinfo {author} {\bibfnamefont {M.~C.}\ \bibnamefont {Marchetti}},\ }\href
  {https://doi.org/10.1073/pnas.2010318117} {\bibfield  {journal} {\bibinfo
  {journal} {Proceedings of the National Academy of Sciences}\ }\textbf
  {\bibinfo {volume} {117}},\ \bibinfo {pages} {19767} (\bibinfo {year}
  {2020})}\BibitemShut {NoStop}%
\bibitem [{\citenamefont {Fruchart}\ \emph {et~al.}(2021)\citenamefont
  {Fruchart}, \citenamefont {Hanai}, \citenamefont {Littlewood},\ and\
  \citenamefont {Vitelli}}]{fruchart_non-reciprocal_2021}%
  \BibitemOpen
  \bibfield  {author} {\bibinfo {author} {\bibfnamefont {M.}~\bibnamefont
  {Fruchart}}, \bibinfo {author} {\bibfnamefont {R.}~\bibnamefont {Hanai}},
  \bibinfo {author} {\bibfnamefont {P.~B.}\ \bibnamefont {Littlewood}},\ and\
  \bibinfo {author} {\bibfnamefont {V.}~\bibnamefont {Vitelli}},\ }\href
  {https://doi.org/10.1038/s41586-021-03375-9} {\bibfield  {journal} {\bibinfo
  {journal} {Nature}\ }\textbf {\bibinfo {volume} {592}},\ \bibinfo {pages}
  {363} (\bibinfo {year} {2021})}\BibitemShut {NoStop}%
\bibitem [{\citenamefont {Loos}\ \emph {et~al.}(2023)\citenamefont {Loos},
  \citenamefont {Klapp},\ and\ \citenamefont
  {Martynec}}]{loos_long-range_2023}%
  \BibitemOpen
  \bibfield  {author} {\bibinfo {author} {\bibfnamefont {S.~A.}\ \bibnamefont
  {Loos}}, \bibinfo {author} {\bibfnamefont {S.~H.}\ \bibnamefont {Klapp}},\
  and\ \bibinfo {author} {\bibfnamefont {T.}~\bibnamefont {Martynec}},\ }\href
  {https://doi.org/10.1103/PhysRevLett.130.198301} {\bibfield  {journal}
  {\bibinfo  {journal} {Physical Review Letters}\ }\textbf {\bibinfo {volume}
  {130}},\ \bibinfo {pages} {198301} (\bibinfo {year} {2023})}\BibitemShut
  {NoStop}%
\bibitem [{\citenamefont {Veenstra}\ \emph
  {et~al.}(2025{\natexlab{b}})\citenamefont {Veenstra}, \citenamefont
  {Gamayun}, \citenamefont {Brandenbourger}, \citenamefont {Van~Gorp},
  \citenamefont {Terwisscha-Dekker}, \citenamefont {Caux},\ and\ \citenamefont
  {Coulais}}]{veenstra_nonreciprocal_2025}%
  \BibitemOpen
  \bibfield  {author} {\bibinfo {author} {\bibfnamefont {J.}~\bibnamefont
  {Veenstra}}, \bibinfo {author} {\bibfnamefont {O.}~\bibnamefont {Gamayun}},
  \bibinfo {author} {\bibfnamefont {M.}~\bibnamefont {Brandenbourger}},
  \bibinfo {author} {\bibfnamefont {F.}~\bibnamefont {Van~Gorp}}, \bibinfo
  {author} {\bibfnamefont {H.}~\bibnamefont {Terwisscha-Dekker}}, \bibinfo
  {author} {\bibfnamefont {J.-S.}\ \bibnamefont {Caux}},\ and\ \bibinfo
  {author} {\bibfnamefont {C.}~\bibnamefont {Coulais}},\ }\href
  {https://doi.org/10.1103/nrv2-9h8z} {\bibfield  {journal} {\bibinfo
  {journal} {Physical Review X}\ }\textbf {\bibinfo {volume} {15}},\ \bibinfo
  {pages} {031045} (\bibinfo {year} {2025}{\natexlab{b}})}\BibitemShut
  {NoStop}%
\bibitem [{\citenamefont {Ramaswamy}(2010)}]{ramaswamy_mechanics_2010}%
  \BibitemOpen
  \bibfield  {author} {\bibinfo {author} {\bibfnamefont {S.}~\bibnamefont
  {Ramaswamy}},\ }\href
  {https://doi.org/10.1146/annurev-conmatphys-070909-104101} {\bibfield
  {journal} {\bibinfo  {journal} {Annual Review of Condensed Matter Physics}\
  }\textbf {\bibinfo {volume} {1}},\ \bibinfo {pages} {323} (\bibinfo {year}
  {2010})}\BibitemShut {NoStop}%
\bibitem [{\citenamefont {Joanny}(1979)}]{joanny_acoustic_1979}%
  \BibitemOpen
  \bibfield  {author} {\bibinfo {author} {\bibfnamefont {J.}~\bibnamefont
  {Joanny}},\ }\href {https://doi.org/10.1016/0021-9797(79)90336-9} {\bibfield
  {journal} {\bibinfo  {journal} {Journal of Colloid and Interface Science}\
  }\textbf {\bibinfo {volume} {71}},\ \bibinfo {pages} {622} (\bibinfo {year}
  {1979})}\BibitemShut {NoStop}%
\bibitem [{\citenamefont {Hurd}\ \emph {et~al.}(1982)\citenamefont {Hurd},
  \citenamefont {Clark}, \citenamefont {Mockler},\ and\ \citenamefont
  {O'Sullivan}}]{hurd_lattice_1982}%
  \BibitemOpen
  \bibfield  {author} {\bibinfo {author} {\bibfnamefont {A.~J.}\ \bibnamefont
  {Hurd}}, \bibinfo {author} {\bibfnamefont {N.~A.}\ \bibnamefont {Clark}},
  \bibinfo {author} {\bibfnamefont {R.~C.}\ \bibnamefont {Mockler}},\ and\
  \bibinfo {author} {\bibfnamefont {W.~J.}\ \bibnamefont {O'Sullivan}},\ }\href
  {https://doi.org/10.1103/PhysRevA.26.2869} {\bibfield  {journal} {\bibinfo
  {journal} {Physical Review A}\ }\textbf {\bibinfo {volume} {26}},\ \bibinfo
  {pages} {2869} (\bibinfo {year} {1982})}\BibitemShut {NoStop}%
\bibitem [{\citenamefont {Dhont}(1996)}]{dhont_introduction_1996}%
  \BibitemOpen
  \bibfield  {author} {\bibinfo {author} {\bibfnamefont {J.~K.~G.}\
  \bibnamefont {Dhont}},\ }\href@noop {} {\emph {\bibinfo {title} {An
  {Introduction} to {Dynamics} of {Colloids}}}},\ \bibinfo {series} {Studies in
  {Interface} {Science}}\ No.~\bibinfo {number} {2}\ (\bibinfo  {publisher}
  {Elsevier},\ \bibinfo {address} {Amsterdam},\ \bibinfo {year}
  {1996})\BibitemShut {NoStop}%
\bibitem [{\citenamefont {Landau}\ and\ \citenamefont
  {Lifshitz}(1980)}]{landau_statistical_1980}%
  \BibitemOpen
  \bibfield  {author} {\bibinfo {author} {\bibfnamefont {L.~D.}\ \bibnamefont
  {Landau}}\ and\ \bibinfo {author} {\bibfnamefont {E.~M.}\ \bibnamefont
  {Lifshitz}},\ }\href@noop {} {\emph {\bibinfo {title} {Statistical
  {Physics}}}},\ \bibinfo {edition} {3rd}\ ed.\ (\bibinfo  {publisher}
  {Butterworth-Heinemann},\ \bibinfo {address} {Oxford, England},\ \bibinfo
  {year} {1980})\BibitemShut {NoStop}%
\bibitem [{\citenamefont {Chaikin}\ and\ \citenamefont
  {Lubensky}(1995)}]{chaikin_principles_1995}%
  \BibitemOpen
  \bibfield  {author} {\bibinfo {author} {\bibfnamefont {P.~M.}\ \bibnamefont
  {Chaikin}}\ and\ \bibinfo {author} {\bibfnamefont {T.~C.}\ \bibnamefont
  {Lubensky}},\ }\href {https://doi.org/10.1017/CBO9780511813467} {\emph
  {\bibinfo {title} {Principles of {Condensed} {Matter} {Physics}}}},\ \bibinfo
  {edition} {1st}\ ed.\ (\bibinfo  {publisher} {Cambridge University Press},\
  \bibinfo {year} {1995})\BibitemShut {NoStop}%
\bibitem [{\citenamefont {Marchetti}\ \emph {et~al.}(2013)\citenamefont
  {Marchetti}, \citenamefont {Joanny}, \citenamefont {Ramaswamy}, \citenamefont
  {Liverpool}, \citenamefont {Prost}, \citenamefont {Rao},\ and\ \citenamefont
  {Simha}}]{marchetti_hydrodynamics_2013}%
  \BibitemOpen
  \bibfield  {author} {\bibinfo {author} {\bibfnamefont {M.~C.}\ \bibnamefont
  {Marchetti}}, \bibinfo {author} {\bibfnamefont {J.~F.}\ \bibnamefont
  {Joanny}}, \bibinfo {author} {\bibfnamefont {S.}~\bibnamefont {Ramaswamy}},
  \bibinfo {author} {\bibfnamefont {T.~B.}\ \bibnamefont {Liverpool}}, \bibinfo
  {author} {\bibfnamefont {J.}~\bibnamefont {Prost}}, \bibinfo {author}
  {\bibfnamefont {M.}~\bibnamefont {Rao}},\ and\ \bibinfo {author}
  {\bibfnamefont {R.~A.}\ \bibnamefont {Simha}},\ }\href
  {https://doi.org/10.1103/RevModPhys.85.1143} {\bibfield  {journal} {\bibinfo
  {journal} {Reviews of Modern Physics}\ }\textbf {\bibinfo {volume} {85}},\
  \bibinfo {pages} {1143} (\bibinfo {year} {2013})}\BibitemShut {NoStop}%
\bibitem [{\citenamefont {Antman}(2004)}]{antman_nonlinear_2004}%
  \BibitemOpen
  \bibfield  {author} {\bibinfo {author} {\bibfnamefont {S.}~\bibnamefont
  {Antman}},\ }\href@noop {} {\emph {\bibinfo {title} {Nonlinear {Problems} of
  {Elasticity}}}}\ (\bibinfo  {publisher} {Springer},\ \bibinfo {address} {New
  York},\ \bibinfo {year} {2004})\BibitemShut {NoStop}%
\bibitem [{\citenamefont {Tan}\ \emph {et~al.}(2022)\citenamefont {Tan},
  \citenamefont {Mietke}, \citenamefont {Li}, \citenamefont {Chen},
  \citenamefont {Higinbotham}, \citenamefont {Foster}, \citenamefont {Gokhale},
  \citenamefont {Dunkel},\ and\ \citenamefont {Fakhri}}]{tan_odd_2022}%
  \BibitemOpen
  \bibfield  {author} {\bibinfo {author} {\bibfnamefont {T.~H.}\ \bibnamefont
  {Tan}}, \bibinfo {author} {\bibfnamefont {A.}~\bibnamefont {Mietke}},
  \bibinfo {author} {\bibfnamefont {J.}~\bibnamefont {Li}}, \bibinfo {author}
  {\bibfnamefont {Y.}~\bibnamefont {Chen}}, \bibinfo {author} {\bibfnamefont
  {H.}~\bibnamefont {Higinbotham}}, \bibinfo {author} {\bibfnamefont {P.~J.}\
  \bibnamefont {Foster}}, \bibinfo {author} {\bibfnamefont {S.}~\bibnamefont
  {Gokhale}}, \bibinfo {author} {\bibfnamefont {J.}~\bibnamefont {Dunkel}},\
  and\ \bibinfo {author} {\bibfnamefont {N.}~\bibnamefont {Fakhri}},\ }\href
  {https://doi.org/10.1038/s41586-022-04889-6} {\bibfield  {journal} {\bibinfo
  {journal} {Nature}\ }\textbf {\bibinfo {volume} {607}},\ \bibinfo {pages}
  {287} (\bibinfo {year} {2022})}\BibitemShut {NoStop}%
\bibitem [{\citenamefont {Healey}(2002)}]{healey_material_2002}%
  \BibitemOpen
  \bibfield  {author} {\bibinfo {author} {\bibfnamefont {T.~J.}\ \bibnamefont
  {Healey}},\ }\href {https://doi.org/10.1177/108128028482} {\bibfield
  {journal} {\bibinfo  {journal} {Mathematics and Mechanics of Solids}\
  }\textbf {\bibinfo {volume} {7}},\ \bibinfo {pages} {405} (\bibinfo {year}
  {2002})}\BibitemShut {NoStop}%
\bibitem [{\citenamefont {Felderhof}(2003)}]{felderhof_mesoscopic_2003}%
  \BibitemOpen
  \bibfield  {author} {\bibinfo {author} {\bibfnamefont {B.~U.}\ \bibnamefont
  {Felderhof}},\ }\href {https://doi.org/10.1103/PhysRevE.68.051402} {\bibfield
   {journal} {\bibinfo  {journal} {Physical Review E}\ }\textbf {\bibinfo
  {volume} {68}},\ \bibinfo {pages} {051402} (\bibinfo {year}
  {2003})}\BibitemShut {NoStop}%
\bibitem [{\citenamefont {Landau}\ \emph {et~al.}(1984)\citenamefont {Landau},
  \citenamefont {Pitaevskii}, \citenamefont {Lifshitz},\ and\ \citenamefont
  {Kosevich}}]{landau_theory_1984}%
  \BibitemOpen
  \bibfield  {author} {\bibinfo {author} {\bibfnamefont {L.~D.}\ \bibnamefont
  {Landau}}, \bibinfo {author} {\bibfnamefont {L.~P.}\ \bibnamefont
  {Pitaevskii}}, \bibinfo {author} {\bibfnamefont {E.~M.}\ \bibnamefont
  {Lifshitz}},\ and\ \bibinfo {author} {\bibfnamefont {A.~M.}\ \bibnamefont
  {Kosevich}},\ }\href@noop {} {\emph {\bibinfo {title} {Theory of
  elasticity}}},\ \bibinfo {edition} {3rd}\ ed.\ (\bibinfo  {publisher}
  {Butterworth-Heinemann},\ \bibinfo {address} {Oxford, England},\ \bibinfo
  {year} {1984})\BibitemShut {NoStop}%
\bibitem [{\citenamefont {Wiggins}\ and\ \citenamefont
  {Goldstein}(1998)}]{wiggins_flexive_1998}%
  \BibitemOpen
  \bibfield  {author} {\bibinfo {author} {\bibfnamefont {C.~H.}\ \bibnamefont
  {Wiggins}}\ and\ \bibinfo {author} {\bibfnamefont {R.~E.}\ \bibnamefont
  {Goldstein}},\ }\href {https://doi.org/10.1103/PhysRevLett.80.3879}
  {\bibfield  {journal} {\bibinfo  {journal} {Physical Review Letters}\
  }\textbf {\bibinfo {volume} {80}},\ \bibinfo {pages} {3879} (\bibinfo {year}
  {1998})}\BibitemShut {NoStop}%
\bibitem [{\citenamefont {Lauga}\ \emph {et~al.}(2021)\citenamefont {Lauga},
  \citenamefont {Nghi~Dang},\ and\ \citenamefont
  {Ishikawa}}]{lauga_zigzag_2021}%
  \BibitemOpen
  \bibfield  {author} {\bibinfo {author} {\bibfnamefont {E.}~\bibnamefont
  {Lauga}}, \bibinfo {author} {\bibfnamefont {T.}~\bibnamefont {Nghi~Dang}},\
  and\ \bibinfo {author} {\bibfnamefont {T.}~\bibnamefont {Ishikawa}},\ }\href
  {https://doi.org/10.1209/0295-5075/133/44002} {\bibfield  {journal} {\bibinfo
   {journal} {Europhysics Letters}\ }\textbf {\bibinfo {volume} {133}},\
  \bibinfo {pages} {44002} (\bibinfo {year} {2021})}\BibitemShut {NoStop}%
\bibitem [{\citenamefont {Montazeri}\ \emph {et~al.}(2025)\citenamefont
  {Montazeri}, \citenamefont {Rahimi},\ and\ \citenamefont
  {Park}}]{montazeri_non-reciprocity_2025}%
  \BibitemOpen
  \bibfield  {author} {\bibinfo {author} {\bibfnamefont {A.}~\bibnamefont
  {Montazeri}}, \bibinfo {author} {\bibfnamefont {M.}~\bibnamefont {Rahimi}},\
  and\ \bibinfo {author} {\bibfnamefont {H.~S.}\ \bibnamefont {Park}},\ }\href
  {https://doi.org/10.1016/j.ijmecsci.2025.109990} {\bibfield  {journal}
  {\bibinfo  {journal} {International Journal of Mechanical Sciences}\ }\textbf
  {\bibinfo {volume} {287}},\ \bibinfo {pages} {109990} (\bibinfo {year}
  {2025})}\BibitemShut {NoStop}%
\bibitem [{\citenamefont {Brandenbourger}\ \emph {et~al.}(2019)\citenamefont
  {Brandenbourger}, \citenamefont {Locsin}, \citenamefont {Lerner},\ and\
  \citenamefont {Coulais}}]{brandenbourger_non-reciprocal_2019}%
  \BibitemOpen
  \bibfield  {author} {\bibinfo {author} {\bibfnamefont {M.}~\bibnamefont
  {Brandenbourger}}, \bibinfo {author} {\bibfnamefont {X.}~\bibnamefont
  {Locsin}}, \bibinfo {author} {\bibfnamefont {E.}~\bibnamefont {Lerner}},\
  and\ \bibinfo {author} {\bibfnamefont {C.}~\bibnamefont {Coulais}},\ }\href
  {https://doi.org/10.1038/s41467-019-12599-3} {\bibfield  {journal} {\bibinfo
  {journal} {Nature Communications}\ }\textbf {\bibinfo {volume} {10}},\
  \bibinfo {pages} {4608} (\bibinfo {year} {2019})}\BibitemShut {NoStop}%
\bibitem [{\citenamefont {Braverman}\ \emph {et~al.}(2021)\citenamefont
  {Braverman}, \citenamefont {Scheibner}, \citenamefont {VanSaders},\ and\
  \citenamefont {Vitelli}}]{braverman_topological_2021}%
  \BibitemOpen
  \bibfield  {author} {\bibinfo {author} {\bibfnamefont {L.}~\bibnamefont
  {Braverman}}, \bibinfo {author} {\bibfnamefont {C.}~\bibnamefont
  {Scheibner}}, \bibinfo {author} {\bibfnamefont {B.}~\bibnamefont
  {VanSaders}},\ and\ \bibinfo {author} {\bibfnamefont {V.}~\bibnamefont
  {Vitelli}},\ }\href {https://doi.org/10.1103/PhysRevLett.127.268001}
  {\bibfield  {journal} {\bibinfo  {journal} {Physical Review Letters}\
  }\textbf {\bibinfo {volume} {127}},\ \bibinfo {pages} {268001} (\bibinfo
  {year} {2021})}\BibitemShut {NoStop}%
\bibitem [{\citenamefont {Bililign}\ \emph {et~al.}(2022)\citenamefont
  {Bililign}, \citenamefont {Balboa~Usabiaga}, \citenamefont {Ganan},
  \citenamefont {Poncet}, \citenamefont {Soni}, \citenamefont {Magkiriadou},
  \citenamefont {Shelley}, \citenamefont {Bartolo},\ and\ \citenamefont
  {Irvine}}]{bililign_motile_2022}%
  \BibitemOpen
  \bibfield  {author} {\bibinfo {author} {\bibfnamefont {E.~S.}\ \bibnamefont
  {Bililign}}, \bibinfo {author} {\bibfnamefont {F.}~\bibnamefont
  {Balboa~Usabiaga}}, \bibinfo {author} {\bibfnamefont {Y.~A.}\ \bibnamefont
  {Ganan}}, \bibinfo {author} {\bibfnamefont {A.}~\bibnamefont {Poncet}},
  \bibinfo {author} {\bibfnamefont {V.}~\bibnamefont {Soni}}, \bibinfo {author}
  {\bibfnamefont {S.}~\bibnamefont {Magkiriadou}}, \bibinfo {author}
  {\bibfnamefont {M.~J.}\ \bibnamefont {Shelley}}, \bibinfo {author}
  {\bibfnamefont {D.}~\bibnamefont {Bartolo}},\ and\ \bibinfo {author}
  {\bibfnamefont {W.~T.~M.}\ \bibnamefont {Irvine}},\ }\href
  {https://doi.org/10.1038/s41567-021-01429-3} {\bibfield  {journal} {\bibinfo
  {journal} {Nature Physics}\ }\textbf {\bibinfo {volume} {18}},\ \bibinfo
  {pages} {212} (\bibinfo {year} {2022})}\BibitemShut {NoStop}%
\bibitem [{\citenamefont {Ishikawa}(2009)}]{ishikawa_suspension_2009}%
  \BibitemOpen
  \bibfield  {author} {\bibinfo {author} {\bibfnamefont {T.}~\bibnamefont
  {Ishikawa}},\ }\href {https://doi.org/10.1098/rsif.2009.0223} {\bibfield
  {journal} {\bibinfo  {journal} {Journal of The Royal Society Interface}\
  }\textbf {\bibinfo {volume} {6}},\ \bibinfo {pages} {815} (\bibinfo {year}
  {2009})}\BibitemShut {NoStop}%
\bibitem [{\citenamefont {Ishikawa}(2024)}]{ishikawa_fluid_2024}%
  \BibitemOpen
  \bibfield  {author} {\bibinfo {author} {\bibfnamefont {T.}~\bibnamefont
  {Ishikawa}},\ }\href {https://doi.org/10.1146/annurev-fluid-121021-042929}
  {\bibfield  {journal} {\bibinfo  {journal} {Annual Review of Fluid
  Mechanics}\ }\textbf {\bibinfo {volume} {56}},\ \bibinfo {pages} {119}
  (\bibinfo {year} {2024})}\BibitemShut {NoStop}%
\bibitem [{\citenamefont {Stone}\ and\ \citenamefont
  {Samuel}(1996)}]{stone_propulsion_1996}%
  \BibitemOpen
  \bibfield  {author} {\bibinfo {author} {\bibfnamefont {H.~A.}\ \bibnamefont
  {Stone}}\ and\ \bibinfo {author} {\bibfnamefont {A.~D.~T.}\ \bibnamefont
  {Samuel}},\ }\href {https://doi.org/10.1103/PhysRevLett.77.4102} {\bibfield
  {journal} {\bibinfo  {journal} {Physical Review Letters}\ }\textbf {\bibinfo
  {volume} {77}},\ \bibinfo {pages} {4102} (\bibinfo {year}
  {1996})}\BibitemShut {NoStop}%
\bibitem [{\citenamefont {Iserles}\ \emph {et~al.}(2000)\citenamefont
  {Iserles}, \citenamefont {Munthe-Kaas}, \citenamefont {Nørsett},\ and\
  \citenamefont {Zanna}}]{iserles_lie-group_2000}%
  \BibitemOpen
  \bibfield  {author} {\bibinfo {author} {\bibfnamefont {A.}~\bibnamefont
  {Iserles}}, \bibinfo {author} {\bibfnamefont {H.~Z.}\ \bibnamefont
  {Munthe-Kaas}}, \bibinfo {author} {\bibfnamefont {S.~P.}\ \bibnamefont
  {Nørsett}},\ and\ \bibinfo {author} {\bibfnamefont {A.}~\bibnamefont
  {Zanna}},\ }\href {https://doi.org/10.1017/S0962492900002154} {\bibfield
  {journal} {\bibinfo  {journal} {Acta Numerica}\ }\textbf {\bibinfo {volume}
  {9}},\ \bibinfo {pages} {215} (\bibinfo {year} {2000})}\BibitemShut {NoStop}%
\bibitem [{\citenamefont {Singh}\ and\ \citenamefont
  {Adhikari}(2020)}]{singh_pystokes_2020}%
  \BibitemOpen
  \bibfield  {author} {\bibinfo {author} {\bibfnamefont {R.}~\bibnamefont
  {Singh}}\ and\ \bibinfo {author} {\bibfnamefont {R.}~\bibnamefont
  {Adhikari}},\ }\href {https://doi.org/10.21105/joss.02318} {\bibfield
  {journal} {\bibinfo  {journal} {Journal of Open Source Software}\ }\textbf
  {\bibinfo {volume} {5}},\ \bibinfo {pages} {2318} (\bibinfo {year}
  {2020})}\BibitemShut {NoStop}%
\bibitem [{\citenamefont {Musy}\ \emph {et~al.}(2025)\citenamefont {Musy},
  \citenamefont {Jacquenot}, \citenamefont {Dalmasso}, \citenamefont {Lee},
  \citenamefont {Pujol}, \citenamefont {Soltwedel}, \citenamefont {de~Bruin},
  \citenamefont {Zhou}, \citenamefont {Tulldahl}, \citenamefont {Poisonous},
  \citenamefont {CorpsSansOrganes}, \citenamefont {RobinEnjalbert},
  \citenamefont {Sol}, \citenamefont {Lu}, \citenamefont {S}, \citenamefont
  {Badger}, \citenamefont {Kunimune}, \citenamefont {Claudi}, \citenamefont
  {Hacha}, \citenamefont {Lee}, \citenamefont {Pollack}, \citenamefont
  {Schneider}, \citenamefont {daizhirui}, \citenamefont {RichardScottOZ},
  \citenamefont {Mitrano}, \citenamefont {Brodersen}, \citenamefont
  {Schlömer}, \citenamefont {mkerrinrapid}, \citenamefont {Linus-Foley},\ and\
  \citenamefont {JohnsWor}}]{marco_musy_marcomusyvedo_2025}%
  \BibitemOpen
  \bibfield  {author} {\bibinfo {author} {\bibfnamefont {M.}~\bibnamefont
  {Musy}}, \bibinfo {author} {\bibfnamefont {G.}~\bibnamefont {Jacquenot}},
  \bibinfo {author} {\bibfnamefont {G.}~\bibnamefont {Dalmasso}}, \bibinfo
  {author} {\bibfnamefont {J.}~\bibnamefont {Lee}}, \bibinfo {author}
  {\bibfnamefont {L.}~\bibnamefont {Pujol}}, \bibinfo {author} {\bibfnamefont
  {J.}~\bibnamefont {Soltwedel}}, \bibinfo {author} {\bibfnamefont
  {R.}~\bibnamefont {de~Bruin}}, \bibinfo {author} {\bibfnamefont {Z.-Q.}\
  \bibnamefont {Zhou}}, \bibinfo {author} {\bibfnamefont {M.}~\bibnamefont
  {Tulldahl}}, \bibinfo {author} {\bibnamefont {Poisonous}}, \bibinfo {author}
  {\bibnamefont {CorpsSansOrganes}}, \bibinfo {author} {\bibnamefont
  {RobinEnjalbert}}, \bibinfo {author} {\bibfnamefont {A.}~\bibnamefont {Sol}},
  \bibinfo {author} {\bibfnamefont {X.}~\bibnamefont {Lu}}, \bibinfo {author}
  {\bibfnamefont {U.~E.}\ \bibnamefont {S}}, \bibinfo {author} {\bibfnamefont
  {C.}~\bibnamefont {Badger}}, \bibinfo {author} {\bibfnamefont
  {J.}~\bibnamefont {Kunimune}}, \bibinfo {author} {\bibfnamefont
  {F.}~\bibnamefont {Claudi}}, \bibinfo {author} {\bibfnamefont
  {B.}~\bibnamefont {Hacha}}, \bibinfo {author} {\bibfnamefont
  {A.}~\bibnamefont {Lee}}, \bibinfo {author} {\bibfnamefont {A.}~\bibnamefont
  {Pollack}}, \bibinfo {author} {\bibfnamefont {O.}~\bibnamefont {Schneider}},
  \bibinfo {author} {\bibnamefont {daizhirui}}, \bibinfo {author} {\bibnamefont
  {RichardScottOZ}}, \bibinfo {author} {\bibfnamefont {P.}~\bibnamefont
  {Mitrano}}, \bibinfo {author} {\bibfnamefont {P.}~\bibnamefont {Brodersen}},
  \bibinfo {author} {\bibfnamefont {N.}~\bibnamefont {Schlömer}}, \bibinfo
  {author} {\bibnamefont {mkerrinrapid}}, \bibinfo {author} {\bibnamefont
  {Linus-Foley}},\ and\ \bibinfo {author} {\bibnamefont {JohnsWor}},\ }\href
  {https://doi.org/10.5281/ZENODO.2561401} {\bibinfo {title} {marcomusy/vedo:
  v2025.5.4}} (\bibinfo {year} {2025})\BibitemShut {NoStop}%
\end{thebibliography}%

\section*{Appendix A: Generalized Stokes laws}

\begingroup\tabcolsep=4pt\def\arraystretch{3}
\begin{table*}
\centering
\begin{tabular}{|c||c|c|c|}
\hline 
$l\sigma$ & $v_{r}^{(l\sigma)}$ & $v_{\theta}^{(l\sigma)}$ & $v_{\phi}^{(l\sigma)}$\tabularnewline
\hline 
\hline 
$2s$ & ${\displaystyle V_{0}^{(2s)}\left(\frac{2}{3}-\sin^{2}\theta\right)}$ & ${\displaystyle -\frac{1}{2}V_{0}^{(2s)}\sin2\theta}$ & $0$\tabularnewline
\hline 
$3a$ & $0$ & $0$ & ${\displaystyle \frac{1}{18}V_{0}^{(3a)}\sin2\theta}$\tabularnewline
\hline 
$3t$ & ${\displaystyle \frac{1}{45}V_{0}^{(3t)}\cos\theta}$ & ${\displaystyle \frac{1}{45}V_{0}^{(3t)}\sin\theta}$ & $0$\tabularnewline
\hline 
$4a$ & $0$ & $0$ & ${\displaystyle \frac{1}{60}V_{0}^{(4a)}\sin\theta\left(\cos^{2}\theta-\frac{1}{5}\right)}$\tabularnewline
\hline 
\end{tabular}

\caption{Components of active slip velocity $\bm{v}^{{\rm slip}}$ in Eq. \eqref{eq:slip}
in spherical polar coordinates $(r,\theta,\phi)$, for leading coefficients
categorized by symmetry. The slip modes are parameterized uniaxially
based on the particle orientation.}
\label{table:slip}
\end{table*}
\endgroup The active particle is modeled using the classical squirmer
model \citep{lighthill_squirming_1952,blake_spherical_1971,ishikawa_suspension_2009,pedley_spherical_2016,ishikawa_fluid_2024}.
This model prescribes a slip velocity $\bm{v}^{\text{slip}}$ on the
surface of a spherical particle of radius $a$, as shown in Eq.\eqref{eq:slip_vel}.
Following \citep{singh_generalized_2018}, we expand the slip velocity
in terms of tensorial spherical harmonics,
\begin{align}
\bm{v}^{{\rm slip}}\left(\bm{r}+\bm{\rho}\right) & =\sum_{l=1}^{\infty}\frac{1}{\left(l-1\right)!\left(2l-3\right)!!}\bm{V}^{(l)}\cdot\bm{Y}^{(l)}\left(\hat{\bm{\rho}}\right),\label{eq:slip_vel}
\end{align}
where the basis functions $\bm{Y}^{(l)}\left(\hat{\bm{\rho}}\right)=(-1)^{l}\rho^{l+1}\bm{\nabla}^{(l)}\rho^{-1}$
represent irreducible tensorial spherical harmonics, with $\bm{\nabla}^{(l)}=\bm{\nabla}_{\alpha_{1}}\cdots\bm{\nabla}_{\alpha_{l}}$.
Here, $\bm{\rho}$ denotes the radius vector from the center of the
particle, and $\hat{\bm{\rho}}=\bm{\rho}/a$. The expansion coefficients
$\bm{V}^{(l)}$ are $l$-th rank reducible Cartesian tensors, which
can be decomposed into three irreducible parts $\bm{V}^{(l\sigma)}$
of ranks $l$, $l-1$, and $l-2$ corresponding respectively to the
symmetric traceless $(\sigma=s)$, antisymmetric $(\sigma=a)$, and
trace $(\sigma=t)$ components. The leading-order slip modes, classified
by their polar and chiral symmetry combinations, are given by 
\begin{equation}
\begin{split}\bm{v}^{{\rm slip}}\left(\bm{\rho}\right) & =\underbrace{\frac{1}{15}\bm{V}^{(3t)}\cdot\bm{Y}^{(2)}\left(\hat{\boldsymbol{\rho}}\right)}_{\text{achiral, polar}}\underbrace{+\bm{V}^{(2s)}\cdot\bm{Y}^{(1)}\left(\hat{\boldsymbol{\rho}}\right)}_{\text{achiral, apolar}}\\
 & \underbrace{-\frac{1}{60}\bm{\epsilon}\cdot\bm{V}^{(4a)}\cdot\bm{Y}^{(3)}\left(\hat{\boldsymbol{\rho}}\right)}_{\text{chiral, polar}}\underbrace{-\frac{1}{9}\bm{\epsilon}\cdot\bm{V}^{(3a)}\cdot\bm{Y}^{(2)}\left(\hat{\boldsymbol{\rho}}\right)}_{\text{chiral, apolar}}.
\end{split}
\label{eq:slip}
\end{equation}
\begin{figure}
\centering
\includegraphics[width=1\linewidth]{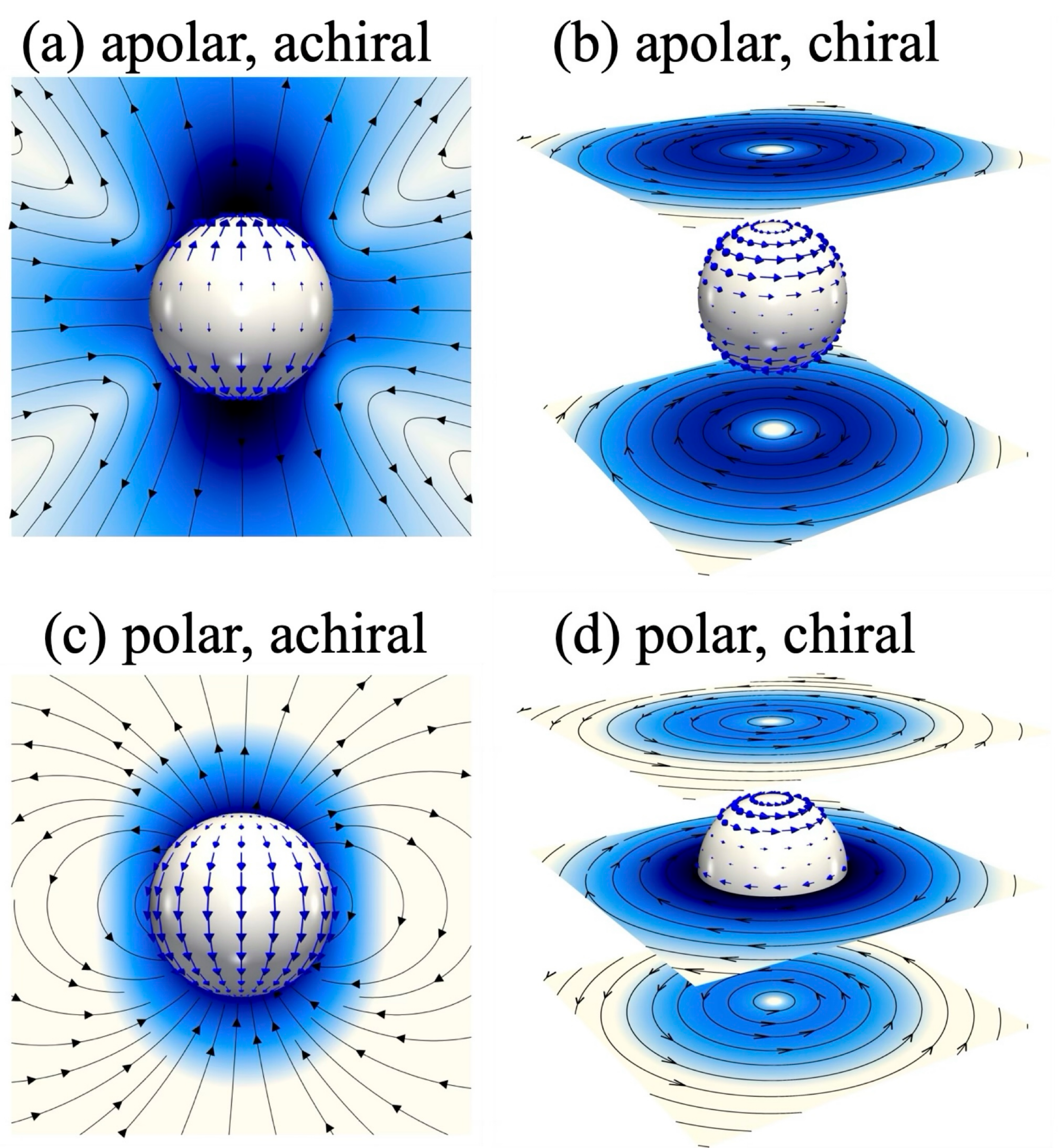}\caption{Slip velocities corresponding to leading order swimming modes in Eq.
\eqref{eq:slip}, and the resulting flow fields.\label{fig:slip}}
\end{figure}
The explicit forms of these modes in spherical coordinates, are presented
in Table \ref{table:slip} and illustrated in \ref{fig:slip}. The
first two modes of the slip, $\bm{V}^{(1s)}\equiv-\bm{V}^{A}$ and
$\bm{V}^{(2a)}\equiv-a\bm{\Omega}_{}^{A}$, are active translational
and angular velocities, respectively. These can be obtained directly
from the surface slip as \citep{stone_propulsion_1996},
\begin{align*}
\bm{V}^{A} & =-\frac{1}{4\pi a^{2}}\int_{\mathcal{S}}\bm{v}^{\text{slip}}\left(\bm{\rho}\right)dS,\\
\bm{\Omega}^{A} & =-\frac{3}{8\pi a^{3}}\int_{\mathcal{S}}\bm{\rho}\times\bm{v}^{\text{slip}}\left(\bm{\rho}\right)dS,
\end{align*}
where $\mathcal{S}$ is the surface of the squirmer. For a collection
of squirmers, given the slip velocity, we seek expressions for the
resulting hydrodynamic forces $\bm{F}_{H}^{i}$ and torques $\bm{T}_{H}^{i}$
on the spheres. By linearity of Stokes equations, it is clear that
these must be of the form (summation on $j$ understood): 
\begin{align*}
\bm{F}_{H}^{i} & =-\bm{\gamma}_{ij}^{TT}\cdot\bm{v}^{j}-\bm{\gamma}_{ij}^{TR}\cdot\bm{\omega}^{j}\underbrace{-\sum_{l\sigma=1s}\bm{\gamma}_{ij}^{(T,l\sigma)}\cdot\bm{V}_{j}^{(l\sigma)}}_{{\displaystyle \bm{F}_{A}^{i}}},\\
\bm{T}_{H}^{i} & =-\bm{\gamma}_{ij}^{RT}\cdot\bm{v}^{j}-\bm{\gamma}_{ij}^{RR}\cdot\bm{\omega}^{j}\underbrace{-\sum_{l\sigma=1s}\bm{\gamma}_{ij}^{(R,l\sigma)}\cdot\bm{V}_{j}^{(l\sigma)}}_{{\displaystyle \bm{T}_{A}^{i}}},
\end{align*}
where the $\bm{V}_{j}^{(l\sigma)}$ are the irreducible parts of the
slip modes of particle $j$. The leading-order active forces and torques
arising from the slip modes can be conveniently expressed using the
Green's function $\bm{G}$ of unbounded Stokes flow \citep{singh_generalized_2018},
as summarized in Table \ref{table:active_force_torque}.\begingroup\tabcolsep=4pt\def\arraystretch{3}
\begin{table*}
\centering
\begin{tabular}{|c||c|c|}
\hline 
$l\sigma$ & $\bm{F}_{ij}^{\mathcal{A}(l\sigma)}$ & $\bm{T}_{ij}^{\mathcal{A}(l\sigma)}$\tabularnewline
\hline 
\hline 
$2s$ & ${\displaystyle -\frac{28}{3}\pi\eta a^{2}\gamma^{T}\mathcal{F}_{i}^{0}\mathcal{F}_{j}^{1}\bm{\nabla}_{\boldsymbol{r}^{j}}\bm{G}\cdot\bm{V}_{j}^{(2s)}}$ & ${\displaystyle -\frac{14}{3}\pi\eta a^{2}\gamma^{R}\bm{\nabla}_{\bm{r}^{i}}\times\left(\bm{\nabla}_{\boldsymbol{r}^{j}}\bm{G}\right)\cdot\bm{V}_{j}^{(2s)}}$\tabularnewline
\hline 
$3a$ & ${\displaystyle -\frac{13}{9}\pi\eta a^{3}\gamma^{T}\bm{\nabla}_{\bm{r}^{j}}\left(\bm{\nabla}_{\bm{r}^{j}}\times\bm{G}\right)\cdot\bm{V}_{j}^{(3a)}}$ & ${\displaystyle -\frac{13}{18}\pi\eta a^{3}\gamma^{R}\bm{\nabla}_{\boldsymbol{r}^{i}}\times\bm{\nabla}_{\boldsymbol{r}^{j}}\left(\bm{\nabla}_{\boldsymbol{r}^{j}}\times\bm{G}\right)\cdot\bm{V}_{j}^{(3a)}}$\tabularnewline
\hline 
$3t$ & ${\displaystyle \frac{4}{5}\pi\eta a^{3}\gamma^{T}\bm{\nabla}_{\bm{r}_{j}}^{2}\bm{G}\cdot\bm{V}_{j}^{(3t)}}$ & $0$\tabularnewline
\hline 
$4a$ & ${\displaystyle \frac{121}{10}\pi\eta a^{4}\gamma^{T}\bm{\nabla}_{\bm{r}^{j}}\bm{\nabla}_{\bm{r}^{j}}\left(\bm{\nabla}_{\bm{r}^{j}}\times\bm{G}\right)\cdot\bm{V}_{j}^{(4a)}}$ & ${\displaystyle \frac{121}{20}\pi\eta a^{4}\gamma^{R}\bm{\nabla}_{\bm{r}^{i}}\times\bm{\nabla}_{\boldsymbol{r}^{j}}\bm{\nabla}_{\boldsymbol{r}^{j}}\left(\bm{\nabla}_{\boldsymbol{r}^{j}}\times\bm{G}\right)\cdot\bm{V}_{j}^{(4a)}}$\tabularnewline
\hline 
\end{tabular}

\caption{Active forces and torques for leading coefficients of polar, apolar
and chiral symmetry in terms of the Green's function of Stokes flow
\citep{singh_generalized_2018}. The operator $\mathcal{F}_{i}^{l}=\left(1+\frac{a^{2}}{4l+6}\nabla_{\boldsymbol{r}^{i}}^{2}\right)$,
which we set to unity in Eq. \eqref{eq:2s_force} as we were in the
dilute limit, corrects for the finite size of the particle.}
\label{table:active_force_torque}
\end{table*}
\endgroup

\section*{Appendix B: Discretization of conservative Cosserat chain}

The conservative forces and torques are derived from an elastic potential
energy 
\begin{equation}
V=\sum_{i=2}^{N-2}U\left(\boldsymbol{r}^{i-1},\dots,\boldsymbol{r}^{i+2},\boldsymbol{e}_{a}^{i-1},\dots,\boldsymbol{e}_{b}^{i+2}\right),\label{eq:discr_pot}
\end{equation}
where $U=U_{\text{stretch}}+U_{\text{shear}}+U_{\text{twist}}+U_{\text{bend}}$
is a conservative four-body potential implementing elastic interactions
along the chain. It is comprised of four terms, with a harmonic energy
cost associated with each of the deformation types as follows: 
\begin{align}
U_{\text{stretch}} & =\frac{\lambda_{\parallel}d^{2}}{2}\left(\mathcal{C}^{0}\boldsymbol{e}_{1}^{i+1/2}\cdot\mathcal{C}^{1}\boldsymbol{r}^{i+1/2}-1\right)^{2}\label{eq:stretch_pot}\\
U_{\text{shear}} & =\frac{\lambda_{\perp}d^{2}}{2}\sum_{a=2}^{3}\left(C^{0}\boldsymbol{e}_{a}^{i+1/2}\cdot\mathcal{C}^{1}\boldsymbol{r}^{i+1/2}\right)^{2}\label{eq:shear_pot}\\
U_{\text{twist}} & =\frac{\mu_{\parallel}}{2}\left(\boldsymbol{e}_{2}^{i}\cdot\boldsymbol{e}_{3}^{i+1}\right)^{2},\label{eq:twist_pot}\\
U_{\text{bend}} & =\frac{\mu_{\perp}}{2}\left(\left(\boldsymbol{e}_{1}^{i}\cdot\boldsymbol{e}_{2}^{i+1}\right)^{2}+\left(\boldsymbol{e}_{3}^{i}\cdot\boldsymbol{e}_{1}^{i+1}\right)^{2}\right),\label{eq:bend_pot}
\end{align}
where 
\[
\mathcal{C}^{0}\boldsymbol{e}_{a}^{i+1/2}=\frac{9}{16}\left(\boldsymbol{e}_{a}^{i}+\boldsymbol{e}_{a}^{i+1}\right)-\frac{1}{16}\left(\boldsymbol{e}_{a}^{i-1}+\boldsymbol{e}_{a}^{i+2}\right)
\]
and
\[
\mathcal{C}^{1}\boldsymbol{r}^{i+1/2}=\frac{9}{8}\frac{\boldsymbol{r}^{i+1}-\boldsymbol{r}^{i}}{d}-\frac{1}{24}\frac{\boldsymbol{r}^{i+2}-\boldsymbol{r}^{i-1}}{d}
\]
are centered difference approximations to the frame vectors and the
tangent vector at the midpoint of the link between bodies $i$ and
$i+1$, respectively. We used difference approximations to ensure
that the discrete conservative potential is also apolar and chose
formulae of high enough order, so that the linearized mode structure
of the discrete system recovers correctly the relaxation rate of the
fourth order bending mode of the continuum theory.

The shear term \eqref{eq:shear_pot} penalizes any misalignment between
the local tangent and the frame vector $\boldsymbol{e}_{1}^{i}$,
while the stretch term penalizes any deviation of the bond length
from the equilibrium distance $d$. The bending term in Eq. \eqref{eq:bend_pot}
penalizes the misalignment of the frame vectors $\boldsymbol{e}_{1}^{i}$,
and finally the twist term in Eq. \eqref{eq:twist_pot} penalizes
misalignment of $\boldsymbol{e}_{a}^{i}$ for $a=2,3$, i.e., any
twist around the $\boldsymbol{e}_{1}^{i}$ frame vectors. In the singular
limit $\lambda_{\perp}\to\infty$, we obtain an unshearable (semiflexible)
chain, where the local tangent is always aligned with the frame vector
$\boldsymbol{e}_{1}^{i}$, but is not always of length $d$. We can
obtain a further reduction if we take $\lambda_{\parallel}\to\infty$
too, which leads to an inextensible chain. For simplicity, we assume
hereafter and in the main part of the paper that $\lambda_{\perp}=\lambda_{\parallel}\equiv\lambda$
and $\mu_{\perp}=\mu_{\parallel}\equiv\mu$, so that there are only
two discrete elastic moduli.

In Sec. \ref{sec:comparison} we also encounter the centered difference
operators
\begin{align*}
\mathcal{D}_{u}^{2}\boldsymbol{u}^{i} & =\mathcal{C}^{1}\mathcal{C}^{1}\boldsymbol{u}^{i}, & \mathcal{D}_{u\varphi}^{1}\boldsymbol{\varphi}^{i} & =\mathcal{C}^{1}\mathcal{C}^{0}\boldsymbol{\varphi}^{i},\\
\mathcal{D}_{\varphi u}^{1}\boldsymbol{u}^{i} & =\mathcal{C}^{0}\mathcal{C}^{1}\boldsymbol{u}^{i}, & \mathcal{D}_{\varphi}^{0}\boldsymbol{\varphi}^{i} & =\mathcal{C}^{0}\mathcal{C}^{0}\boldsymbol{\varphi}^{i},
\end{align*}
together with 
\[
D_{\varphi}^{2}\boldsymbol{\varphi}^{i}=\frac{\boldsymbol{\varphi}^{i+1}-2\boldsymbol{\varphi}^{i}+\boldsymbol{\varphi}^{i-1}}{d^{2}},
\]
all of which are finite difference approximations of derivatives along
the chain, following from linearization of the conservative forces
and torques.

\section*{Appendix C: Discrete and continuum linearized Cosserat equations}

\begingroup\tabcolsep=4pt\def\arraystretch{2}
\begin{table*}
\centering
\begin{tabular}{|c||c|c|}
\hline 
 & \multicolumn{1}{c|}{achiral} & chiral\tabularnewline
\hline 
\hline 
\multirow{1}{*}{apolar} & \multicolumn{1}{c|}{$\left(\mathbb{I}-3\hat{\bm{x}}\hat{\bm{x}}\right)\cdot\left(\bm{u}^{i+1}-2\bm{u}^{i}+\bm{u}^{i-1}\right)$} & $\hat{\bm{x}}\times\left(\bm{u}^{i+1}-\bm{u}^{i-1}\right)-d\left(\bm{\varphi}^{i+1}+\bm{\varphi}^{i-1}\right)_{\perp}$\tabularnewline
\hline 
\multirow{1}{*}{polar} & \multicolumn{1}{c|}{$\left(\mathbb{I}-3\hat{\bm{x}}\hat{\bm{x}}\right)\cdot\left(\bm{u}^{i+1}-\bm{u}^{i-1}\right)+\frac{d}{3}\hat{\bm{x}}\times\left(\bm{\varphi}^{i+1}+\bm{\varphi}^{i-1}\right)$} & $\hat{\bm{x}}\times\left(\bm{u}^{i+1}-2\bm{u}^{i}+\bm{u}^{i-1}\right)-d\left(\bm{\varphi}^{i+1}+\bm{\varphi}^{i-1}\right)_{\perp}$\tabularnewline
\hline 
\end{tabular}\caption{Linearized forces entering the discrete equations of motion coming
from the active squirming forces and torques up dimensional prefactors.\label{tab:discr_forces}}
\end{table*}
\endgroup\begingroup\tabcolsep=4pt\def\arraystretch{2}
\begin{table*}
\centering
\begin{tabular}{|c||c|c|}
\hline 
 & \multicolumn{1}{c|}{achiral} & chiral\tabularnewline
\hline 
\hline 
\multirow{1}{*}{apolar} & \multicolumn{1}{c|}{$C_{\parallel}^{F\varepsilon}\boldsymbol{u}_{\parallel}''+C_{\perp}^{F\varepsilon}\boldsymbol{u}_{\perp}''+\left(C_{\perp}^{F\varepsilon}+H_{\circlearrowleft}^{f\tau}-\tilde{F}_{\parallel}\right)\hat{\boldsymbol{x}}\times\boldsymbol{\varphi}_{\perp}'$} & $H_{\circlearrowleft}^{f\varepsilon}\left(\hat{\boldsymbol{x}}\times\boldsymbol{u}{}_{\perp}'-\boldsymbol{\varphi}_{\perp}\right)+C_{\parallel}^{F\tau}\boldsymbol{\varphi}_{\parallel}''+C_{\perp}^{F\tau}\boldsymbol{\varphi}_{\perp}''$\tabularnewline
\hline 
\multirow{1}{*}{polar} & \multicolumn{1}{c|}{$H_{\parallel}^{f\varepsilon}\boldsymbol{u}_{\parallel}'+H_{\perp}^{f\varepsilon}\boldsymbol{u}_{\perp}'+\left(H_{\perp}^{f\varepsilon}-\tilde{f}_{\parallel}\right)\hat{\boldsymbol{x}}\times\boldsymbol{\varphi}_{\perp}+C_{\circlearrowleft}^{F\tau}\hat{\boldsymbol{x}}\times\boldsymbol{\varphi}_{\perp}''$} & $H_{\parallel}^{f\tau}\boldsymbol{\varphi}'_{\parallel}+C_{\circlearrowleft}^{F\varepsilon}\hat{\boldsymbol{x}}\times\boldsymbol{u}_{\perp}''+\left(H_{\perp}^{f\tau}-C_{\circlearrowleft}^{F\varepsilon}\right)\boldsymbol{\varphi}_{\perp}'$\tabularnewline
\hline 
\end{tabular}\caption{Linearized forces entering the continuum equations of motion.\label{tab:cont_forces}}
\end{table*}
\endgroup\begingroup\tabcolsep=4pt\def\arraystretch{2}
\begin{table*}
\centering
\begin{tabular}{|c||c|c|}
\hline 
 & \multicolumn{1}{c|}{achiral} & chiral\tabularnewline
\hline 
\hline 
\multirow{1}{*}{apolar} & $\hat{\bm{x}}\times\left(\bm{u}^{i+1}-\bm{u}^{i-1}\right)-d\left(\bm{\varphi}^{i+1}+\bm{\varphi}^{i-1}\right)_{\perp}$ & $\left(\mathbb{I}-3\hat{\bm{x}}\hat{\bm{x}}\right)\cdot\left(\bm{u}^{i+1}-2\bm{u}^{i}+\bm{u}^{i-1}\right)+\frac{d}{2}\hat{\bm{x}}\times\left(\bm{\varphi}^{i+1}-\bm{\varphi}^{i-1}\right)$\tabularnewline
\hline 
\multirow{1}{*}{polar} & $\boldsymbol{0}$ & $\left(\mathbb{I}-3\hat{\bm{x}}\hat{\bm{x}}\right)\cdot\left(\bm{u}^{i+1}-\bm{u}^{i-1}\right)+d\hat{\bm{x}}\times\left(\bm{\varphi}^{i+1}+\bm{\varphi}^{i-1}\right)$\tabularnewline
\hline 
\end{tabular}\caption{Linearized torques entering the discrete equations of motion coming
from the active squirming forces and torques up dimensional prefactors.\label{tab:discr_torques}}
\end{table*}
\endgroup\begingroup\tabcolsep=4pt\def\arraystretch{2}
\begin{table*}
\centering
\begin{tabular}{|c||c|c|}
\hline 
 & \multicolumn{1}{c|}{achiral} & chiral\tabularnewline
\hline 
\hline 
\multirow{1}{*}{apolar} & $C_{\parallel}^{M\tau}\boldsymbol{\varphi}_{\parallel}''+C_{\perp}^{M\tau}\boldsymbol{\varphi}_{\perp}''+\left(H_{\circlearrowleft}^{m\varepsilon}+C_{\parallel}^{F\varepsilon}-\tilde{F}_{\parallel}\right)\left(\hat{\boldsymbol{x}}\times\boldsymbol{u}{}_{\perp}'-\boldsymbol{\varphi}_{\perp}\right)$ & $C_{\perp}^{M\varepsilon}\boldsymbol{u}_{\perp}''+C_{\parallel}^{M\varepsilon}\boldsymbol{u}_{\parallel}''+\left(C_{\perp}^{M\varepsilon}+C_{\perp}^{F\tau}+H_{\circlearrowleft}^{m\tau}-\tilde{M}_{\parallel}\right)\hat{\boldsymbol{x}}\times\boldsymbol{\varphi}_{\perp}'$\tabularnewline
\hline 
\multirow{1}{*}{polar} & $C_{\circlearrowleft}^{M\varepsilon}\hat{\boldsymbol{x}}\times\boldsymbol{u}_{\perp}''+\left(H_{\perp}^{m\tau}-C_{\circlearrowleft}^{F\tau}-C_{\circlearrowleft}^{M\varepsilon}\right)\boldsymbol{\varphi}_{\perp}'+H_{\parallel}^{m\tau}\boldsymbol{\varphi}_{\parallel}'$ & $H_{\parallel}^{m\varepsilon}\boldsymbol{u}_{\parallel}'+\left(H_{\perp}^{m\varepsilon}-C_{\circlearrowleft}^{F\varepsilon}\right)\left(\boldsymbol{u}_{\perp}'+\boldsymbol{x}\times\boldsymbol{\varphi}_{\perp}\right)+C_{\circlearrowleft}^{M\tau}\left(\boldsymbol{x}\times\boldsymbol{\varphi}_{\perp}''\right)$\tabularnewline
\hline 
\end{tabular}\caption{Linearized torques entering the continuum equations of motion.\label{tab:cont_torques}}
\end{table*}
\endgroup In this section of the appendix, we provide details of
the linearization of both the continuum equations of motion \eqref{eq:balance_fixed_frame}
of a Cosserat rod and the discrete equations of motion \eqref{eq:pos_eom}-\eqref{eq:rot_eom}
for each symmetry combination. In each case, we assume isotropic dissipation.

\textbf{Passive chain.} For a passive chain with an isotropic hyperelastic
constitutive law \eqref{eq:hyperelastic1}-\eqref{eq:hyperelastic2}
we have 
\[
\boldsymbol{F}^{E}=\kappa_{T}\boldsymbol{\varepsilon},\quad\boldsymbol{M}^{E}=\kappa_{R}\boldsymbol{\tau},\quad\boldsymbol{f}^{E}=\boldsymbol{m}^{E}=\boldsymbol{0},
\]
where we used vector notation $\boldsymbol{\varepsilon}=\varepsilon_{a}\boldsymbol{e}_{a},\boldsymbol{\tau}=\tau_{a}\boldsymbol{e}_{a}$.
The reference configuration about which we linearize is taken to be
a straight chain along the $x$ axis with parallel frames so that
\[
\tilde{\boldsymbol{r}}(s)=s\hat{\boldsymbol{x}},\quad\tilde{\boldsymbol{e}}_{1}\left(s\right)=\hat{\boldsymbol{x}},\quad\tilde{\boldsymbol{e}}_{2}\left(s\right)=\hat{\boldsymbol{y}},\quad\tilde{\boldsymbol{e}}_{3}\left(s\right)=\hat{\boldsymbol{z}}.
\]
We find:
\[
\Gamma^{T}\dot{\boldsymbol{u}}=\left(\kappa_{T}\boldsymbol{\varepsilon}\right)',\quad\Gamma^{R}\dot{\boldsymbol{\varphi}}=\left(\kappa_{R}\boldsymbol{\tau}\right)'+\hat{\boldsymbol{x}}\times\left(\kappa_{T}\boldsymbol{\varepsilon}\right).
\]
Using the expressions for the linearized strain measures \eqref{eq:rod_strain_lin},
we obtain Eqs. \eqref{eq:passive_cont_tr}-\eqref{eq:passive_cont_rot}
of the main text.

To adiabatically eliminate $\boldsymbol{\varphi}_{\perp}$ from the
transverse dynamics \eqref{eq:tr_part}-\eqref{eq:rot_part}, we consider
a perturbative expansion
\begin{equation}
\boldsymbol{\varphi}_{\perp}=\boldsymbol{\varphi}_{\perp}^{(0)}+\frac{1}{\kappa_{T}}\boldsymbol{\varphi}_{\perp}^{(1)}+\mathcal{O}\left(\frac{1}{\kappa_{T}^{2}}\right)\label{eq:phi_pert}
\end{equation}
in powers of $1/\kappa_{T}$ for large $\kappa_{T}$, and substitute
this into the right hand side of \eqref{eq:rot_part} to find:
\begin{align*}
\Gamma^{R}\dot{\boldsymbol{\varphi}}_{\perp}\approx0 & =\kappa_{T}\hat{\boldsymbol{x}}\times\left(\boldsymbol{u}_{\perp}+\hat{\boldsymbol{x}}\times\boldsymbol{\varphi}_{\perp}^{(0)}\right)\\
 & -\boldsymbol{\varphi}_{\perp}^{(1)}+\kappa_{R}\boldsymbol{\varphi}_{\perp}^{(0)}{}''+\mathcal{O}\left(\frac{1}{\kappa_{T}}\right).
\end{align*}
By setting the coefficients of $\kappa_{T}$ zero in this expansion,
to leading order we find 
\[
\boldsymbol{\varphi}_{\perp}^{(0)}=\hat{\boldsymbol{x}}\times\boldsymbol{u}_{\perp}'\quad\boldsymbol{\varphi}_{\perp}^{(1)}=\kappa_{R}\boldsymbol{\varphi}_{\perp}^{(0)}{}''=\kappa_{R}\hat{\boldsymbol{x}}\times\boldsymbol{u}_{\perp}'''.
\]
Substituting these expressions for $\boldsymbol{\varphi}_{\perp}$
into Eq. \eqref{eq:tr_part}, we obtain Eq. \eqref{eq:eb} of the
main text.

\textbf{Apolar, achiral chain. }Small perturbations about the same
straight chain reference state $\left(\tilde{\bm{r}}_{i},\tilde{\boldsymbol{e}}_{i}^{a}\right)$
are governed by the linearization of the discrete equations of motion
\eqref{eq:pos_eom}-\eqref{eq:rot_eom} with active forces and torques
\eqref{eq:2s_force}-\eqref{eq:2s_torque}:
\begin{align}
\gamma^{T}\dot{\bm{u}}^{i} & =\lambda\mathcal{D}_{u}^{2}\bm{u}^{i}+\lambda d\hat{\bm{x}}\times\mathcal{D}_{u\varphi}^{1}\bm{\varphi}^{i}\nonumber \\
 & -\frac{7a^{2}}{3d^{3}}\gamma^{T}V_{0}^{(2s)}\left(\mathbb{I}-3\hat{\bm{x}}\hat{\bm{x}}\right)\cdot\left(\bm{u}^{i+1}-2\bm{u}^{i}+\bm{u}^{i-1}\right),\label{eq:2s_micro_tr}\\
\gamma^{R}\dot{\boldsymbol{\varphi}}^{i} & =\lambda d\hat{\bm{x}}\times\left(\mathcal{D}_{\varphi u}^{1}\boldsymbol{u}^{i}+d\hat{\bm{x}}\times\mathcal{D}^{0}\bm{\varphi}^{i}\right)+\mu\mathcal{D}_{\varphi}^{2}\boldsymbol{\varphi}^{i}\nonumber \\
 & +\frac{7a^{2}}{2d^{4}}\gamma^{R}V_{0}^{(2s)}\hat{\bm{x}}\times\left(\bm{u}^{i+1}-\bm{u}^{i-1}\right)\nonumber \\
 & -\frac{7a^{2}}{2d^{3}}\gamma^{R}V_{0}^{(2s)}\left(\bm{\varphi}^{i+1}+\bm{\varphi}^{i-1}\right)_{\perp}.\label{eq:2s_micro_rot}
\end{align}
For a continuum active apolar, achiral rod, the active contributions
to the passive stresses and sources are given by
\begin{align*}
\boldsymbol{F}_{AP,AC}^{E} & =\tilde{F}_{\parallel}\boldsymbol{e}_{1}+C_{\parallel}^{F\varepsilon}\boldsymbol{\varepsilon}_{\parallel}+C_{\perp}^{F\varepsilon}\boldsymbol{\varepsilon}_{\perp},\\
\boldsymbol{M}_{AP,AC}^{E} & =C_{\parallel}^{M\tau}\boldsymbol{\tau}_{\parallel}+C_{\perp}^{M\tau}\boldsymbol{\tau}_{\perp},\\
\boldsymbol{f}_{AP,AC}^{E} & =H_{\circlearrowleft}^{f\tau}\boldsymbol{e}_{1}\times\boldsymbol{\tau},\\
\boldsymbol{m}_{AP,AC}^{E} & =H_{\circlearrowleft}^{m\varepsilon}\boldsymbol{e}_{1}\times\boldsymbol{\varepsilon},
\end{align*}
where we have used subscripts $\parallel$ and $\perp$ to denote
components of strain measures along and perpendicular to $\hat{\boldsymbol{x}}$.
We obtain the following equations:
\begin{align}
\Gamma^{T}\dot{\boldsymbol{u}} & =\left(\kappa_{T}\boldsymbol{\varepsilon}+C_{\parallel}^{F\varepsilon}\boldsymbol{\varepsilon}_{\parallel}+C_{\perp}^{F\varepsilon}\boldsymbol{\varepsilon}_{\perp}+\tilde{F}_{\parallel}\boldsymbol{\varphi}\times\hat{\boldsymbol{x}}\right)'\nonumber \\
 & +H_{\circlearrowleft}^{f\tau}\hat{\boldsymbol{x}}\times\boldsymbol{\tau},\label{eq:tr_cont_eom_2s}\\
\Gamma^{R}\dot{\boldsymbol{\varphi}} & =\left(\kappa_{R}\boldsymbol{\tau}+C_{\parallel}^{M\tau}\boldsymbol{\tau}_{\parallel}+C_{\perp}^{M\tau}\boldsymbol{\tau}_{\perp}\right)'\nonumber \\
 & +\hat{\boldsymbol{x}}\times\left(\tilde{F}_{\parallel}\boldsymbol{\varphi}\times\hat{\boldsymbol{x}}+C_{\parallel}^{F\varepsilon}\boldsymbol{\varepsilon}_{\parallel}+C_{\perp}^{F\varepsilon}\boldsymbol{\varepsilon}_{\perp}\right)\nonumber \\
 & +\boldsymbol{u}'\times\tilde{F}_{\parallel}\hat{\boldsymbol{x}}+\left(H_{\circlearrowleft}^{m\varepsilon}+\kappa_{T}\right)\hat{\boldsymbol{x}}\times\boldsymbol{\varepsilon}.\label{eq:rot_cont_eom_2s}
\end{align}
Note, in particular, the subtle effects of the prestress $\tilde{F}_{\parallel}$,
which contributes both to the linear and angular momentum equation
as an effective force or moment density proportional to strain.

To derive the inextensible and unshearable limit, we take $\kappa_{T}\to\infty$,
in which case $\boldsymbol{\varphi}_{\perp}$ is a fast variable and
we can neglect its time derivative in \eqref{eq:rot_cont_eom_2s},
and substitute a series expansion in powers of $1/\kappa_{T}$. We
find that, to leading order in $1/\kappa_{T}$,
\[
\boldsymbol{\varphi}_{\perp}=\hat{\boldsymbol{x}}\times\boldsymbol{u}_{\perp}'+\frac{\kappa_{R}+C_{\perp}^{M\tau}}{\kappa_{T}}\hat{\boldsymbol{x}}\times\boldsymbol{u}_{\perp}''',
\]
which we substitute back to the perpendicular part of \eqref{eq:tr_cont_eom_2s}.
The leading order equations (in $1/\kappa_{T}$) constitute the beam
limit of the dynamics, given in Eq. \eqref{eq:2s_beam_lim} of the
main text.

\textbf{Apolar, chiral chain. }Linearizing the discrete equations
of motion of an apolar, chiral chain about the straight configuration,
we obtain 
\begin{align}
\gamma^{T}\dot{\bm{u}}^{i} & =\lambda\mathcal{D}_{u}^{2}\bm{u}^{i}+\lambda d\hat{\bm{x}}\times\mathcal{D}_{u\varphi}^{1}\bm{\varphi}^{i}\nonumber \\
 & +\frac{13a^{3}}{12d^{4}}\gamma^{T}V_{0}^{(3a)}\hat{\bm{x}}\times\left(\bm{u}^{i+1}-\bm{u}^{i-1}\right)\nonumber \\
 & -\frac{13a^{3}}{12d^{3}}\gamma^{T}V_{0}^{(3a)}\left(\bm{\varphi}^{i+1}+\bm{\varphi}^{i-1}\right)_{\perp},\label{eq:3a_discr_tr}\\
\gamma^{R}\dot{\bm{\varphi}}^{i} & =\lambda d\hat{\bm{x}}\times\left(\mathcal{D}_{\varphi u}^{1}\boldsymbol{u}^{i}+d\hat{\bm{x}}\times\mathcal{D}^{0}\bm{\varphi}^{i}\right)+\mu\mathcal{D}_{\varphi}^{2}\boldsymbol{\varphi}^{i}\nonumber \\
 & -\frac{13a^{3}}{6d^{5}}\gamma^{R}V_{0}^{(3a)}\left(\mathbb{I}-3\hat{\bm{x}}\hat{\bm{x}}\right)\cdot\left(\bm{u}^{i+1}-2\bm{u}^{i}+\bm{u}^{i-1}\right)\nonumber \\
 & -\frac{13a^{3}}{12d^{4}}\gamma^{R}V_{0}^{(3a)}\hat{\bm{x}}\times\left(\bm{\varphi}^{i+1}-\bm{\varphi}^{i-1}\right).\label{eq:3a_discr_rot}
\end{align}
For a continuum active apolar and chiral chain rod, the active contributions
to the passive stresses and sources are given by
\begin{align*}
\boldsymbol{F}_{AP,C}^{E} & =C_{\parallel}^{F\tau}\boldsymbol{\tau}_{\parallel}+C_{\perp}^{F\tau}\boldsymbol{\tau}_{\perp},\\
\boldsymbol{M}_{AP,C}^{E} & =\tilde{M}_{\parallel}\boldsymbol{e}_{1}+C_{\parallel}^{M\varepsilon}\boldsymbol{\varepsilon}_{\parallel}+C_{\perp}^{M\varepsilon}\boldsymbol{\varepsilon}_{\perp},\\
\boldsymbol{f}_{AP,C}^{E} & =H_{\circlearrowleft}^{f\varepsilon}\boldsymbol{e}_{1}\times\boldsymbol{\varepsilon},\\
\boldsymbol{m}_{AP,C}^{E} & =H_{\circlearrowleft}^{m\tau}\boldsymbol{e}_{1}\times\boldsymbol{\tau}.
\end{align*}
We obtain the following equations of motion:
\begin{align}
\Gamma^{T}\dot{\boldsymbol{u}} & =\left(\kappa_{T}\boldsymbol{\varepsilon}+C_{\parallel}^{F\tau}\boldsymbol{\tau}_{\parallel}+C_{\perp}^{F\tau}\boldsymbol{\tau}_{\perp}\right)'+H_{\circlearrowleft}^{f\varepsilon}\hat{\boldsymbol{x}}\times\boldsymbol{\varepsilon},\label{eq:tr_cont_eom_3a}\\
\Gamma^{R}\dot{\boldsymbol{\varphi}} & =\left(\kappa_{R}\boldsymbol{\tau}+\tilde{M}_{\parallel}\boldsymbol{\varphi}\times\hat{\boldsymbol{x}}+C_{\parallel}^{M\varepsilon}\boldsymbol{\varepsilon}_{\parallel}+C_{\perp}^{M\varepsilon}\boldsymbol{\varepsilon}_{\perp}\right)'\nonumber \\
 & +\hat{\boldsymbol{x}}\times\left(C_{\parallel}^{F\tau}\boldsymbol{\tau}_{\parallel}+C_{\perp}^{F\tau}\boldsymbol{\tau}_{\perp}+\kappa_{T}\boldsymbol{\varepsilon}\right)+H_{\circlearrowleft}^{m\tau}\hat{\boldsymbol{x}}\times\boldsymbol{\tau}.\nonumber 
\end{align}
Note, in particular, that the pre-torque $\tilde{M}_{\parallel}$
gives rise to a moment density proportional to the rotational strain.

The derivation of the inextensible and unshearable limit proceeds
in the same way as before. For large $\kappa_{T}$, we find that,
to leading order in $1/\kappa_{T}$,
\[
\boldsymbol{\varphi}_{\perp}=\hat{\boldsymbol{x}}\times\boldsymbol{u}_{\perp}'+\frac{\left(\tilde{M}_{\parallel}-H_{\circlearrowleft}^{m\tau}-C_{\perp}^{F\tau}\right)\boldsymbol{u}_{\perp}''+\kappa_{R}\hat{\boldsymbol{x}}\times\boldsymbol{u}_{\perp}'''}{\kappa_{T}}.
\]
Upon substituting this into Eq. \ref{eq:tr_cont_eom_3a}, we find
that in the beam limit, the leading effect of nonreciprocal activity
is a term proportional to $\hat{\boldsymbol{x}}\times\boldsymbol{u}_{\perp}'''$,
as shown in Eq. \eqref{eq:3a_beam_lim} of the main text. Special
care needs to be taken to find the correct rate in Eq. \eqref{eq:3a_trans_ac_disp},
as from the continuum limit of \eqref{eq:3a_discr_tr}, we get an
additional term proportional to $\hat{\boldsymbol{x}}\times\boldsymbol{u}_{\perp}'''$
when Taylor expanding the discrete force proportional to $\hat{\bm{x}}\times\left(\bm{u}^{i+1}-\bm{u}^{i-1}\right)$,
leading to Eq. \eqref{eq:3a_trans_ac_disp} of the main text.\begingroup\tabcolsep=4pt\def\arraystretch{2}
\begin{table*}
\centering
\begin{tabular}{|c||c|c|c|c|}
\hline 
 & \multicolumn{2}{c|}{Achiral} & \multicolumn{2}{c|}{Chiral}\tabularnewline
\hline 
\hline 
\multirow{4}{*}{Apolar} & \multicolumn{2}{c|}{$\tilde{F}_{\parallel}=-\frac{7a^{2}}{3d}\Gamma^{T}V_{0}^{(2s)}$} & \multicolumn{2}{c|}{$\tilde{M}_{\parallel}=\frac{13a^{3}}{12d^{3}}\Gamma^{R}V_{0}^{(3a)}$}\tabularnewline
\cline{2-5}
 & $C_{\parallel}^{F\varepsilon}=\frac{14a^{2}}{3d}\Gamma^{T}V_{0}^{(2s)}$ & $C_{\parallel}^{M\tau}=0$ & $C_{\parallel}^{F\tau}=0$ & $C_{\parallel}^{M\varepsilon}=\frac{13a^{3}}{3d^{3}}\Gamma^{R}V_{0}^{(3a)}$\tabularnewline
\cline{2-5}
 & $C_{\perp}^{F\varepsilon}=-\frac{7a^{2}}{3d}\Gamma^{T}V_{0}^{(2s)}$ & $C_{\perp}^{M\tau}=-\frac{7a^{2}}{2d}\Gamma^{R}V_{0}^{(2s)}$ & $C_{\perp}^{F\tau}=-\frac{13a^{3}}{12d}\Gamma^{T}V_{0}^{(3a)}$ & $C_{\perp}^{M\varepsilon}=-\frac{13a^{3}}{6d^{3}}\Gamma^{R}V_{0}^{(3a)}$\tabularnewline
\cline{2-5}
 & $H_{\circlearrowleft}^{f\tau}=0$ & $H_{\circlearrowleft}^{m\varepsilon}=\frac{7a^{2}}{d^{3}}\Gamma^{R}V_{0}^{(2s)}$ & $H_{\circlearrowleft}^{f\varepsilon}=\frac{13a^{3}}{6d^{3}}\Gamma^{T}V_{0}^{(3a)}$ & $H_{\circlearrowleft}^{m\tau}=\frac{13a^{3}}{12d}\Gamma^{T}V_{0}^{(3a)}$\tabularnewline
\hline 
\multirow{4}{*}{Polar} & \multicolumn{2}{c|}{$\tilde{f}_{\parallel}=\frac{4a^{3}}{5d^{3}}\Gamma^{T}V_{0}^{(3t)}$} & \multicolumn{2}{c|}{$\tilde{m}_{\parallel}=-\frac{363a^{4}}{5d^{5}}\Gamma^{R}V_{0}^{(4a)}$}\tabularnewline
\cline{2-5}
 & $H_{\parallel}^{f\varepsilon}=-\frac{12a^{3}}{5d^{3}}\Gamma^{T}V_{0}^{(3t)}$ & $H_{\parallel}^{m\tau}=0$ & $H_{\parallel}^{f\tau}=0$ & $H_{\parallel}^{m\varepsilon}=\frac{363a^{4}}{d^{5}}\Gamma^{R}V_{0}^{(4a)}$\tabularnewline
\cline{2-5}
 & $C_{\circlearrowleft}^{F\tau}=\frac{a^{3}}{5d}\Gamma^{T}V_{0}^{(3t)}$ & $C_{\circlearrowleft}^{M\varepsilon}=0$ & $C_{\circlearrowleft}^{F\varepsilon}=\frac{363a^{4}}{10d^{3}}\Gamma^{T}V_{0}^{(4a)}$ & $C_{\circlearrowleft}^{M\tau}=\frac{-1089a^{4}}{20d^{3}}\Gamma^{R}V_{0}^{(4a)}$\tabularnewline
\cline{2-5}
 & $H_{\perp}^{f\varepsilon}=\frac{6a^{3}}{5d^{3}}\Gamma^{T}V_{0}^{(3t)}$ & $H_{\perp}^{m\tau}=\frac{a^{3}}{5d}\Gamma^{T}V_{0}^{(3t)}$ & $H_{\perp}^{f\tau}=-\frac{363a^{4}}{10d^{3}}\Gamma^{T}V_{0}^{(4a)}$ & $H_{\perp}^{m\varepsilon}=\frac{363a^{4}}{10d^{3}}\Gamma^{T}V_{0}^{(4a)}$\tabularnewline
\hline 
\end{tabular}\caption{Effective elastic moduli from active forces and torques.\label{tab:moduli}}
\end{table*}
\endgroup

\textbf{Polar, achiral chain. }As a polar, achiral chain experiences
a net force, the shape of the reference configuration is still the
straight chain, but it is uniformly translating along its axis (which
we still take to be the $x$ axis). Linearizing the discrete equations
of motion about the translating reference configuration $\left(\tilde{\bm{r}}^{i},\tilde{\bm{e}}_{a}^{i}\right)$
yields
\begin{align}
\begin{split}\gamma^{T}\dot{\bm{u}}^{i} & =\lambda\mathcal{D}_{u}^{2}\bm{u}^{i}+\lambda d\hat{\bm{x}}\times\mathcal{D}_{u\varphi}^{1}\bm{\varphi}^{i}\\
 & +\frac{3a^{3}}{5d^{4}}\gamma^{T}V_{0}^{(3t)}\left(\mathbb{I}-3\hat{\bm{x}}\hat{\bm{x}}\right)\cdot\left(\bm{u}^{i+1}-\bm{u}^{i-1}\right)\\
 & +\frac{a^{3}}{5d^{3}}\gamma^{T}V_{0}^{(3t)}\hat{\bm{x}}\times\left(\bm{\varphi}^{i+1}+\bm{\varphi}^{i-1}\right),
\end{split}
\label{eq:3t_discr_tr}\\
\begin{split}\gamma^{R}\dot{\bm{\varphi}}^{i} & =\lambda d\hat{\bm{x}}\times\left(\mathcal{D}_{\varphi u}^{1}\boldsymbol{u}^{i}+d\hat{\bm{x}}\times\mathcal{D}^{0}\bm{\varphi}^{i}\right)+\mu\mathcal{D}_{\varphi}^{2}\boldsymbol{\varphi}^{i}\end{split}
\label{eq:3t_discr_rot}
\end{align}
For a continuum active polar and achiral chain, the active contributions
to the passive stresses and sources are given by
\begin{align*}
\boldsymbol{F}_{P,AC}^{E} & =C_{\circlearrowleft}^{F\tau}\boldsymbol{e}_{1}\times\boldsymbol{\tau},\\
\boldsymbol{M}_{P,AC}^{E} & =C_{\circlearrowleft}^{M\varepsilon}\boldsymbol{e}_{1}\times\boldsymbol{\varepsilon},\\
\boldsymbol{f}_{P,AC}^{E} & =\tilde{f}_{\parallel}\boldsymbol{e}_{1}+H_{\parallel}^{f\varepsilon}\boldsymbol{\varepsilon}_{\parallel}+H_{\perp}^{f\varepsilon}\boldsymbol{\varepsilon}_{\perp},\\
\boldsymbol{m}_{P,AC}^{E} & =H_{\parallel}^{m\tau}\boldsymbol{\tau}_{\parallel}+H_{\perp}^{m\tau}\boldsymbol{\tau}_{\perp}.
\end{align*}
The steady state velocity can be computed from force balance $\Gamma^{T}\boldsymbol{V}^{(0)}=\bar{f}_{\parallel}\hat{\boldsymbol{x}}$.
The linearized equations of motion are given by:
\begin{align*}
\Gamma^{T}\dot{\boldsymbol{u}} & =\left(\kappa_{T}\boldsymbol{\varepsilon}+C_{\circlearrowleft}^{F\tau}\hat{\boldsymbol{x}}\times\boldsymbol{\tau}\right)'+\tilde{f}_{\parallel}\boldsymbol{\varphi}\times\hat{\boldsymbol{x}}+H_{\parallel}^{f\varepsilon}\boldsymbol{\varepsilon}_{\parallel}+H_{\perp}^{f\varepsilon}\boldsymbol{\varepsilon}_{\perp},\\
\Gamma^{R}\dot{\boldsymbol{\varphi}} & =\left(\kappa_{R}\boldsymbol{\tau}+C_{\circlearrowleft}^{M\varepsilon}\hat{\boldsymbol{x}}\times\boldsymbol{\varepsilon}\right)'+\hat{\boldsymbol{x}}\times\left(C_{\circlearrowleft}^{F\tau}\hat{\boldsymbol{x}}\times\boldsymbol{\tau}\right)\\
 & +H_{\parallel}^{m\tau}\boldsymbol{\tau}_{\parallel}+H_{\perp}^{m\tau}\boldsymbol{\tau}_{\perp}+\kappa_{T}\hat{\boldsymbol{x}}\times\boldsymbol{\varepsilon}.
\end{align*}
The expressions for the strain measures are still given by Eq. \eqref{eq:rod_strain_lin},
as in the reference configuration the chain is globally translating
in a straight configuration without deformation. Note, in particular,
that the term $\tilde{f}_{\parallel}$ induces a force proportional
to the rotational displacement $\boldsymbol{\varphi}$ itself.

The derivation of the inextensible and unshearable limit proceeds
in the same way as before. For large $\kappa_{T}$, we find that,
to leading order in $1/\kappa_{T}$,
\[
\boldsymbol{\varphi}_{\perp}=\hat{\boldsymbol{x}}\times\boldsymbol{u}_{\perp}'+\frac{\left(H_{\perp}^{m\tau}-C_{\circlearrowleft}^{F\tau}\right)\hat{\boldsymbol{x}}\times\boldsymbol{u}_{\perp}''+\kappa_{R}\hat{\boldsymbol{x}}\times\boldsymbol{u}_{\perp}'''}{\kappa_{T}}.
\]
The leading order equations (in $1/\kappa_{T}$) constitute the beam
limit of the dynamics, given in Eq. \eqref{eq:3t_beam_lim} of the
main text.

\textbf{Polar, chiral chain}. As a polar, chiral chain experiences
a net torque, the shape of the reference configuration is still the
straight chain, but it will be uniformly rotating along its axis (which
we still take to be the $x$-axis). Linearizing the discrete equations
of motion about the rotating reference configuration $\left(\tilde{\boldsymbol{r}}^{i},\tilde{\bm{e}}_{a}^{i}\right)$,
we obtain the following equations of motion
\begin{align}
\gamma^{T}\dot{\bm{u}}^{i} & =\lambda\mathcal{D}_{u}^{2}\bm{u}^{i}+\lambda d\hat{\bm{x}}\times\mathcal{D}_{u\varphi}^{1}\bm{\varphi}^{i}\nonumber \\
 & +\frac{363a^{4}}{10d^{5}}\gamma^{T}V_{0}^{(4a)}\hat{\bm{x}}\times\left(\bm{u}^{i+1}-2\bm{u}^{i}+\bm{u}^{i-1}\right)\nonumber \\
 & -\frac{363a^{4}}{10d^{4}}\gamma^{T}V_{0}^{(4a)}\left(\bm{\varphi}^{i+1}-\bm{\varphi}^{i-1}\right)_{\perp},\label{eq:4a_discr_tr}\\
\gamma^{R}\dot{\bm{\varphi}}^{i} & =\lambda d\hat{\bm{x}}\times\left(\mathcal{D}_{\varphi u}^{1}\boldsymbol{u}^{i}+d\hat{\bm{x}}\times\mathcal{D}^{0}\bm{\varphi}^{i}\right)\nonumber \\
 & +\mu\mathcal{D}_{\varphi}^{2}\boldsymbol{\varphi}^{i}+\gamma^{R}\bm{\Omega}^{(0)}\times\boldsymbol{\varphi}^{i}\\
 & -\frac{363a^{4}}{4d^{6}}\gamma^{R}V_{0}^{(4a)}\left(\mathbb{I}-3\hat{\bm{x}}\hat{\bm{x}}\right)\cdot\left(\bm{u}^{i+1}-\bm{u}^{i-1}\right)\nonumber \\
 & -\frac{1089a^{4}}{20d^{5}}\gamma^{R}V_{0}^{(4a)}\hat{\bm{x}}\times\left(\bm{\varphi}^{i+1}+\bm{\varphi}^{i-1}\right).\label{eq:4a_discr_rot}
\end{align}
\textbf{ }For an active polar and chiral chain, the active contributions
to the passive stresses, moment stresses, forces and torques are given
by
\begin{align*}
\boldsymbol{F}_{P,C}^{E} & =C_{\circlearrowleft}^{F\varepsilon}\boldsymbol{e}_{1}\times\boldsymbol{\varepsilon},\\
\boldsymbol{M}_{P,C}^{E} & =C_{\circlearrowleft}^{M\tau}\boldsymbol{e}_{1}\times\boldsymbol{\tau},\\
\boldsymbol{f}_{P,C}^{E} & =H_{\parallel}^{f\tau}\boldsymbol{\tau}_{\parallel}+H_{\perp}^{f\tau}\boldsymbol{\tau}_{\perp},\\
\boldsymbol{m}_{P,C}^{E} & =\tilde{m}_{\parallel}\boldsymbol{e}_{1}+H_{\parallel}^{m\varepsilon}\boldsymbol{\varepsilon}_{\parallel}+H_{\perp}^{m\varepsilon}\boldsymbol{\varepsilon}_{\perp}.
\end{align*}
The steady state angular velocity can be computed from moment balance
$\Gamma^{R}\boldsymbol{\Omega}^{(0)}=\tilde{m}_{\parallel}\hat{\boldsymbol{x}}$.
The linearized equations of motion are given by:
\begin{align*}
\Gamma^{T}\dot{\boldsymbol{u}} & =\left(\kappa_{T}\boldsymbol{\varepsilon}+C_{\circlearrowleft}^{F\varepsilon}\hat{\boldsymbol{x}}\times\boldsymbol{\varepsilon}\right)'+H_{\parallel}^{f\tau}\boldsymbol{\tau}_{\parallel}+H_{\perp}^{f\tau}\boldsymbol{\tau}_{\perp},\\
\Gamma^{R}\dot{\boldsymbol{\varphi}} & =\left(\kappa_{R}\boldsymbol{\tau}+C_{\circlearrowleft}^{M\tau}\hat{\boldsymbol{x}}\times\boldsymbol{\tau}\right)'+\Gamma^{R}\left(\boldsymbol{\Omega}^{(0)}\times\boldsymbol{\varphi}\right)\\
 & +\hat{\boldsymbol{x}}\times\left(C_{\circlearrowleft}^{F\varepsilon}\hat{\boldsymbol{x}}\times\boldsymbol{\varepsilon}\right)+\tilde{m}_{\parallel}\boldsymbol{\varphi}\times\hat{\boldsymbol{x}}\\
 & +H_{\parallel}^{m\varepsilon}\boldsymbol{\varepsilon}_{\parallel}+H_{\perp}^{m\varepsilon}\boldsymbol{\varepsilon}_{\perp}+\kappa_{T}\hat{\boldsymbol{x}}\times\boldsymbol{\varepsilon}.
\end{align*}
The expressions for the strain measures are still given by Eq. \eqref{eq:rod_strain_lin},
as in the reference configuration the chain is globally rotating in
a straight configuration without deformation. Interestingly, the terms
involving $\boldsymbol{\Omega}^{(0)}$ on the left hand side and $\tilde{m}_{\parallel}$
of the on the right hand side of the moment balance equation cancel.

The derivation of the inextensible and unshearable limit proceeds
in the same way as before. For large $\kappa_{T}$, we find that,
to leading order in $1/\kappa_{T}$,
\[
\boldsymbol{\varphi}_{\perp}=\hat{\boldsymbol{x}}\times\boldsymbol{u}_{\perp}'+\frac{-C_{\circlearrowleft}^{M\tau}\boldsymbol{u}_{\perp}'''+\kappa_{R}\hat{\boldsymbol{x}}\times\boldsymbol{u}_{\perp}'''}{\kappa_{T}}.
\]
The leading order equations (in $1/\kappa_{T}$) constitute the beam
limit of the dynamics, given in Eq. \eqref{eq:4a_beam_lim} of the
main text.

\section*{Appendix D: Details of numerical simulations}

To verify our theoretical predictions, we have numerically simulated
the discrete equations of motion Eqs. \eqref{eq:pos_eom}-\eqref{eq:rot_eom}
in the overdamped limit. To maintain the orthonormality of frames
at all times, we have employed a geometric Lie group integrator \citep{iserles_lie-group_2000}
the special Euclidean group $\mathrm{SE}\left(3\right)$: we have
parametrized the configuration of each squirmer using the Cayley map.
As opposed to the original equations in terms of the frame vectors,
the Cayley-transformed equations of motion live in a vector space
(the Lie algebra of the group), hence during their numerical integration,
errors will be accrued in the Lie algebra and the orthonormality constraint
will automatically be preserved.

For our simulations, we have chosen units in which the radii of the
squirmers are $a=1$ and the viscosity of the surrounding fluid $\eta=1$.
We have fixed $d=10$ and chose the discrete bending modulus to be
$\mu=\gamma^{T}d^{2}$. Since we are interested in the beam limit
$\kappa_{T}\to\infty$, we have chosen $\lambda$ to be large enough,
as follows. For a chain of $N$ squirmers of length $L=\left(N-1\right)d$,
we chose $\lambda$ such that
\[
\frac{\kappa_{T}\big/\Gamma^{R}}{\kappa_{R}\big/\Gamma^{T}L^{4}}=\frac{\lambda d^{2}\big/\gamma^{R}}{\mu d^{2}\big/\gamma^{T}d^{4}\left(N-1\right)^{4}}\gg1,
\]
i.e., the ratio of the relaxation timescales of shear and stretch
modes is much greater than unity. We set this ratio to $10^{7}$ to
produce simulation snapshots of the chain and $10^{12}$ to produce
the plots of dispersion relations. As the existence of such vastly
different timescales make our system stiff, we used an implicit BDF
scheme to integrate the equations. In the bulk of the chain, the passive
elastic restoring forces and torques were computed by differentiating
the potential in Eq. \eqref{eq:discr_pot}, while at the boundaries,
we have used two-body potentials of the form $\lambda_{\parallel}d^{2}\left[\left(\boldsymbol{e}_{1}^{i}+\boldsymbol{e}_{1}^{i+1}\right)/2\cdot\left(\boldsymbol{r}_{i+1}-\boldsymbol{r}_{i}\right)/d-1\right]^{2}$
and $\sum_{a=1}^{2}\lambda_{\perp}d^{2}\left[\left(\boldsymbol{e}_{a}^{i}+\boldsymbol{e}_{a}^{i+1}\right)/2\cdot\left(\boldsymbol{r}_{i+1}-\boldsymbol{r}_{i}\right)/d\right]^{2}$
to penalize stretch and shear, respectively. The active forces and
torques were obtained using the \textsc{PyStokes} library \citep{singh_pystokes_2020}.

To measure the dispersion relations numerically, we looked at varying
chain lengths with fixed $d$. We started the chain from an initial
condition that was either a plane sine wave or a helix of amplitude
$a/10$ and wavenumber $q=\pi/L$. The frames were initially aligned
with the Frenet-Serret framings of the curves. After an initial transient,
the fast shear and stretch modes decayed, and then to the shape of
the chain we fitted either a growing, decaying, traveling or rotating
sine wave or helix depending on the symmetry of the mode, giving us
either the growth rate or the phase velocity of the collective mode.
By increasing the number of squirmers (the length of the chain), we
were able to probe longer wavelengths.

In Fig. 4., panels (d) and (e), we set $V_{0}^{(2s)}=\pm100$, in
Fig. 5., panels (d) and (e) we set $V_{0}^{(3a)}=\pm5000$ , in Fig.
6., panel (d) we set $V_{0}^{(3t)}=1000$ and in Fig. 7., panels (d)-(e)
we set $V_{0}^{(4a)}=\pm10000$. Visualization of chains were produced
using the \textsc{vedo} library in Python \citep{marco_musy_marcomusyvedo_2025}.
\end{document}